\documentclass{elsart}
\usepackage{epsfig}
\usepackage{rotate}
\usepackage{amssymb}
\usepackage{amsmath}

\newcommand{\vo}{\mbox{$V^0$}}
\newcommand{\lam}{\mbox{$\rm \Lambda$}}
\newcommand{\lamdecay}{\mbox{$\rm \Lambda \to p \pi^-$}}
\newcommand{\alam}{\mbox{$\rm \bar \Lambda$}}
\newcommand{\alamdecay}{\mbox{$\rm \bar \Lambda \to \bar p \pi^+$}}
\newcommand{\ko}{\mbox{$\rm K^0_s$}}
\newcommand{\kodecay}{\mbox{$\rm K^0_s \to \pi^+ \pi^-$}}

\newcommand{\gam}{\mbox{$\rm \gamma$}}

\begin{document}
\begin{frontmatter}
\title{\boldmath 
A Study of Strange Particle Production \\
in $\nu_\mu$ Charged Current Interactions \\
in the NOMAD Experiment}
\centerline{\bf NOMAD Collaboration}
\vskip 0.5cm

\author[Paris]             {P.~Astier}
\author[CERN]              {D.~Autiero}
\author[Saclay]            {A.~Baldisseri}
\author[Padova]            {M.~Baldo-Ceolin}
\author[Paris]             {M.~Banner}
\author[LAPP]              {G.~Bassompierre}
\author[Lausanne]          {K.~Benslama}
\author[Saclay]            {N.~Besson}
\author[CERN,Lausanne]     {I.~Bird}
\author[Johns Hopkins]     {B.~Blumenfeld}
% ...

\address[LAPP]           {LAPP, Annecy, France}                               
\address[Johns Hopkins]  {Johns Hopkins Univ., Baltimore, MD, USA}            
% ...
\address[CERN]           {CERN, Geneva, Switzerland}                          
\address[Lausanne]       {University of Lausanne, Lausanne, Switzerland}      
% ...
\address[Padova]         {Univ. of Padova and INFN, Padova, Italy}            
\address[Paris]          {LPNHE, Univ. of Paris VI and VII, Paris, France}    
% ...
\address[Saclay]         {DAPNIA, CEA Saclay, France}                         
% ...

\thanks[Perugia]         {Now at Univ. of Perugia and INFN, Perugia, Italy}

\clearpage
\begin{abstract}

A study of strange particle production in $\nu_\mu$ charged current interactions 
has been performed using the data from the NOMAD experiment. 
Yields of neutral strange particles 
($\ko$, $\lam$, $\alam$) have been measured. Mean
multiplicities are reported as a function of the event kinematic variables
$E_\nu$, $W^2$ and $Q^2$ as well as of the variables describing 
particle behaviour within a hadronic jet: $x_F$, $z$ and $p_T^2$.
Decays of resonances and heavy hyperons with identified 
$\ko$ and $\lam$ in the final state have been analyzed. 
Clear signals corresponding to $\rm {K^\star}^\pm$, $\rm {\Sigma^\star}^\pm$,
$\rm \Xi^-$ and $\rm \Sigma^0$ have been observed.

\end{abstract}
\end{frontmatter}

\section{INTRODUCTION\label{sec:intro}}

The production of strange particles in neutrino interactions 
can provide a testing ground for the 
quark-parton as well as for hadronization models. 
Neutral strange particles can be reliably identified 
using the $\vo$-like signature of their decays 
($\kodecay$, $\lamdecay$ and $\alamdecay$) in contrast to most
other hadrons which require particle identification hardware.
It is noteworthy that
all previous investigations of strange particle production
by neutrinos have come from bubble chamber 
experiments~\cite{Barish,Deden,Berge,Deden2,Berge2,Erriquez,ammosov,Baker,ammosov2,Brock,grassler,Bosetti,son1,allasia,chang,son2,allasia2,Jones,FNAL,mann,ammosov3,ammosov4,Willocq,BEBC,Prospo}.
No other technique has so far yielded results on this subject.
However, previous bubble chamber experiments with (anti)neutrino beams suffered 
from the low statistics of their $\vo$ samples.

The NOMAD experiment~\cite{NOMAD_NIM} has collected a large number of neutrino interactions 
with a reconstruction quality similar to that of bubble chamber experiments. The order of magnitude 
increase in statistics can be used to improve our knowledge of strange 
particle production in neutrino interactions. In this paper 
we present measurements of the yields
of neutral strange particles ($\ko$, $\lam$ and $\alam$),
as well as the yields of $\rm {K^\star}^\pm$,
$\rm {\Sigma^\star}^\pm$,
$\rm \Xi^-$ and
$\rm \Sigma^0$
in $\nu_\mu$ charged current (CC) interactions.
These results are compared to the predictions of the NOMAD 
Monte Carlo (MC) simulation program.

The results of the present analysis 
are the measurements of:
\begin{enumerate}
\item 
the production properties of neutral strange particles 
in $\nu_\mu$ CC interactions. This study will allow
tuning of the parameters of the MC simulation
programs in order to correctly reproduce the production of strange particles 
by neutrinos;
\item 
the contribution 
of
strange resonances 
and heavy hyperons 
to the total number of observed $\ko$, $\lam$ and $\alam$. This will allow
a quantitative theoretical interpretation of the $\lam$ and $\alam$ polarization
measurements in $\nu_\mu$ CC deep inelastic
scattering (DIS) reported in our previous articles~\cite{NOMAD_lam,NOMAD_alam}.
\end{enumerate}

\section{EXPERIMENTAL PROCEDURE} 
\label{sec:identification}

\subsection{The NOMAD experiment} \label{sec:nomad}
 
The main goal of the NOMAD experiment~\cite{NOMAD_NIM} was the search for
$\nu_\mu \rightarrow \nu_\tau$ oscillations in a wide-band neutrino
beam from the CERN SPS. The main characteristics of the beam are given in 
Table~\ref{tab:beam_info}. This search uses kinematic criteria 
to identify $\nu_\tau$ CC interactions~\cite{NOMAD_OSC} 
and requires a very good quality of event reconstruction similar to that 
of bubble chamber experiments. This has indeed been achieved by 
the NOMAD detector, and, moreover, the large data sample collected
during four years of data taking (1995-1998) allows for a detailed
study of neutrino interactions. The full data sample from the NOMAD
experiment 
corresponding to about 1.3 million $\nu_\mu$ CC interactions in the detector fiducial volume
is used in the present analysis.
The data are compared to the results
of a Monte Carlo simulation based on modified versions of LEPTO 6.1~\cite{LEPTO} and
JETSET 7.4~\cite{JETSET} generators for neutrino interactions 
(with $Q^2$ and $W^2$ cutoff parameters removed) and 
on a GEANT~\cite{GEANT} based program for the detector response. 
Strange particle production is described by the set of default
parameters in JETSET.
To define the parton content of the nucleon for the cross-section calculation
we have used the GRV-HO parametrization~\cite{GRV} of the parton
density functions available in PDFLIB~\cite{PDFLIB}. 
The above description of the MC will be referred to as the default MC.
For the analysis reported below we used a MC sample consisting of
about 3 million events.
\begin{table}[htb]
\begin{center}
\caption{\it The CERN SPS neutrino beam composition at the position of
  the NOMAD detector (as predicted by the beam simulation program~\cite{nomad_beam}).
}
\vspace*{0.5cm}
\begin{tabular}{||c|c|c|c|c||}
\hline
\hline
\multicolumn{1}{||c|}{Neutrino} & \multicolumn{2}{c}{Flux} 
& \multicolumn{2}{|c||}{CC interactions in NOMAD} \\
\cline{2-5}
flavours & $\langle E_\nu \rangle$ [GeV] & rel.abund. & $\langle E_\nu \rangle$ [GeV] & rel.abund. \\
\hline
\hline
$\nu_\mu$         & 24.2 & 1    & 45.3 &  1 \\
$\bar{\nu}_\mu$ &  18.5 &  0.0637  &  40.9 &  0.0244 \\
$\nu_e$          &  36.6 &  0.0102  &  57.1 &  0.0153 \\
$\bar{\nu}_e$  & 28.7 &  0.0025 & 49.9 &  0.0015 \\
\hline
\hline
\end{tabular}
\label{tab:beam_info}
\end{center}
\end{table}

\subsection{The NOMAD detector} 

\begin{figure}[h]
\begin{center}
   \mbox{
     \epsfig{file=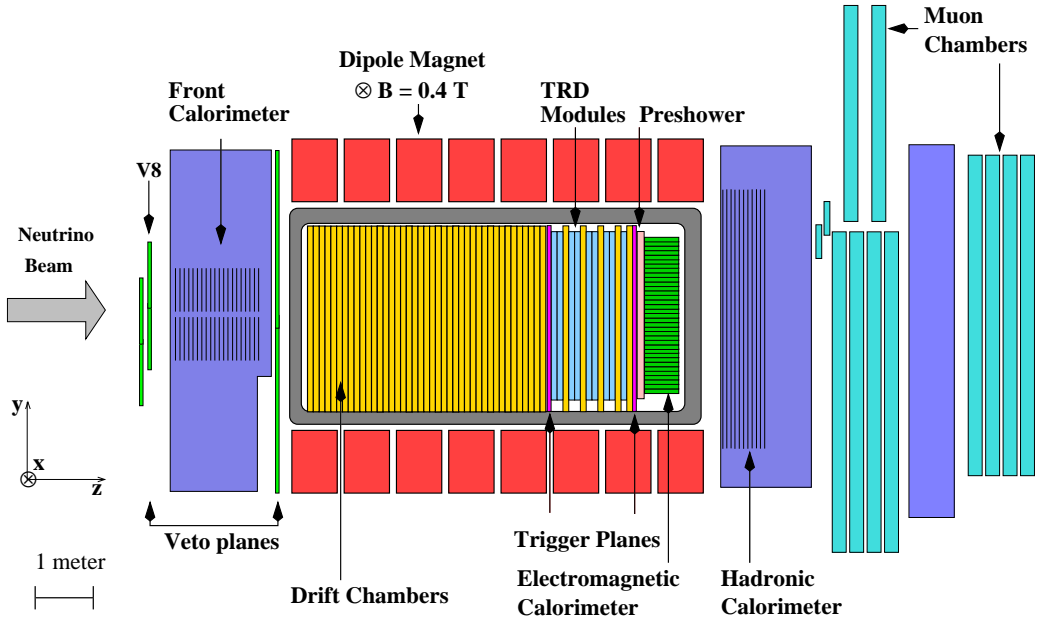,width=1.0\linewidth}}
     \caption{A sideview of the NOMAD detector.}
      \label{fig:nomad_detector}
   \end{center}
\end{figure}

For a study of strange particle production the tracking capabilities of a detector are 
of paramount importance. The NOMAD detector (Fig.~\ref{fig:nomad_detector})
is especially well suited for this. It consists of an active target of 
44 drift chambers, with a total fiducial mass of 2.7~tons, located in
a 0.4~Tesla dipole magnetic field. 
The drift chambers (DC)~\cite{NOMAD_NIM_DC}, made of low $Z$ material (mainly carbon)
serve the dual role of a nearly isoscalar target for neutrino interactions
and of the tracking medium. The average density of the drift chamber volume is 
0.1~$\mbox{g}/\mbox{cm}^3$, very close to that of liquid hydrogen. 
These drift chambers provide an overall efficiency 
for charged particle reconstruction of better than 95\% and a momentum 
resolution which can be parametrized as
$$
\frac{\sigma_p}{p} = \frac{0.05}{\sqrt{L}}\oplus \frac{0.008 \cdot p}{\sqrt{L^5}},
$$
where the track length $L$ is in metres and the track momentum $p$
is in $\rm GeV/c$. This amounts to a resolution of approximately 3.5\% 
in the momentum range of interest (less than 10~$\mbox{GeV}/\mbox{c}$).
Reconstructed tracks are used to determine the event topology 
(the assignment of tracks to vertices), to reconstruct the vertex position and 
the track parameters at each vertex and, finally, to
identify the vertex type (primary, secondary, $\vo$, etc.).
A transition radiation detector~\cite{NOMAD_TRD} is used 
for electron identification.
The pion rejection achieved for isolated tracks is $10^3$ with 
a 90\% electron identification efficiency.
A lead-glass electromagnetic calorimeter~\cite{NOMAD_ECAL} located
downstream of the tracking region provides 
an energy resolution of $3.2\%/\sqrt{E \mbox{[GeV]} } \oplus 1\%$
for electromagnetic showers and is essential to measure 
the total energy flow in neutrino interactions.
In addition, an iron absorber and a set of muon chambers located after 
the electromagnetic calorimeter are used for muon
identification, providing a muon detection 
efficiency of 97\% for momenta greater than 5~GeV/c.

The large statistics of the data combined with 
the good quality of event reconstruction in the NOMAD detector allows 
a detailed study of strange particle production in neutrino
interactions to be performed.

\subsection{Event selection and $\vo$ identification procedure}

\begin{figure}[htb]
\begin{center}
\epsfig{file=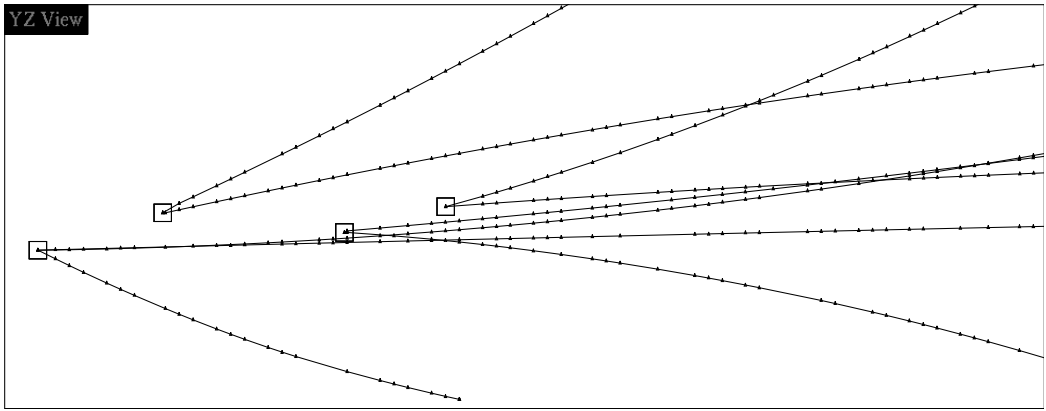,angle=-90,width=1.0\linewidth}
\protect\caption{\it A reconstructed data event containing 
3 $\vo$ vertices identified as $\ko$ decays 
by our identification procedure.
The scale on this plot is given by the size of the vertex boxes  
($3\times3$ cm$^2$).
}
\label{fig:lambda-antilambda}
\end{center}
\end{figure}

The NOMAD experiment has collected $1.3 \times 10^6$ $\nu_\mu$ CC events 
and has observed an unprecedented number of neutral strange particle decays. 
Such a decay appears in the detector as a $\vo$-like vertex: two tracks of opposite charge
emerging from a common vertex separated from the primary neutrino interaction 
vertex (Fig.~\ref{fig:lambda-antilambda}).
The $\vo$-like signature is expected also for photon conversions. 

The selection procedure for the $\nu_\mu$ CC event sample used in this analysis
has been described in~\cite{NOMAD_lam}. 

Since the NOMAD detector is unable to distinguish (anti)protons 
from pions in the momentum range relevant to this analysis,
our $\vo$ identification procedure relies on the kinematic properties of 
a $\vo$ decay.

For the $\vo$ identification a kinematic fit method has been used 
as described in~\cite{NOMAD_lam,Dima_thesis}.
This fit has been performed for three decay hypotheses: 
$\kodecay$, $\lamdecay$, $\alamdecay$ and for 
the hypothesis of a photon conversion to $\rm e^+ e^-$.
The output of the kinematic fits applied to a given $\vo$ vertex
consists of four values of $\chi^2_{V^0}$ describing the goodness of
these fits. Different
regions in the four-dimensional $\chi^2_{V^0}$ space
populated by particles identified as $\ko$, $\lam$ and $\alam$
have been selected. 
Identified $\vo$ are of two types:
\begin{itemize}
\item {\em uniquely} identified $\vo$, which, in 
the four-dimensional $\chi^2_{V^0}$ space described above, populate regions 
in which decays of only a single particle type are present;
\item {\em ambiguously} identified $\vo$, which populate 
regions in which 
decays of different particle types are present.
\end{itemize}
The treatment of ambiguities aims at selecting a given $\vo$ decay with the highest efficiency 
and the lowest background contamination from other $\vo$ types.
An optimum compromise between high statistics of the identified $\vo$ sample and well 
understood background contamination is
the aim of our identification strategy.
The MC simulation program has been used to 
define the criteria for the kinematic $\vo$ selection and to determine
the purity of the final
$\ko$, $\lam$ and $\alam$ samples.
We selected a sample consisting of more than 90\% of uniquely
identified $\vo$.
Results are reported in Table~\ref{tab:TAB_SAMPLE}.

\begin{table}[htb]
\caption{\it 
Efficiency ($\epsilon$) and purity ($P$) for each selected $\vo$ category.
Numbers of identified neutral strange particles in the data 
are also shown in the last column.}
\vspace*{0.5cm}
\begin{center}
\begin{tabular}{||c|c|c|c||}
\hline 
\hline
\vo   & $\epsilon$ (\%) & $P$ (\%)       & Data \\
\hline
\hline
\ko   & $22.1 \pm 0.1$  & $97.2 \pm 0.1$ & 15074  \\
\lam  & $16.4 \pm 0.1$  & $95.9 \pm 0.1$ & 8087 \\
\alam & $18.6 \pm 0.5$  & $89.7 \pm 0.7$ & 649 \\
\hline
\hline
\end{tabular}
\label{tab:TAB_SAMPLE}
\end{center}
\end{table}

The total $\vo$ sample in our data contains 15074 identified $\ko$, 8087 identified $\lam$
and 649 identified $\alam$ decays, representing  
significantly larger numbers than in previous (anti)neutrino experiments performed 
with bubble chambers~\cite{Barish,Deden,Berge,Deden2,Berge2,Erriquez,ammosov,Baker,ammosov2,Brock,grassler,Bosetti,son1,allasia,chang,son2,allasia2,Jones,FNAL,mann,ammosov3,ammosov4,Willocq,BEBC,Prospo}.

\begin{figure}[htb]
\begin{center}
\vspace*{-0.5cm}
\begin{tabular}{cc}
\mbox{\epsfig{file=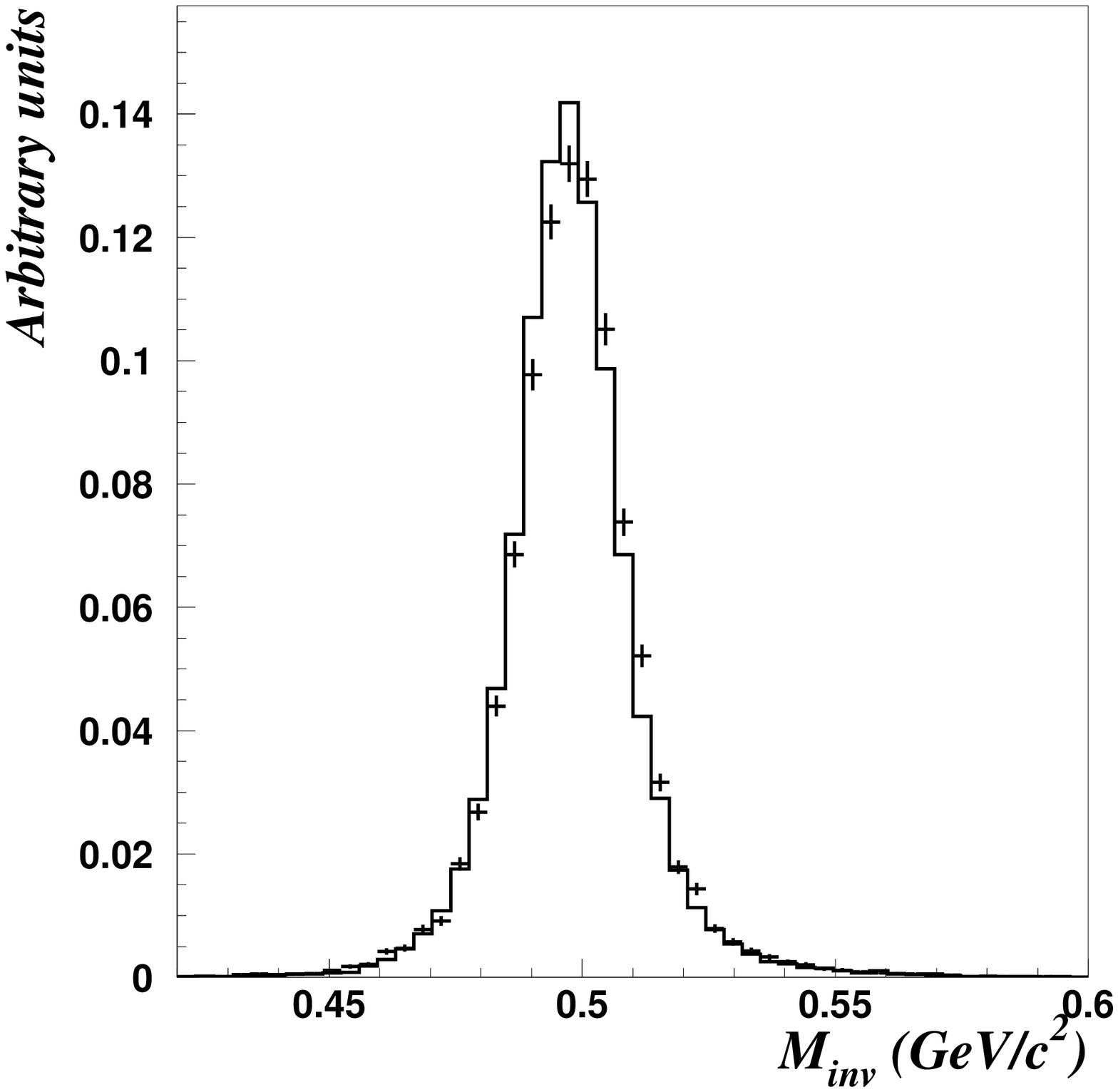,width=0.5\linewidth}}&
\mbox{\epsfig{file=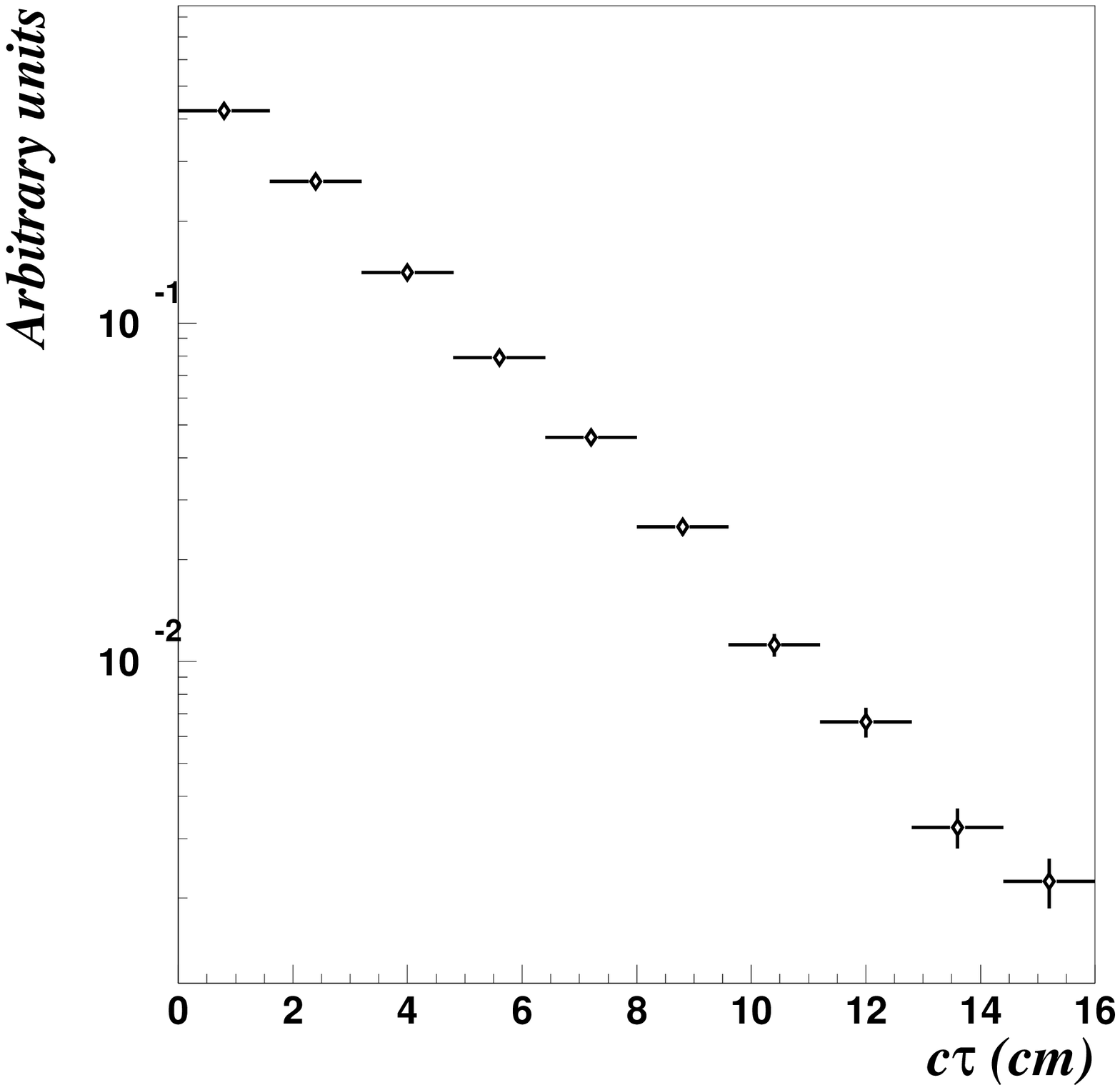,width=0.5\linewidth}}\\
\mbox{\epsfig{file=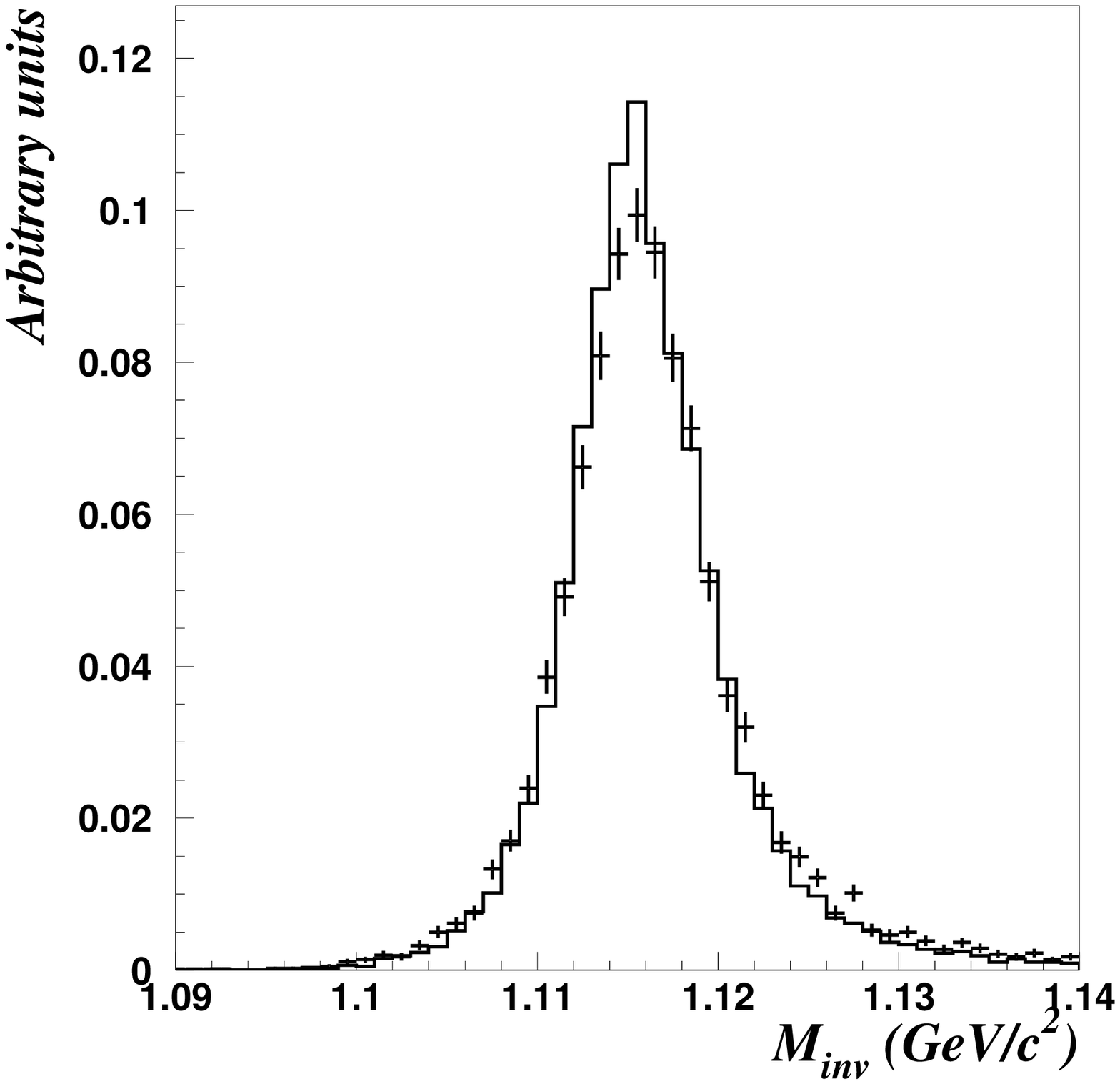,width=0.5\linewidth}}&
\mbox{\epsfig{file=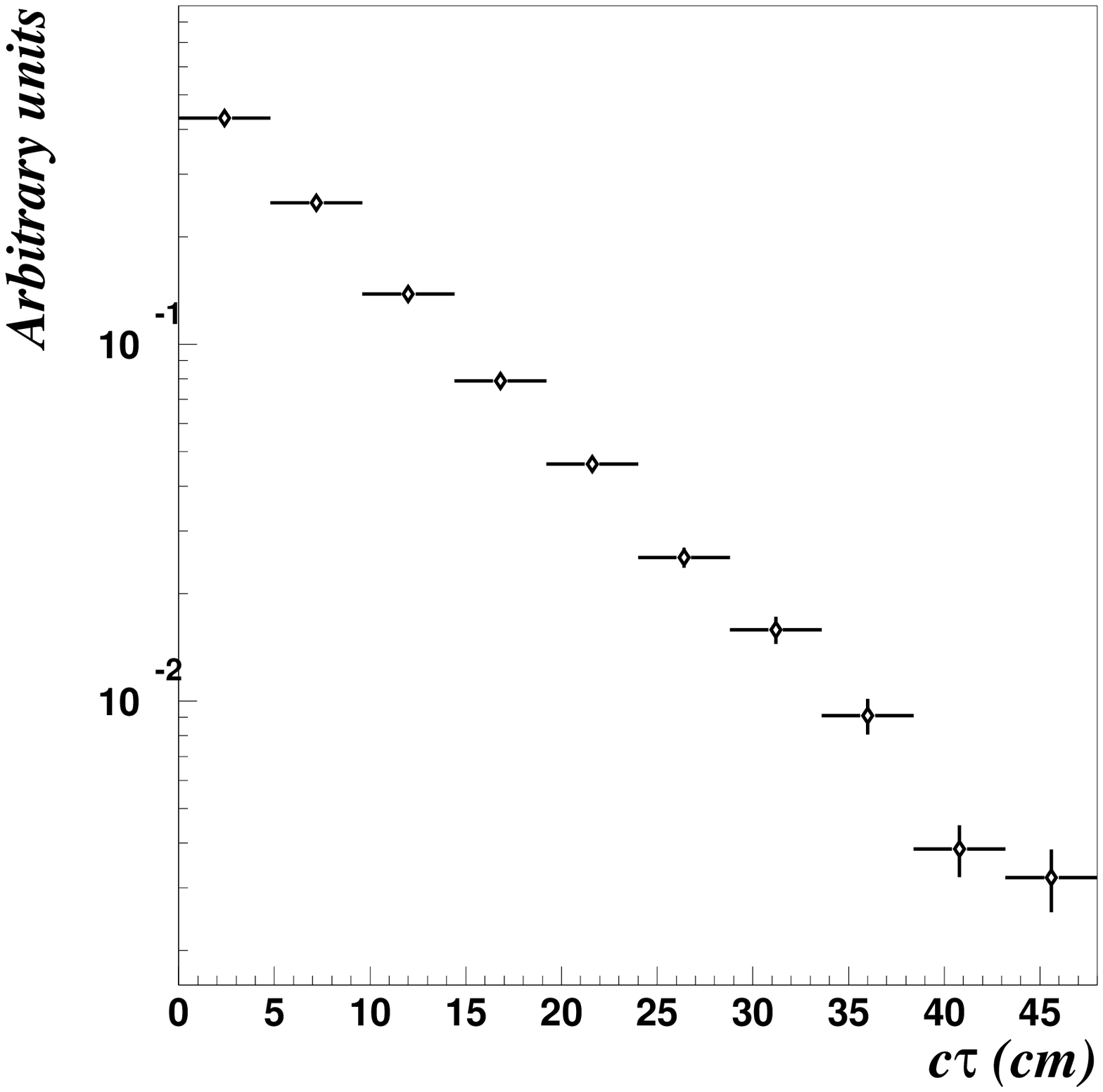,width=0.5\linewidth}}\\
\mbox{\epsfig{file=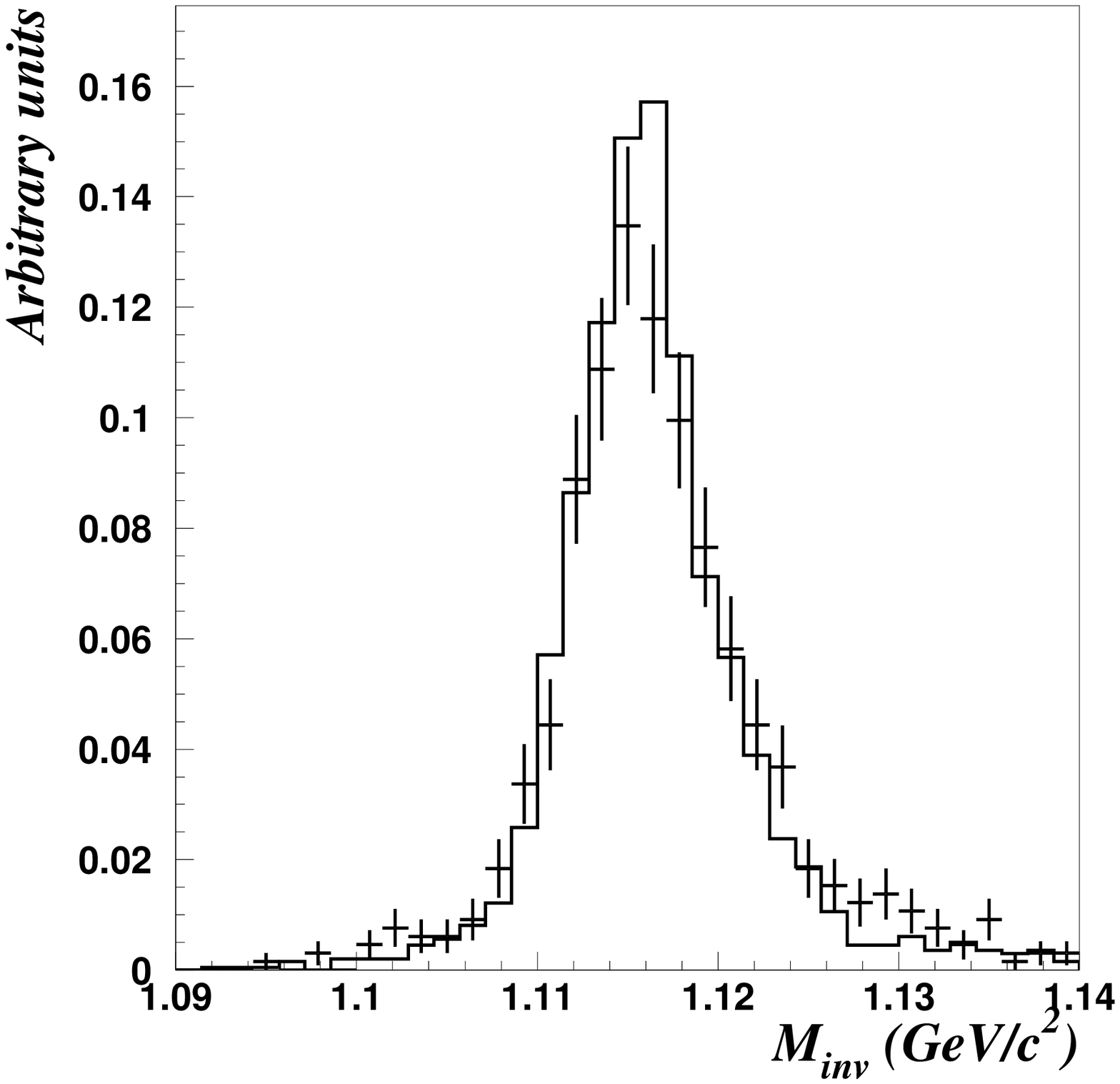,width=0.5\linewidth}}&
\mbox{\epsfig{file=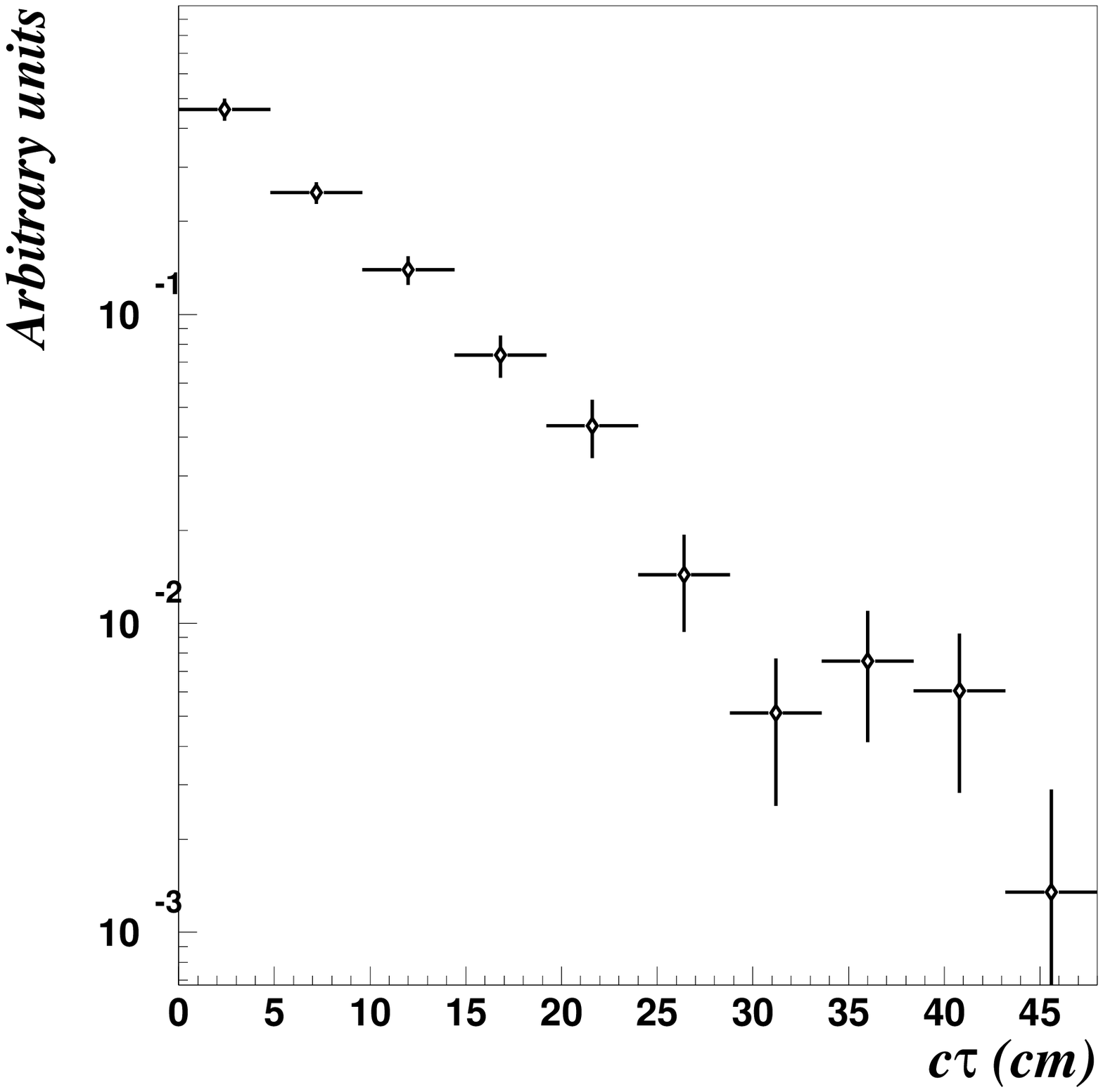,width=0.5\linewidth}}\\
\end{tabular}
\caption{
\it
Invariant mass distributions and efficiency corrected $c\tau$ distributions 
for identified $\ko$ (top), $\lam$ (center) and $\alam$ (bottom) in $\nu_\mu$ 
CC DIS events for both data (points with error bars) and MC (histogram). 
}
\label{fig:invmass}
\end{center}
\end{figure}

Fig.~\ref{fig:invmass} shows the invariant mass and $c\tau$
distributions for identified $\ko$, $\lam$ and $\alam$.
The measured 
mass and the lifetime of identified neutral
strange particles are in agreement with the world averages~\cite{PDG}.
The corresponding results are given in Tables~\ref{tab:mass_v0} and~\ref{tab:ctau_v0}.

The efficiencies and purities reported in Table~\ref{tab:TAB_SAMPLE}
are momentum dependent. However, we have checked that they are
applicable to the data because the momenta distributions of identified
$\vo$ and of their decay products are identical in the data and MC simulation.

In the rest of this paper we will always present efficiency corrected distributions.

\begin{table}[htb]
\begin{center}
\caption {
{\it Measured $\vo$ mass and resolution
(in MeV/c$^2$) for both MC and data.}}
\begin{tabular}{||c||c|c||c|c||c||}
\hline
\hline
\vo &\multicolumn{2}{c||}{MC}&\multicolumn{2}{c||}{Data} & {PDG \cite{PDG}}\\
\cline{2-6}
& Mass & Resolution & Mass & Resolution & Mass\\
\hline
\hline
\ko & 497.9 $\pm$ 0.1 & 11.0 $\pm$ 0.1 & 497.9 $\pm$ 0.1 & 11.3 $\pm$ 0.1 & 497.672 $\pm$ 0.031 \\
\hline
\lam & 1\,115.7 $\pm$ 0.02 & 4.0 $\pm$ 0.03 & 1\,116.0 $\pm$ 0.05 &
4.45 $\pm$ 0.05 & 1115.683 $\pm$ 0.006 \\
\hline
\alam & 1\,116.0 $\pm$ 0.1 & 4.1 $\pm$ 0.1 & 1\,116.3 $\pm$ 0.1 & 
4.8 $\pm$ 0.2 & 1115.683 $\pm$ 0.006 \\
\hline
\hline
\end{tabular}
\label{tab:mass_v0}
\end{center}
\end{table}
 
\begin{table}[htb]
\begin{center}
\caption {
{\it Measured $c\tau$ (in cm) for a given $\vo$ type 
for both MC and data.}}
\begin{tabular}{||c||ccc||ccc||ccc||}
\hline
\hline
\vo &\multicolumn{3}{c||}{MC}&\multicolumn{3}{c||}{Data}&\multicolumn{3}{c||}{PDG \cite{PDG}}\\
\hline
\hline
\ko  &\multicolumn{3}{c||}{2.60 $\pm$ 0.01} &\multicolumn{3}{c||}{2.72 $\pm$ 0.03}&\multicolumn{3}{c||}{2.6786}\\
\hline
\lam &\multicolumn{3}{c||}{7.91 $\pm$ 0.02} &\multicolumn{3}{c||}{8.07 $\pm$ 0.12} &\multicolumn{3}{c||}{7.89}\\
\hline
\alam&\multicolumn{3}{c||}{7.82 $\pm$ 0.06} &\multicolumn{3}{c||}{7.33 $\pm$ 0.33}  &\multicolumn{3}{c||}{7.89}\\
\hline
\hline
\end{tabular}
\label{tab:ctau_v0}
\end{center}
\end{table}

\section{YIELDS OF NEUTRAL STRANGE PARTICLES\label{sec:v0_yields}}

We have studied the production rates of 
neutral strange particles ($\ko$, $\lam$, $\alam$) in $\nu_\mu$ CC interactions.
The particles can be produced 
at the primary vertex and also from
secondary interactions of primary particles with the
detector material. 
Neutral strange particles produced via resonance or heavier hyperon decays are
classified as primary $\vo$.
We applied a correction obtained from the Monte Carlo
to the yields of neutral strange particles
in the data in order to extract the yields at the primary vertex. 

\subsection{Integral yields of neutral strange particles}

The measured yield per $\nu_\mu$ CC interaction for each $\vo$ type is defined as:
\begin{equation}
{T}\rm_{\vo} = \xi \cdot \frac{N_{\vo}}{N_{\nu_\mu CC}},
\end{equation}
where $N_{\vo}$ is the number of reconstructed and identified $\vo$ 
in the number $N_{\nu_\mu CC}$ of reconstructed $\nu_\mu$ CC events and 
$\xi$ is a correction factor calculated as:
$$
\xi = \frac{P_{\vo} \times \epsilon_{\nu_\mu CC}}{\epsilon_{\vo} \times Br(\vo \to h^+ h^-)},
$$
where $\epsilon_{\nu_\mu CC} = (85.30 \pm 0.02) \%$ is the efficiency 
(reconstruction and identification) of $\nu_\mu$ CC events
in the fiducial volume, and $\epsilon_{\vo}$ is the global $\vo$ efficiency 
which takes into account the contribution from 
particles produced 
in the fiducial volume, but decaying outside. $P_{\vo}$ is the purity of 
the final $\vo$ sample, and $Br(\vo \to h^+ h^-)$ is the branching ratio for
a given $\vo$ type decaying to a pair of charged hadrons.
\begin{table}[htbp]
\caption{\it Integral yields 
of primary $\vo$ in $\nu_\mu$ CC interactions 
in both the data and in the default MC.
The errors are only statistical.
}
\vspace*{0.5cm}
\begin{center}
\begin{tabular}{||c|c|c|c||}
\hline
\hline
$\vo$ type & ${T}\rm_{\vo}^{DATA}$ (\%) & ${T}\rm_{\vo}^{MC}$ (\%) & 
${T}\rm_{\vo}^{MC}/{T}\rm_{\vo}^{DATA}$ \\
\hline
\hline
$\ko$&  $6.76 \pm 0.06$ &$9.50 \pm 0.02$  & $1.40 \pm 0.01$\\
\hline
$\lam$& $5.04 \pm 0.06$ &$8.10 \pm 0.02$  &$1.61 \pm 0.02$\\
\hline
$\alam$& $0.37 \pm 0.02$ & $0.60 \pm 0.01$ & $1.62 \pm 0.03$\\
\hline
\hline
\end{tabular}
\end{center}
\label{tab:global_production_numu} 
\end{table}

Table~\ref{tab:global_production_numu} shows the overall inclusive
production rates for $\ko$, $\lam$ and $\alam$ in $\nu_\mu$ CC
interactions compared to the MC predictions. 
Note that 
the production rates in the default MC 
are
a factor of 1.4 to 1.6 higher 
than in the data. 
This could be explained in part by the choice of the LEPTO~\cite{LEPTO}
and JETSET~\cite{JETSET} parameters in the NOMAD event generator.
We have to emphasize that the so-called $s \bar s$
suppression factor (PARJ(2) parameter in JETSET) - 
defined as the ratio of the probability 
$\gamma_s$ of producing an $s \bar s$ pair to the probability $\gamma_u$
($\gamma_d$) of producing a $u \bar u$ ($d \bar d$) pair in the fragmentation 
chain - was set to its default value of 0.3 in the 
default
NOMAD
MC production\footnote{The value 0.3 has been suggested by the authors of
JETSET as the default for this parameter.}. 
However, this parameter was measured to be about 0.2 in 
previous bubble chamber experiments: for example, the values obtained
by the BEBC WA21 Collaboration~\cite{ssbar_bebc} in a
neutrino beam similar to ours are $0.200 \pm 0.022(stat) \pm 0.010(sys)$
for $\bar \nu p$ and $0.207 \pm 0.018(stat) \pm 0.020(sys)$ for $\nu p$
interactions. Moreover, later results from 
OPAL (0.245)~\cite{ssbar_opal_new}, 
DELPHI (0.23)~\cite{ssbar_delphi_new}, 
E665 (0.2)~\cite{ssbar_e665}, ZEUS~\cite{ssbar_zeus} and H1~\cite{ssbar_h1} 
collaborations support a value close to 0.2 for this parameter.
\begin{table}[htb]
\caption{\it Numbers of $\nu_\mu$ CC events 
with a specified combination of observed neutral strange particle decays 
for both the default MC and  data. $X$ indicates 
the
hadronic system 
accompanying the
observed 
$\vo$.
} 
\vspace*{0.5cm}
\begin{center}
\begin{tabular}{||c|c|c|c||}
\hline
\hline
 & \multicolumn{2}{c|}{Number of observed events} & \\ \cline{2-3}
\ Channel & MC & Data & MC/DATA \\
\hline
\hline
$\lam X$                  & 11686 & 7778 & $1.50 \pm 0.02$\\
$\ko X$                   & 18971 & 14228& $1.33 \pm 0.01$\\
$\alam X$                 & 831   & 594  & $1.40 \pm 0.07$\\
$\ko \ko X$               & 485   & 284  & $1.7 \pm 0.1$\\
$\lam \ko X$              & 617   & 247  & $2.5 \pm 0.2$\\
$\lam \alam X$            & 98    & 40   & $2.5 \pm 0.4$\\
$\ko \alam X$             & 24    & 15   & $1.6 \pm 0.5$\\
$\lam \lam X$             & 19    & 10   & $1.9 \pm 0.7$\\
$\lam \ko \ko X$          & 7     & 2    & $3.4 \pm 2.6$\\
$\ko \ko \ko X$           & 2     & 4    & $0.6 \pm 0.4$\\
\hline
\hline
\end{tabular}
\label{tab:multiV0}
\end{center}
\end{table}

However, the problem of the 
inaccurate
description of neutral strange particle
production in the MC is a more complex one.
This is illustrated in Table~\ref{tab:multiV0}, where 
we give the observed numbers of $\nu_\mu$ CC
events for 10 exclusive multi-$\vo$ channels in the data compared to
the default MC predictions. The number of MC events in Table~\ref{tab:multiV0}
is renormalized to the same number of $\nu_\mu$ CC events as in
the data. From this comparison one can conclude that it is
not possible to rescale just a single parameter (the $s \bar s$ suppression
factor) in order to describe the neutral strange particle production
observed in the data.
Rather, the discrepancy is due to a combination of several parameters which describe
the probability that an
$s$($\bar s$)-quark appears as a meson/baryon(antibaryon), 
the probability that a strange meson/baryon(antibaryon) 
appears electrically neutral, etc. A tuning of the JETSET parameters
to reproduce the yields of neutral strange particles observed in the NOMAD
data is a subject of an analysis currently in progress.

The integral yields reported in Table~\ref{tab:global_production_numu} 
can be 
compared to previous measurements
summarized in Table~\ref{tab:v0_production_numu}. The $\ko$ rates from
Table~\ref{tab:global_production_numu} have been converted into 
$K^0 (=K^0 +\bar{K^0})$ rates by multiplying by a factor of 2.

For 
completeness
we have 
recalculated the integral overall
yields taking into account contributions from both primary and secondary
$\vo$. These results are given in Table~\ref{tab:v0_production_numu}
and denoted by a star ($\star$). 
Our overall yields are consistent with the results of the $\nu$-Ne 
experiment~\cite{FNAL} performed in a similar neutrino beam. However,
our primary yields of $\ko$ and $\lam$ are 
about 30\%
lower and
the primary yield of $\alam$ is $\sim$20\% lower. 

\begin{table}[htbp]
\caption{\it Inclusive 
yields
of neutral strange particles in $\nu_\mu$ CC
  interactions 
  measured in this analysis and in previous bubble
  chamber experiments. $N_K$, $N_\Lambda$ and
  $N_{\bar \Lambda}$ are the observed numbers of $\ko$, $\lam$ and $\alam$,
  respectively.
  $K^0$ stands for $K^0 + \bar K^0$. See text for explanation of a
  star ($\star$).}
\vspace*{0.5cm}
\begin{center}
\begin{tabular}{||c|c|c|c|c|c|c|c||}
\hline
\hline
Reaction & $\langle E_\nu \rangle$ & $N_{\ko}$ & $K^0$ rate &
$N_\Lambda$ & $\lam$ rate & $N_{\bar \Lambda}$ & $\alam$ rate\\
$\rm [Ref]$ & (GeV) & & (\%) &
 & (\%) & & (\%)\\
\hline
\hline
NOMAD
&\raisebox{-1.5ex}{45}   & \raisebox{-1.5ex}{15075} & $13.52 \pm 0.12$ & \raisebox{-1.5ex}{8087} & $5.04
\pm 0.06$ & \raisebox{-1.5ex}{649} & $0.37 \pm 0.02$\\
NOMAD$^\star$
&   &  & $18.22 \pm 0.16$ &
& $6.66 \pm 0.08$ &  & $0.45 \pm 0.02$\\ \hline
$\nu$ - Ne~\cite{FNAL} & 46 & 2279 & $16.8 \pm 1.2$ & 1843 & 
$6.5 \pm 0.5$ & 93 & $0.46 \pm 0.08$\\
$\nu$ - p~\cite{BEBC} & 51 & 831 & $19.0 \pm 0.9$ & 491 & 
$5.2 \pm 0.3$ & 27 & $0.34 \pm 0.07$\\
$\nu$ - Ne~\cite{Prospo} & 150 & 502 & $40.8 \pm 4.8$ & 285 & 
$12.7 \pm 1.4$ & 27 & $1.5 \pm 0.5$\\
$\nu$ - p~\cite{grassler} & 43 & 359 & $17.5 \pm 0.9$ & 180 & 
$4.5 \pm 0.4$ & 13 & $0.3 \pm 0.1$ \\
$\nu$ - Ne~\cite{Bosetti} & 103 & 203 & $23.0 \pm 1.7$ & 98 & 
$5.7 \pm 0.7$ & &\\
$\nu$ - n~\cite{allasia} & 62 & 234 & $20.8 \pm 1.6$ & 157 & 
$7.1 \pm 0.7$ & &\\
$\nu$ - p~\cite{allasia} & 62 & 154 & $17.7 \pm 1.6$ & 77 & 
$4.3 \pm 0.6$ & &\\
$\nu$ - n~\cite{allasia2} & 62 &  & $20.5 \pm 1.1$ &  & 
$6.6 \pm 0.7$ & &\\
$\nu$ - p~\cite{allasia2} & 62 &  & $17.4 \pm 1.2$ &  & 
$4.4 \pm 0.5$ & &\\
$\nu$ - A~\cite{ammosov3} & $\sim$10 & 82 & $7.1 \pm 0.8$ & 76 & $3.1 \pm 0.4$ & &\\
$\nu$ - p~\cite{Brock} & $\sim$45 & 23 & $15 \pm 4$ &  &  & &\\
$\nu$ - p~\cite{chang} & $\sim$50 &  &  &  & $7.0 \pm 0.8$ & &\\
$\nu$ - n~\cite{chang} & $\sim$50 &  &  &  & $7.0 \pm 1.2$ & &\\
\hline
\hline
\end{tabular}
\end{center}
\label{tab:v0_production_numu} 
\end{table}

\subsection{Yields of neutral strange particles 
as a function of kinematic variables} \label{sec:diff_rates}

To investigate neutral strange particle production mechanisms we have
measured the average $\ko$, $\lam$ and $\alam$ 
yields
as a function of the neutrino energy $E_\nu$, the invariant effective 
mass squared $W^2$ of the hadronic system, the invariant square
of the four-momentum transfer from the neutrino to the target $Q^2$, and 
the Bjorken scaling variable $x_{Bj}$.

\begin{figure}[htb]
\center{%
\begin{tabular}{cc}
\mbox{\epsfig{file=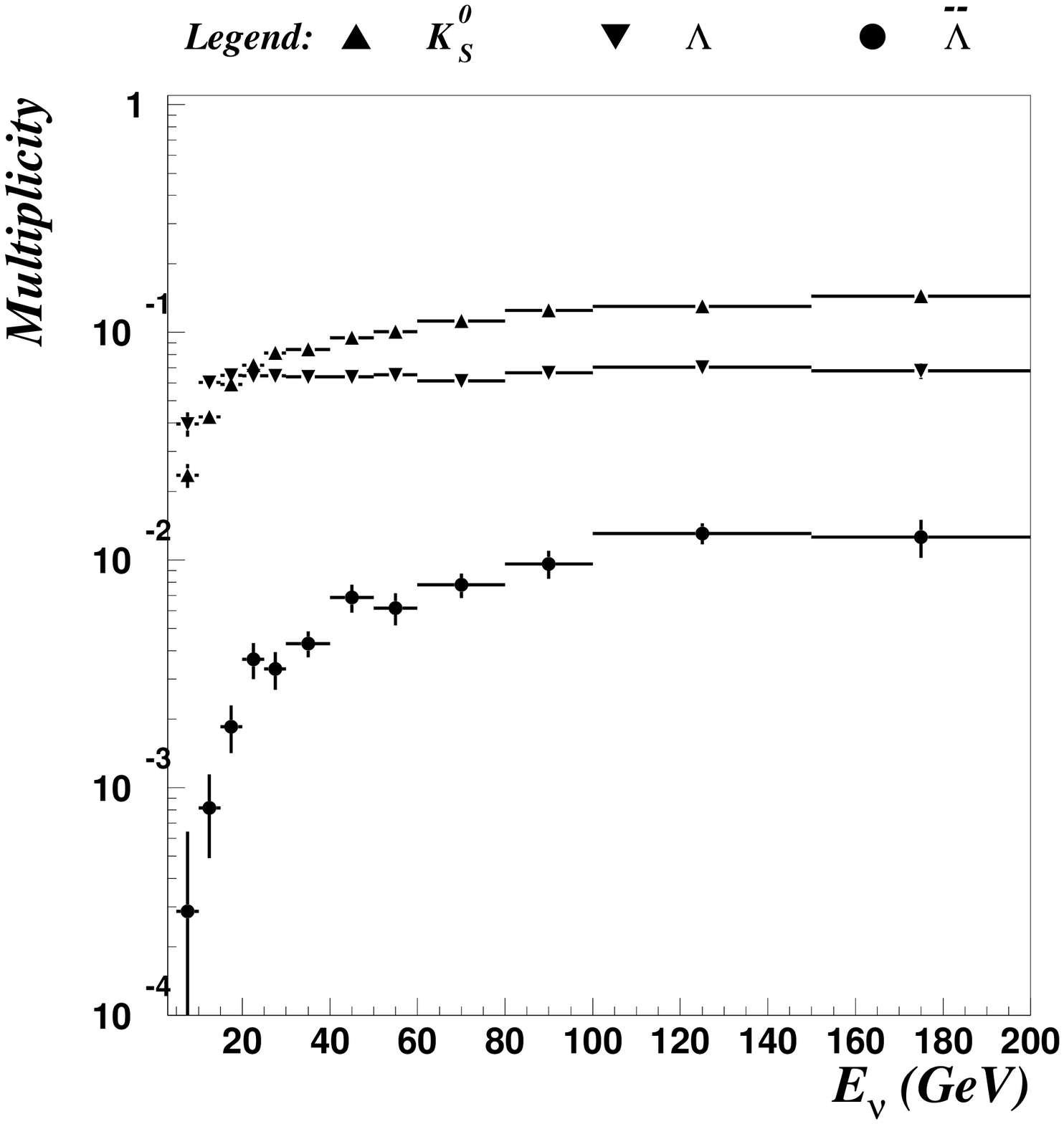,width=0.5\linewidth}}&
\mbox{\epsfig{file=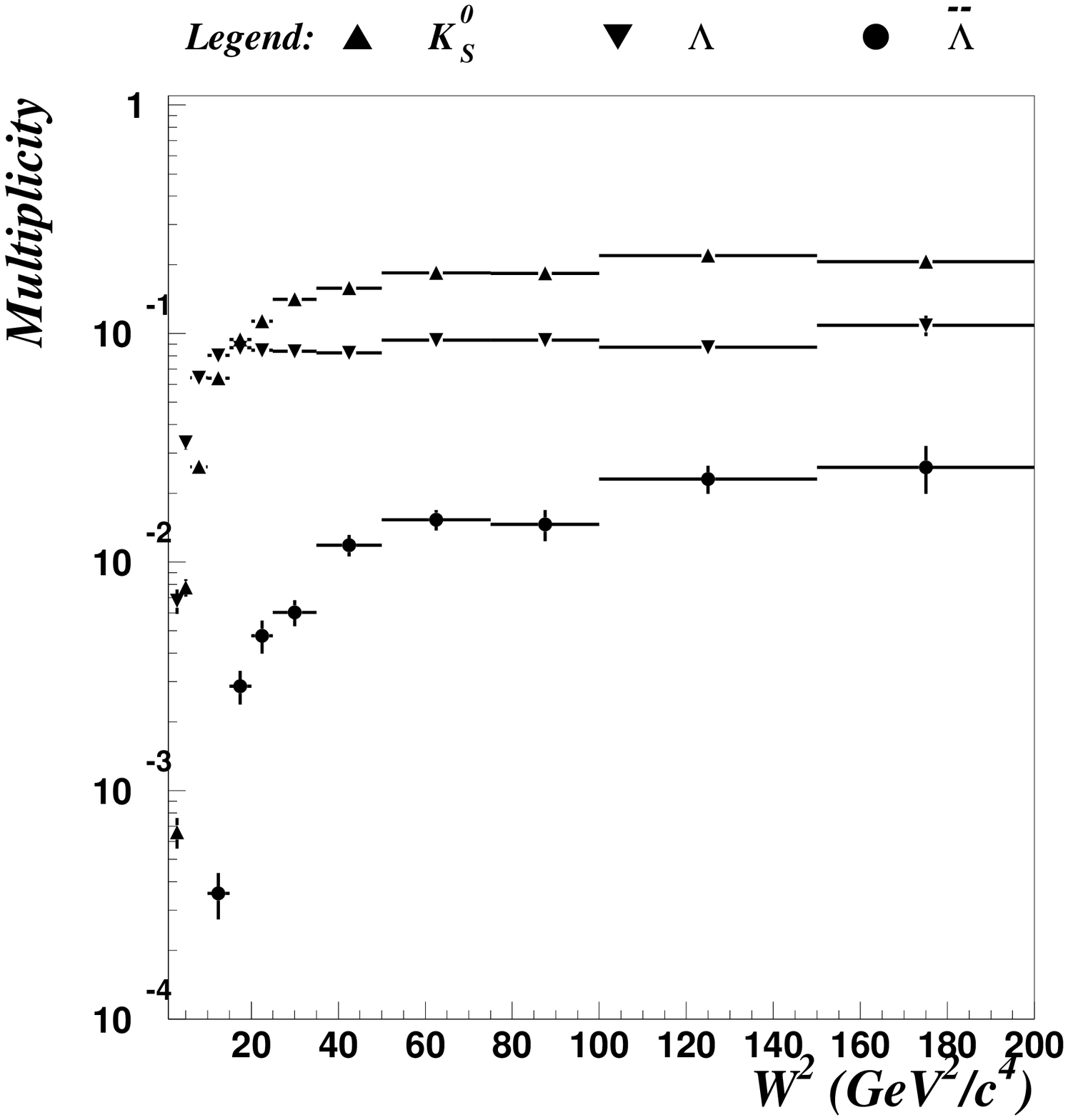,width=0.5\linewidth}}\\
\mbox{\epsfig{file=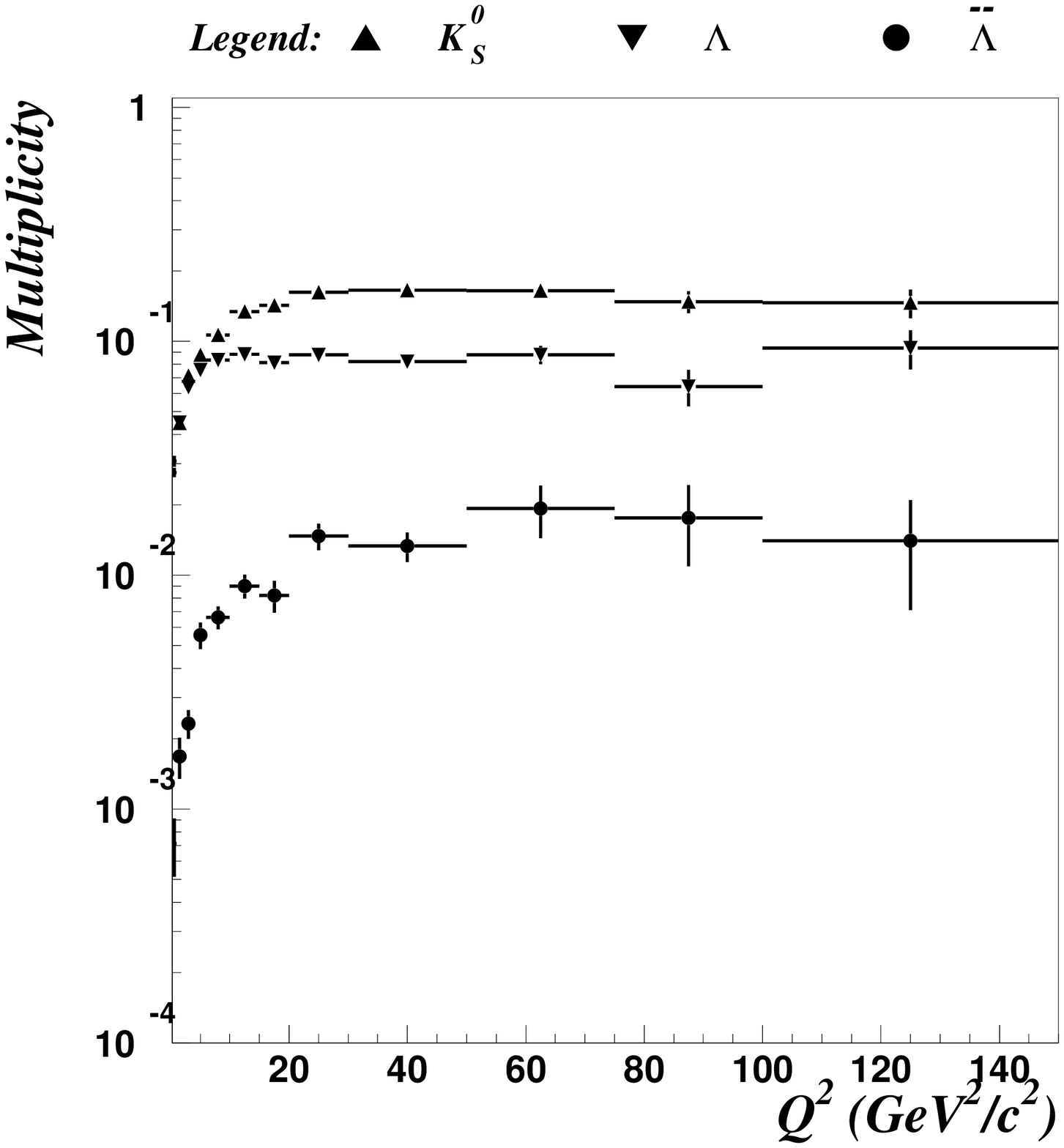,width=0.5\linewidth}}&
\mbox{\epsfig{file=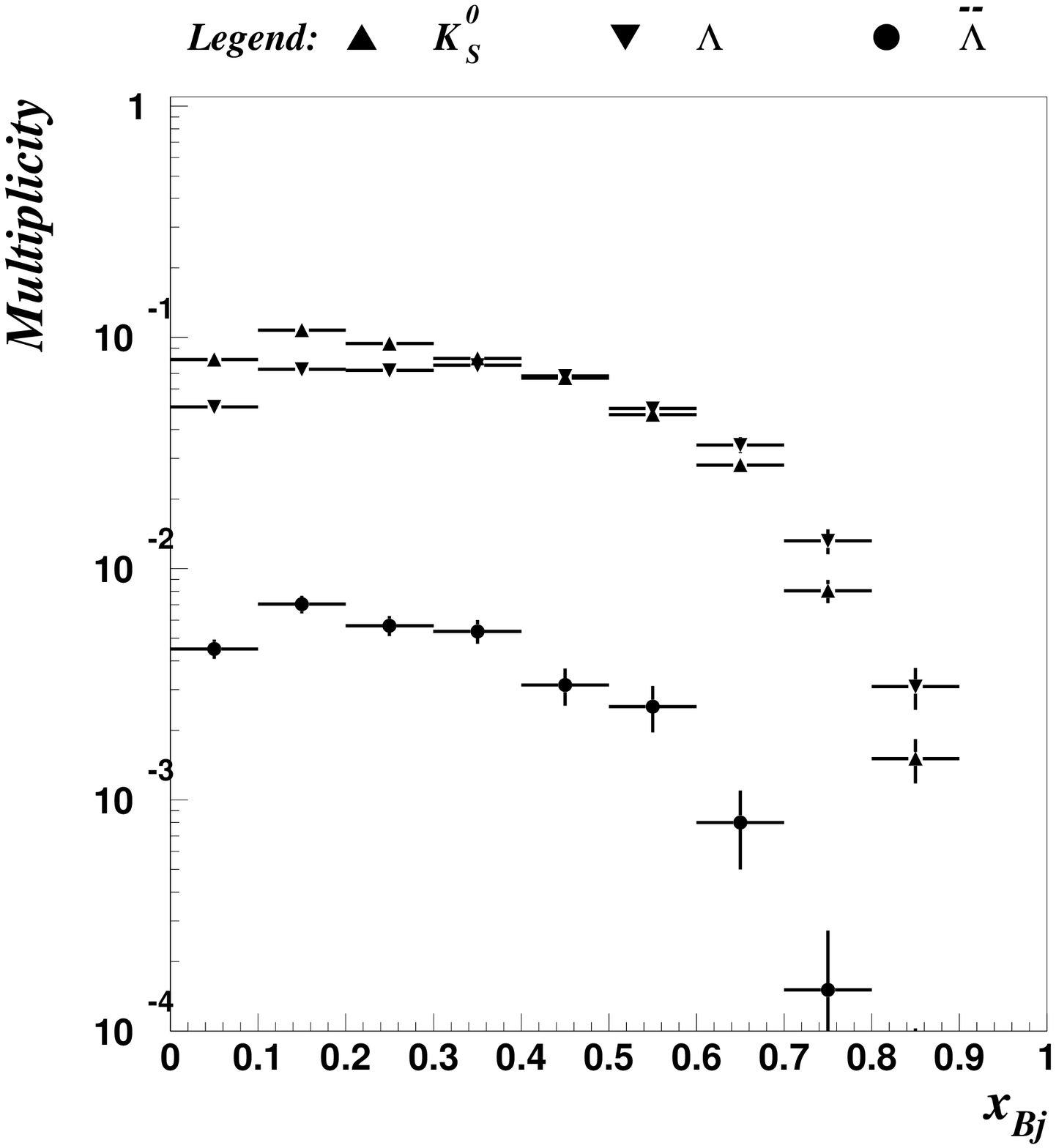,width=0.5\linewidth}}\\
\end{tabular}
}
\protect\caption{\it 
Measured yields of the
$\ko$, $\lam$ and $\alam$ as a function of
$E_\nu$, $W^2$, $Q^2$ and $x_{Bj}$.
}
\label{fig:v0_yields}
\end{figure}

The 
yields of
$\ko$, $\lam$ and $\alam$
are shown in Fig.~\ref{fig:v0_yields}.
The 
$\lam$ 
yield
shows a behaviour which is
almost independent of $E_\nu$, $W^2$ and $Q^2$ after a sharp initial
rise. It drops at large values of $x_{Bj}$. On the other hand, 
the 
yield
of 
$\ko$ rises steadily with $E_\nu$ and $W^2$, reaches a plateau at large $Q^2$ 
and falls with increasing $x_{Bj}$.
Similar observations have been made by previous experiments, but with
larger statistical uncertainties.
The 
$\alam$ 
yields 
as a function of kinematic variables
are measured for the first
time in a neutrino experiment. 
In general they show a behaviour similar to that of the $\ko$.
However, as expected, clear $W^2$ and $E_\nu$ thresholds are present
in the $\alam$ production.

\begin{figure}[htb]
\center{%
\begin{tabular}{cc}
\mbox{\epsfig{file=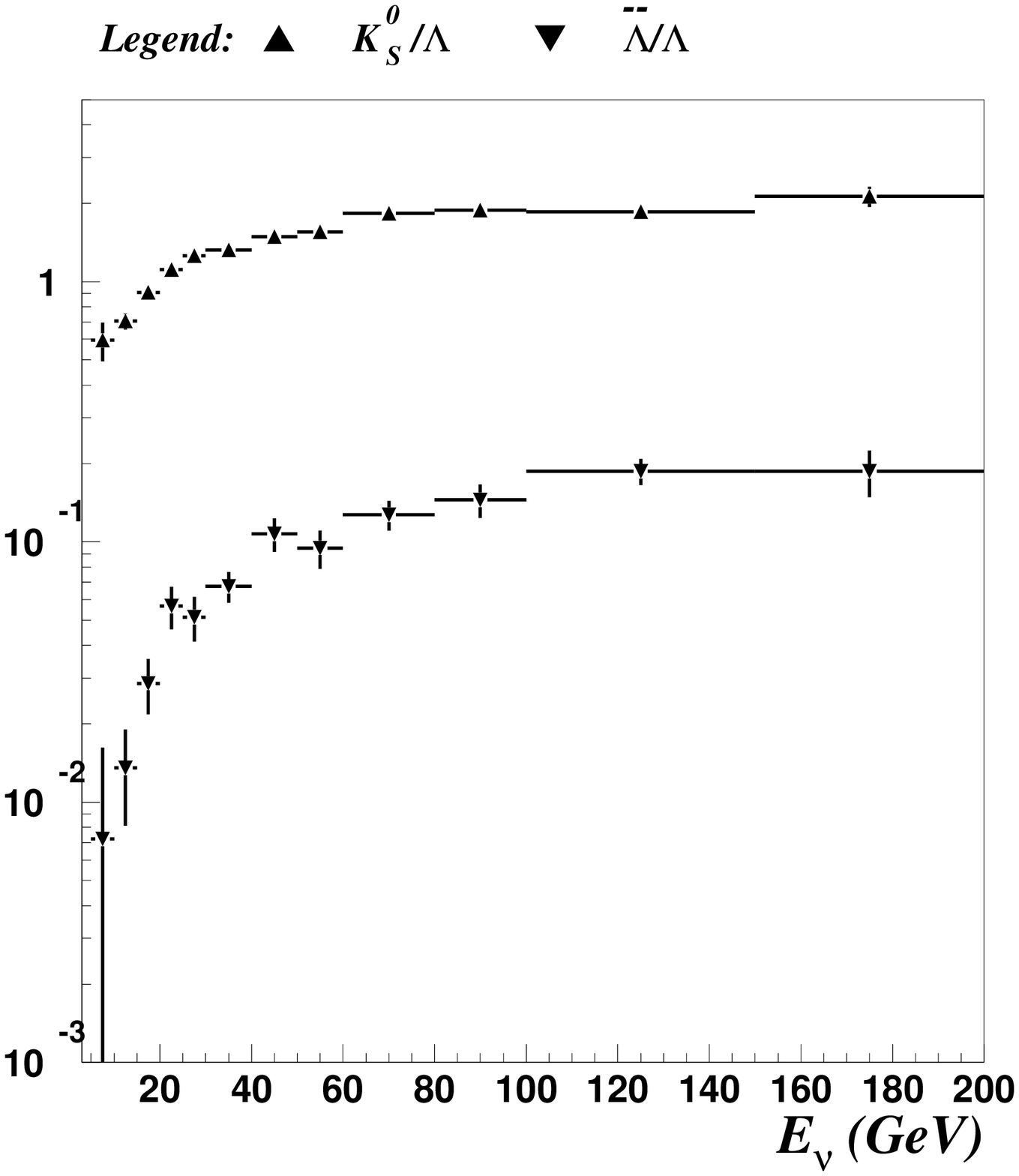,width=0.5\linewidth}}&
\mbox{\epsfig{file=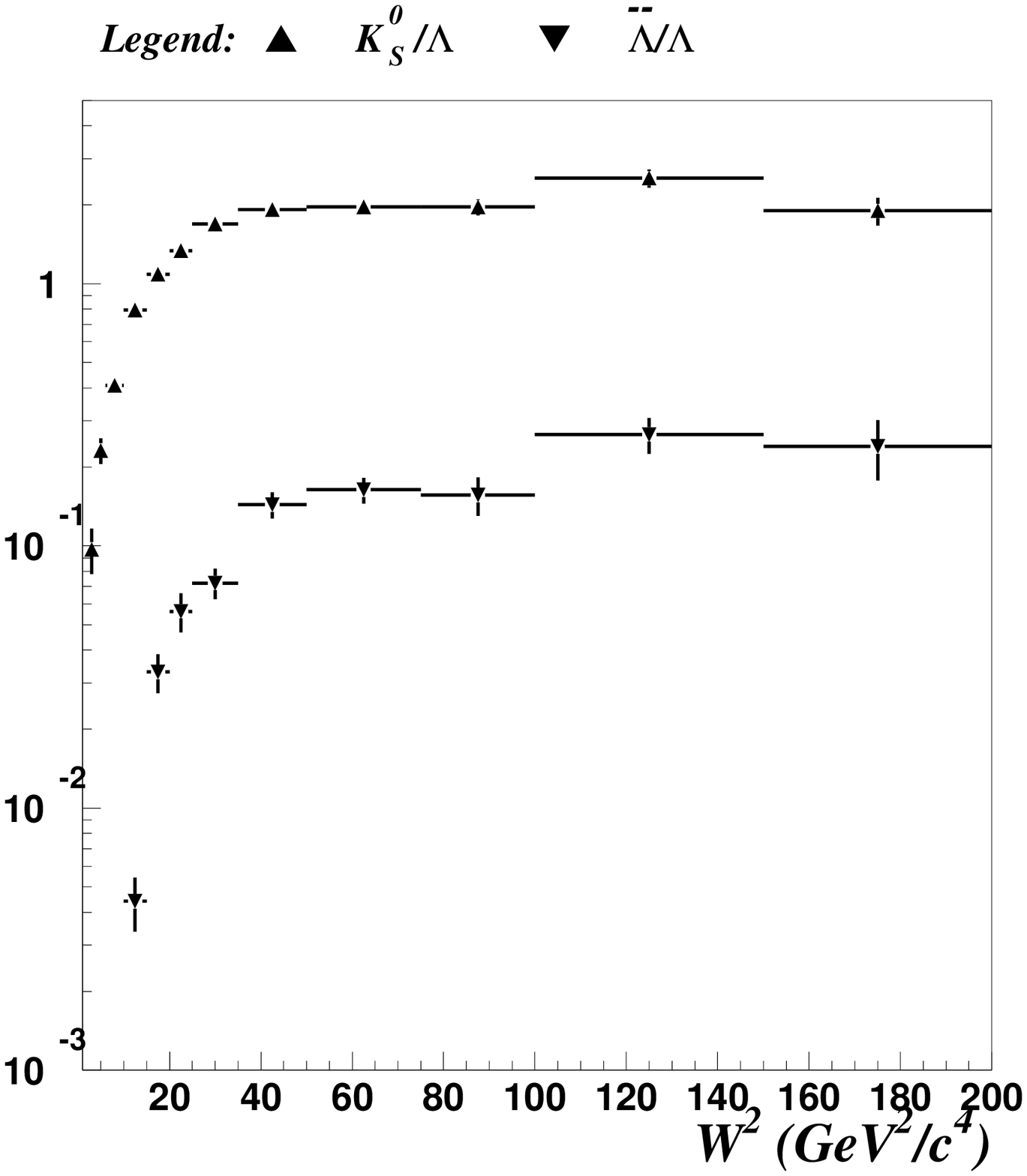,width=0.5\linewidth}}\\
\mbox{\epsfig{file=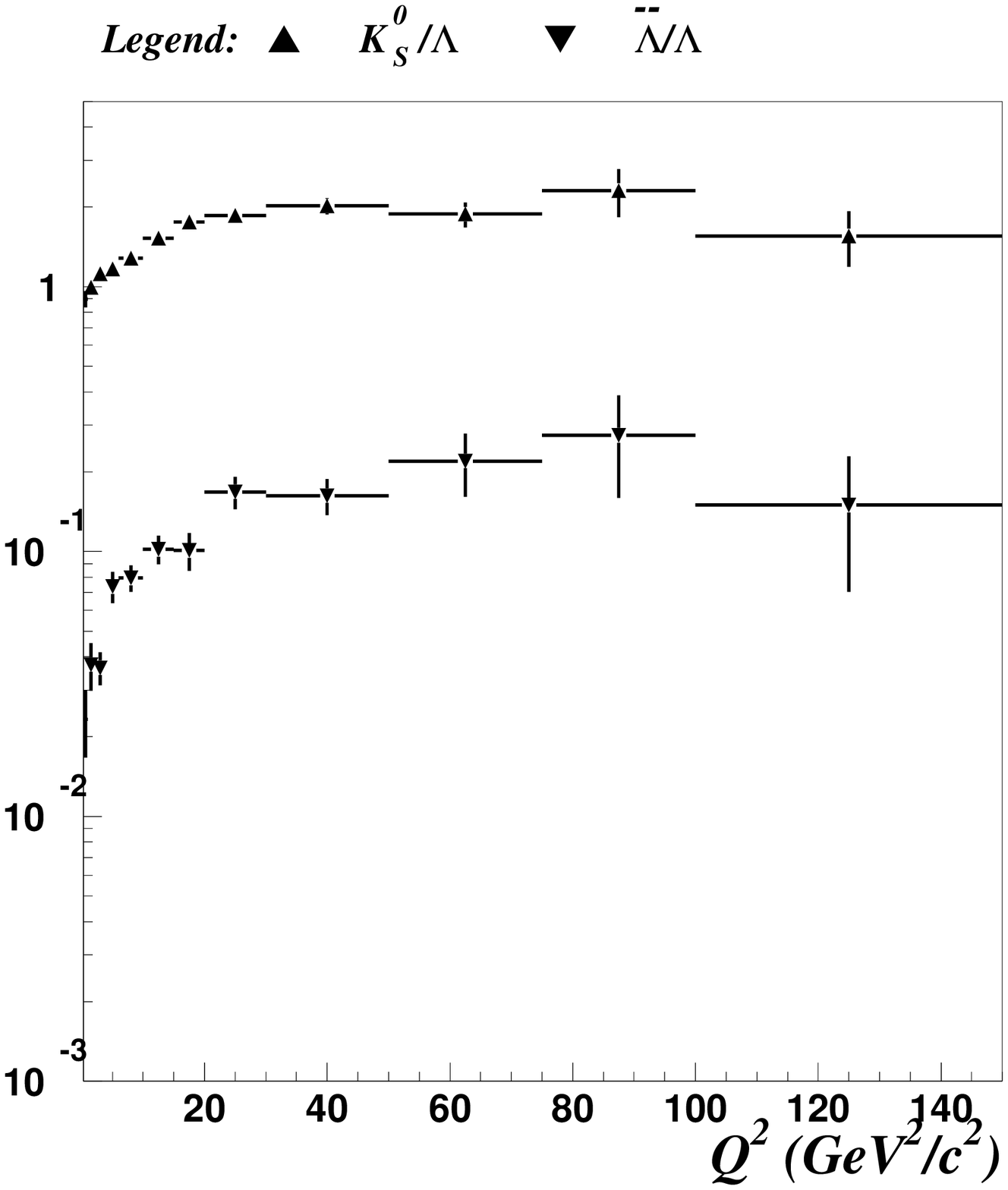,width=0.5\linewidth}}&
\mbox{\epsfig{file=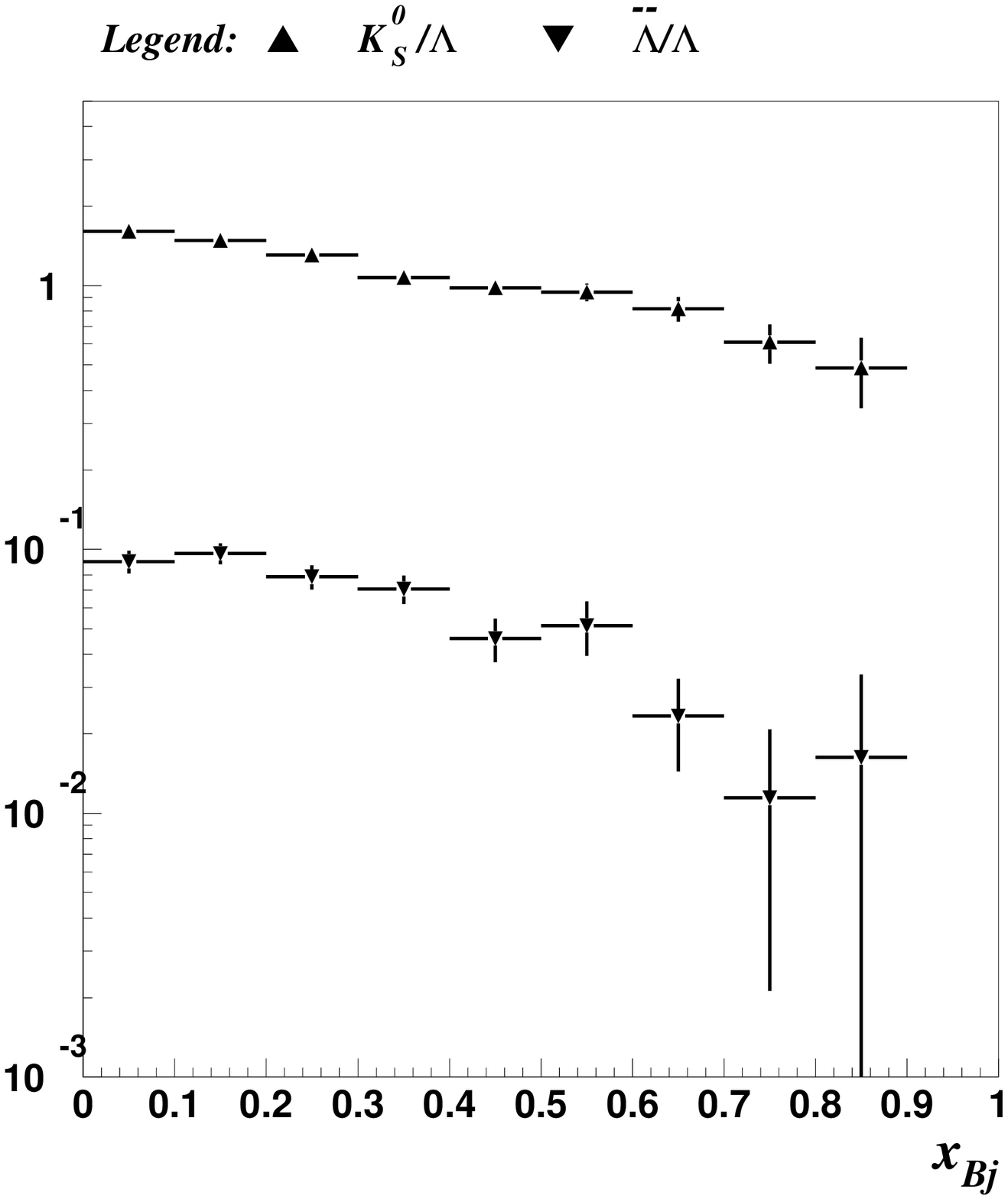,width=0.5\linewidth}}\\
\end{tabular}
}
\protect\caption{\it 
Ratios of 
measured yields
for $\ko$/$\lam$ and
$\alam$/$\lam$ 
as a function of
$E_\nu$, $W^2$, $Q^2$
and $x_{Bj}$.
}
\label{fig:v0_yields_ratios}
\end{figure}

The ratios of 
yields
for $\ko$/$\lam$ and
$\alam$/$\lam$ are presented in Fig.~\ref{fig:v0_yields_ratios}.

\subsection{Comparison with LUND model predictions}

\begin{figure}[htb]
\center{%
\begin{tabular}{cc}
\mbox{\epsfig{file=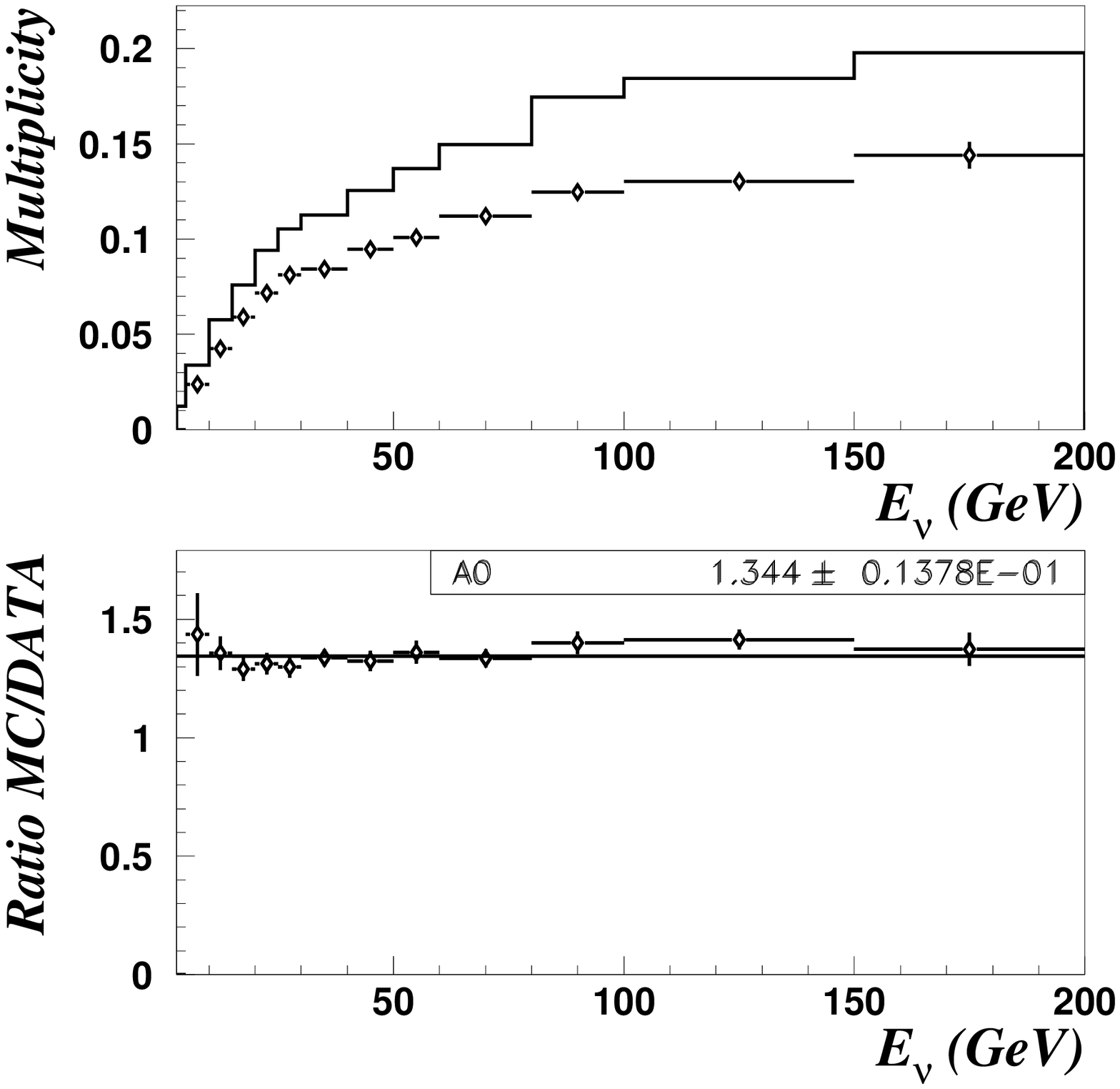,width=0.5\linewidth}}&
\mbox{\epsfig{file=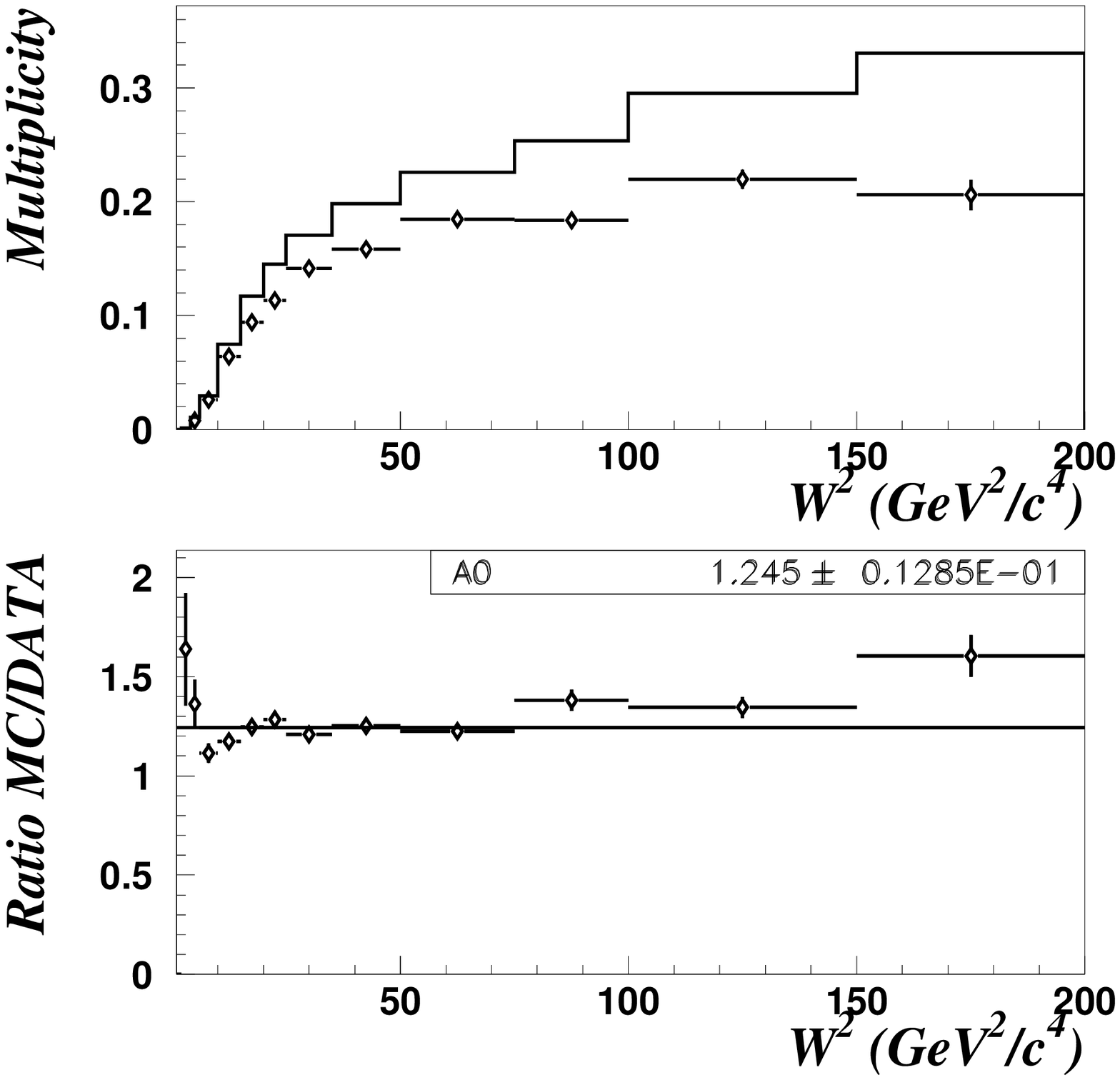,width=0.5\linewidth}}\\
\mbox{\epsfig{file=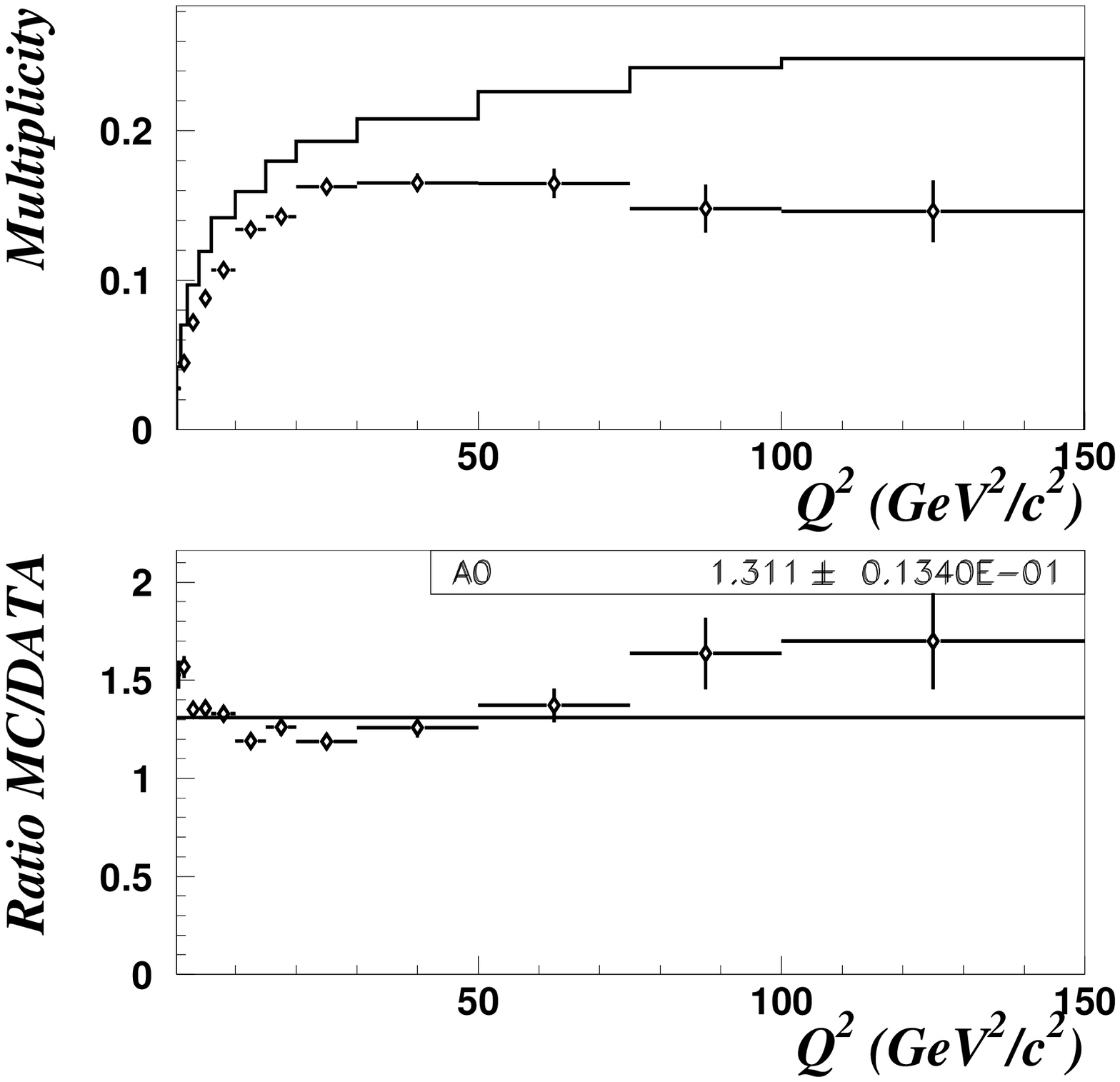,width=0.5\linewidth}}&
\mbox{\epsfig{file=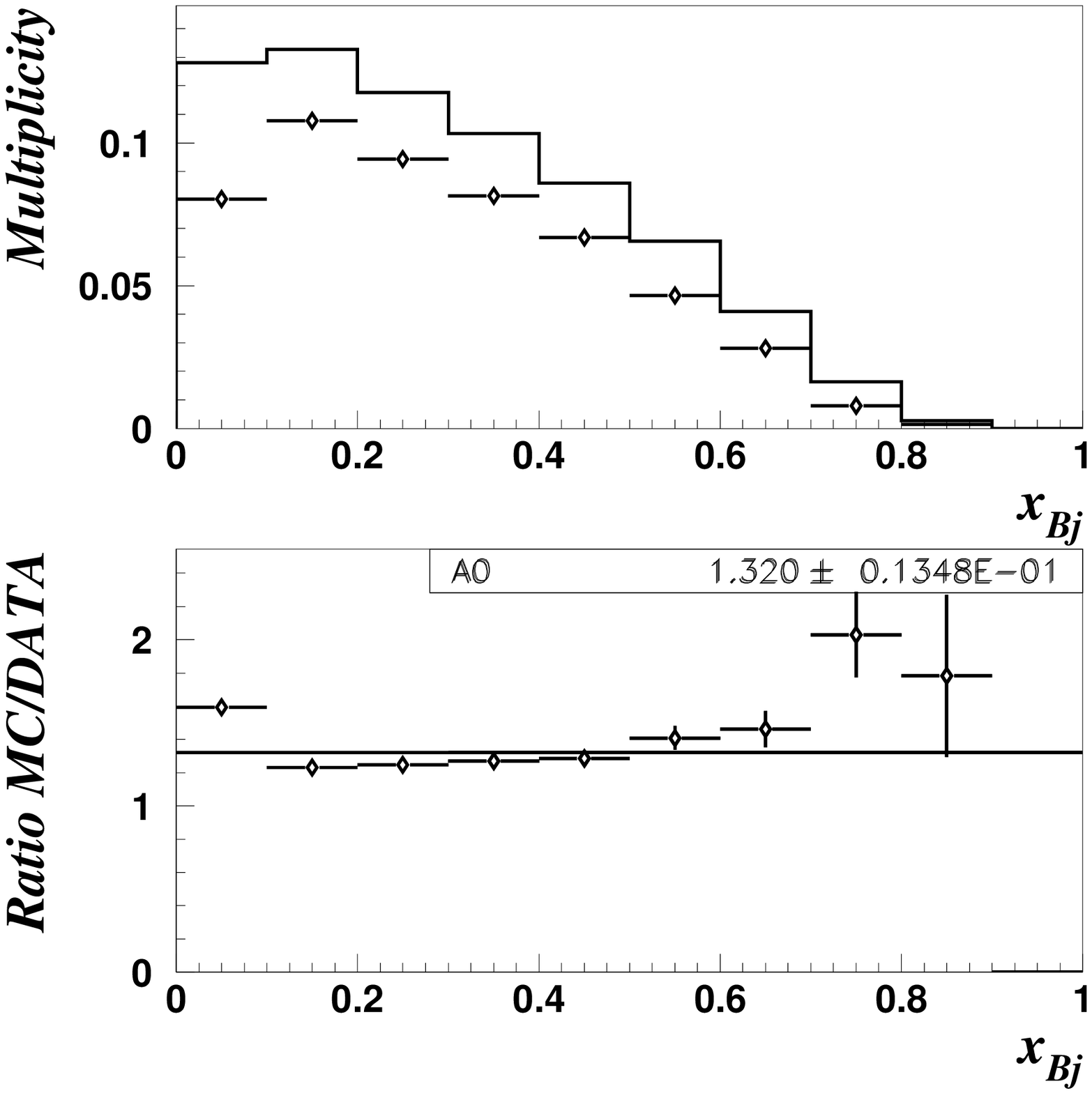,width=0.5\linewidth}}\\
\end{tabular}
}
\protect\caption{\it 
Yields
in the default MC (histogram)
and in the data (points with error bars) for $\ko$ as a function of
$E_\nu$, $W^2$, $Q^2$
and $x_{Bj}$. The MC/Data ratios and their fit to a constant are also shown.
}
\label{fig:k0_yields}
\end{figure}

\begin{figure}[htb]
\center{%
\begin{tabular}{cc}
\mbox{\epsfig{file=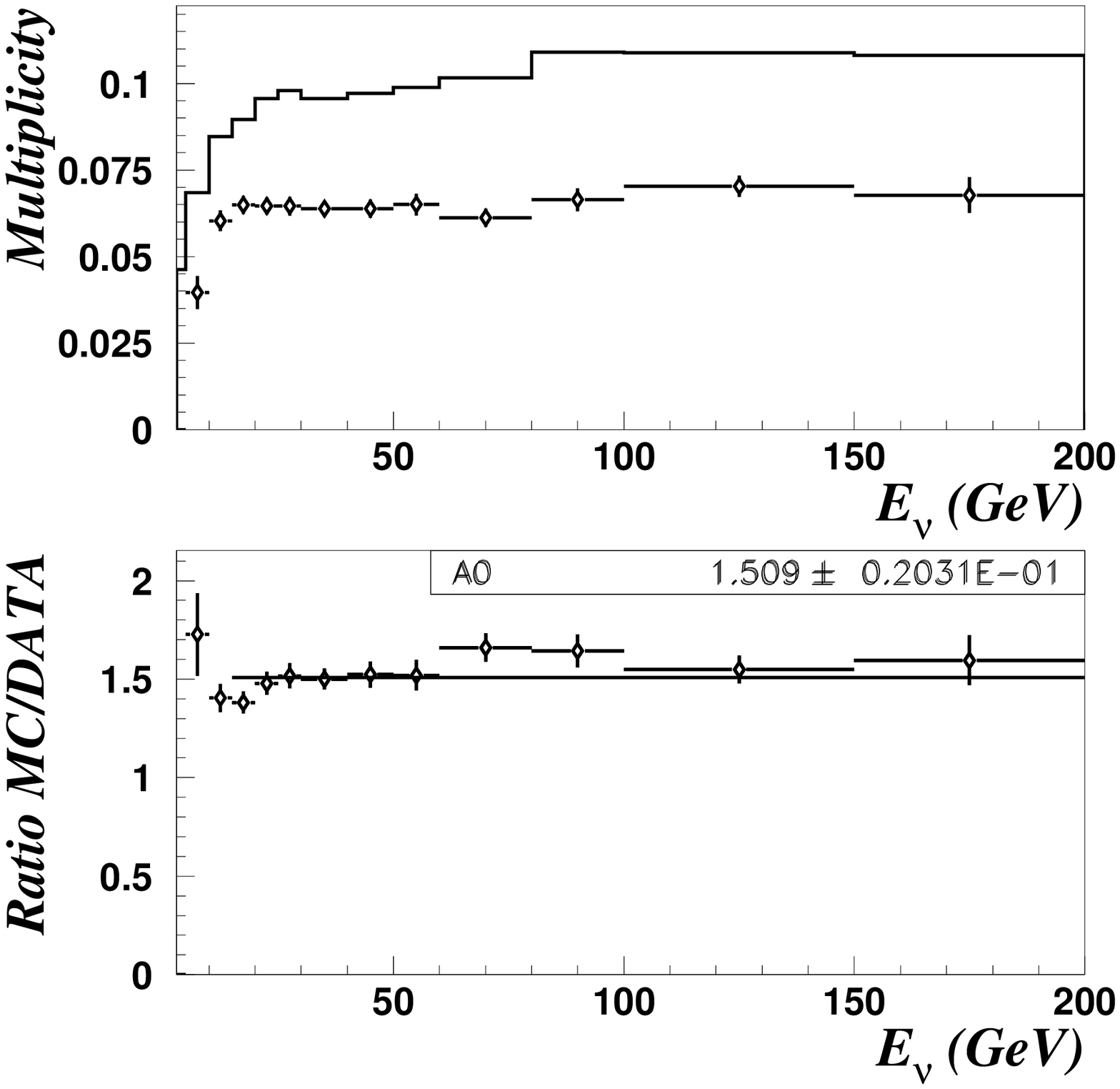,width=0.5\linewidth}}&
\mbox{\epsfig{file=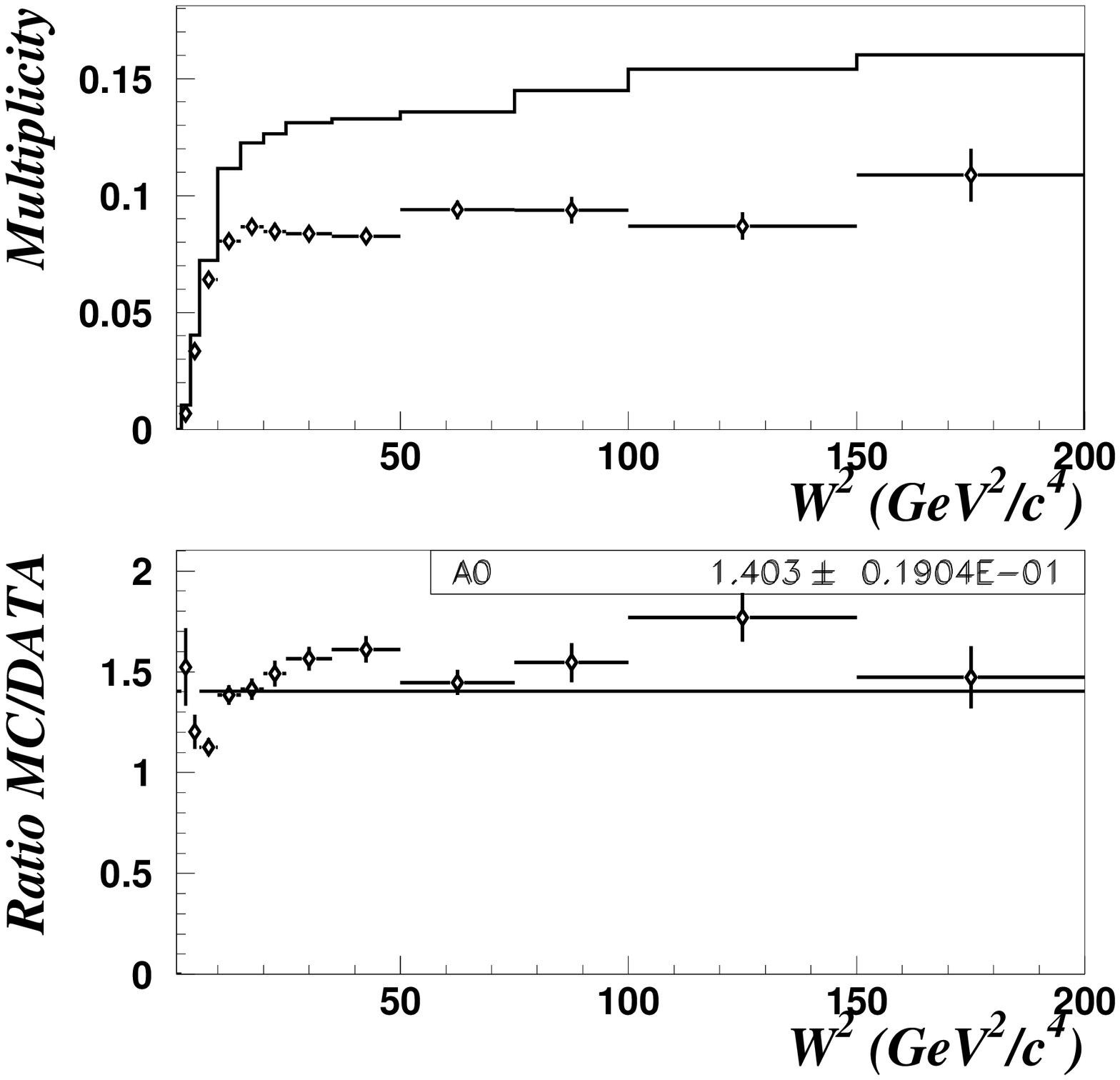,width=0.5\linewidth}}\\
\mbox{\epsfig{file=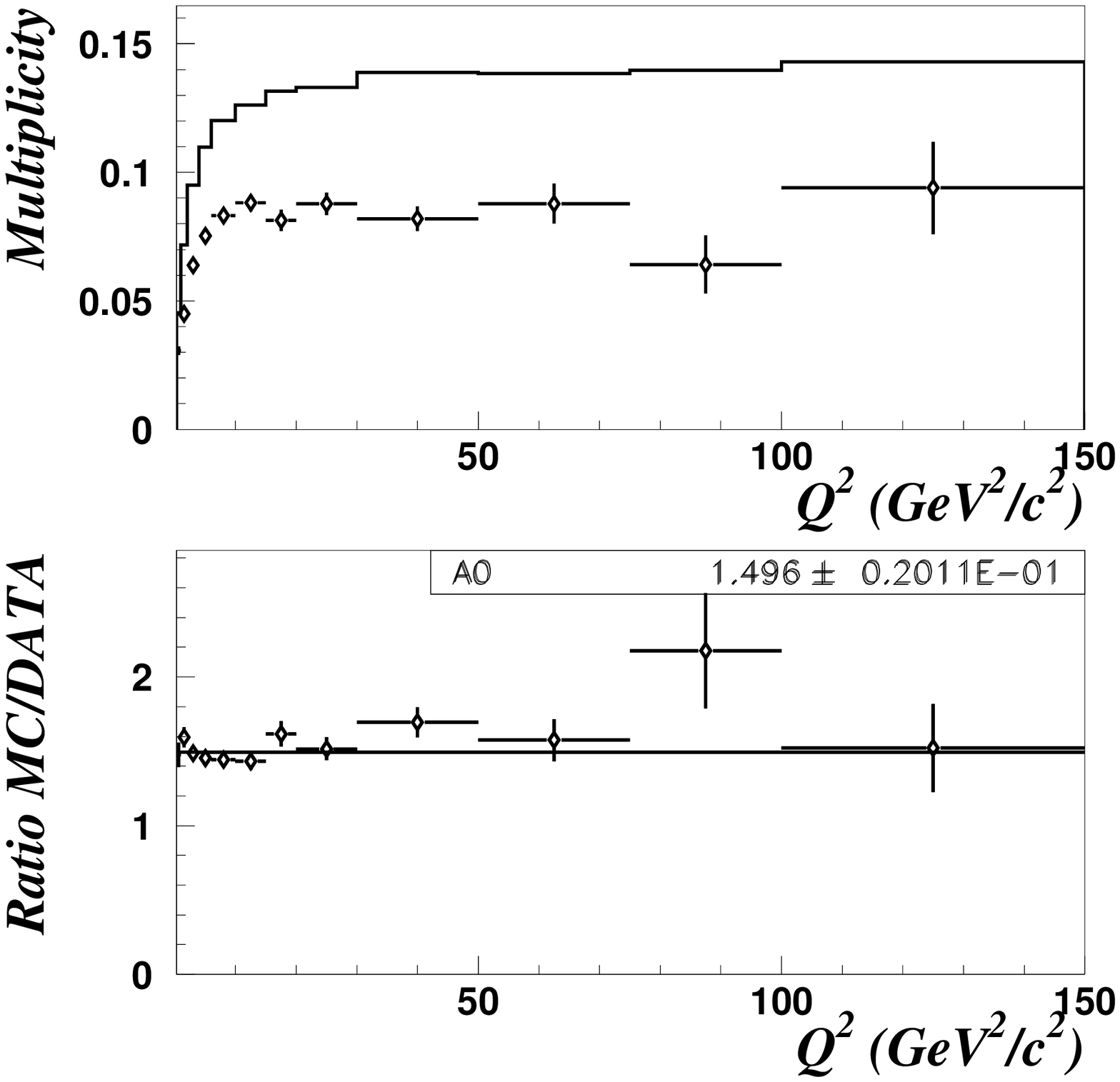,width=0.5\linewidth}}&
\mbox{\epsfig{file=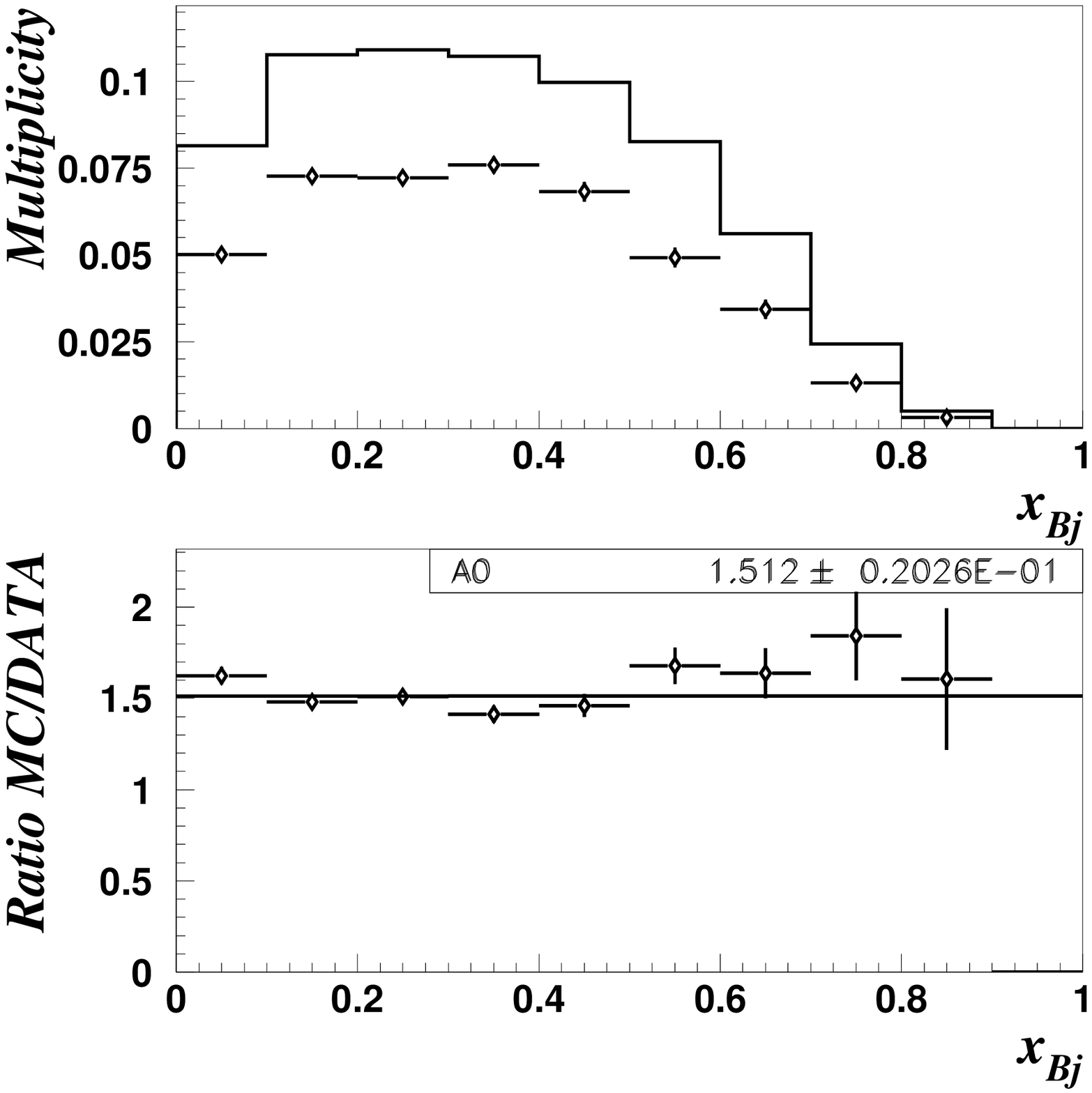,width=0.5\linewidth}}\\
\end{tabular}
}
\protect\caption{\it 
Yields
in the default MC (histogram)
and in the data (points with error bars) for $\lam$ as a function of
$E_\nu$, $W^2$, $Q^2$
and $x_{Bj}$. The MC/Data ratios and their fit to a constant are also shown.
}
\label{fig:lambda_yields}
\end{figure}

\begin{figure}[htb]
\center{%
\begin{tabular}{cc}
\mbox{\epsfig{file=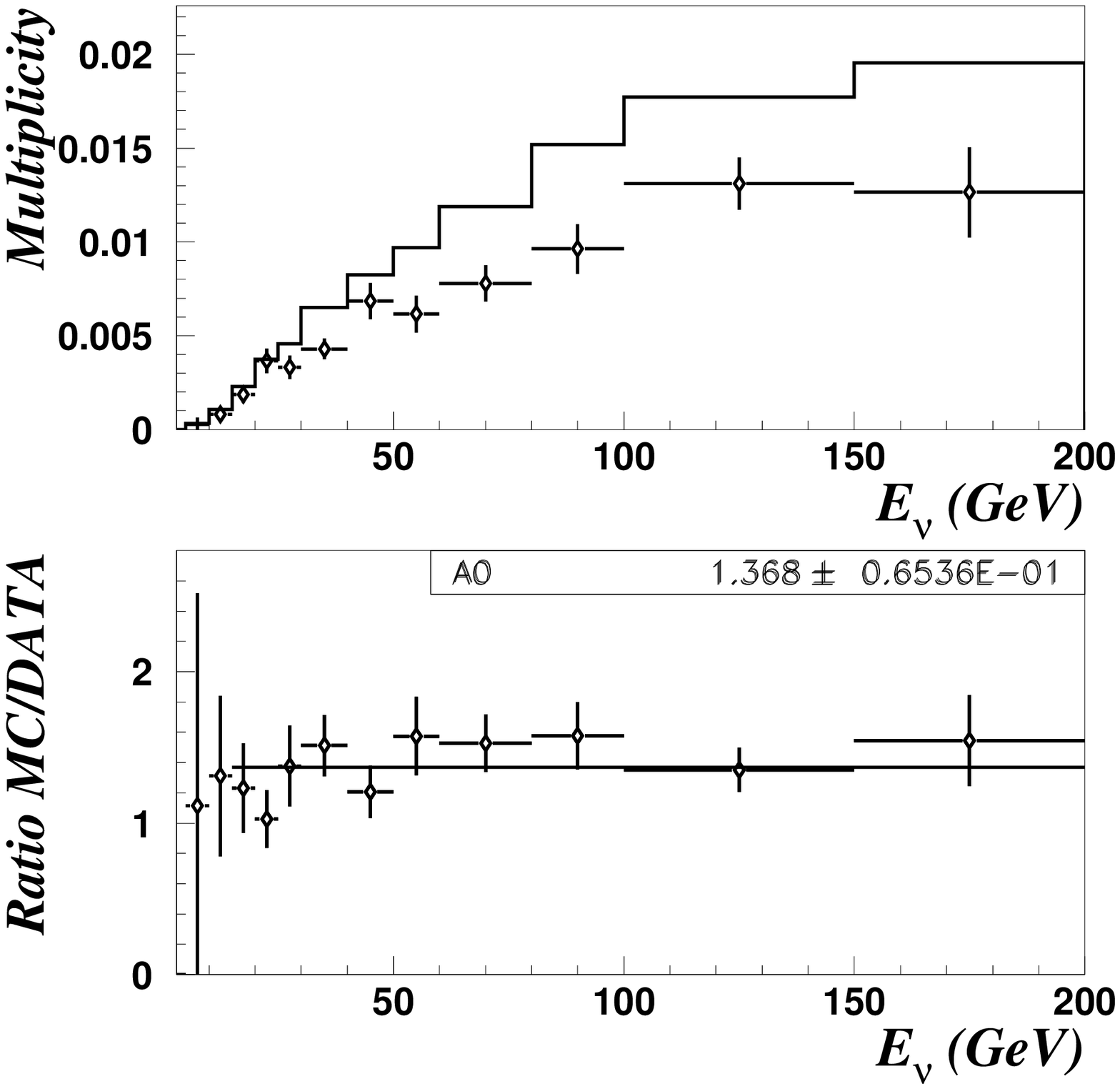,width=0.5\linewidth}}&
\mbox{\epsfig{file=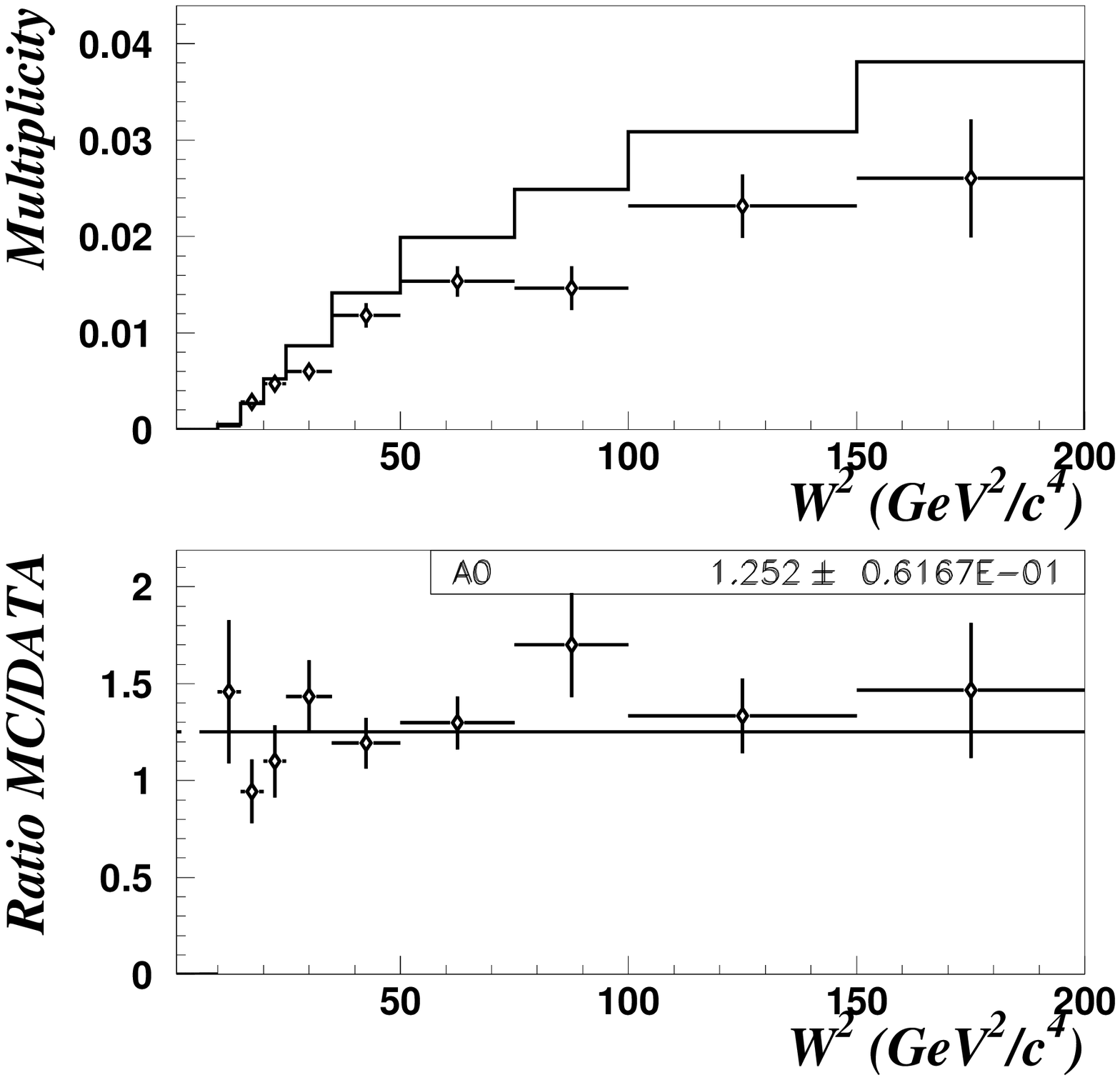,width=0.5\linewidth}}\\
\mbox{\epsfig{file=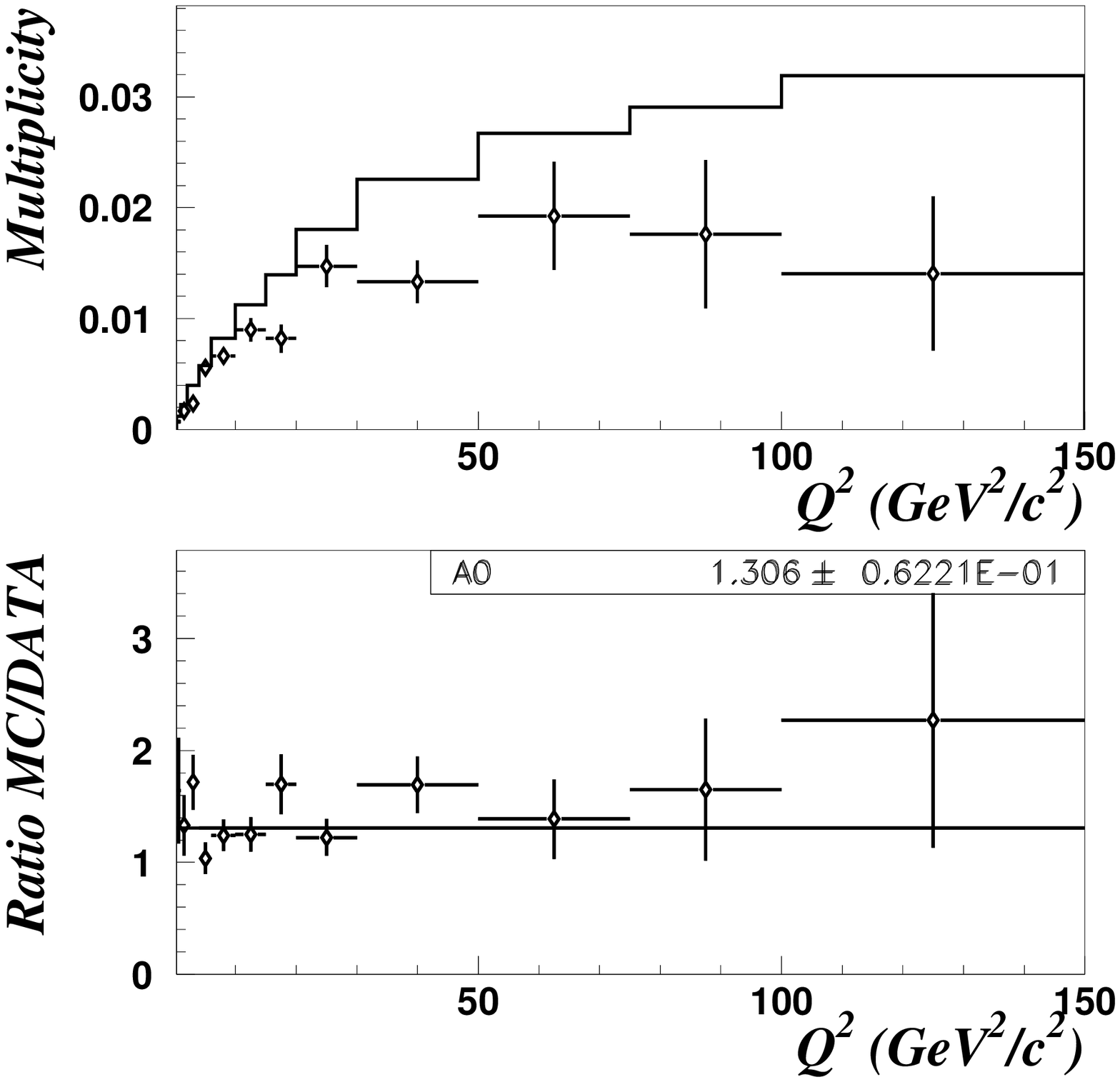,width=0.5\linewidth}}&
\mbox{\epsfig{file=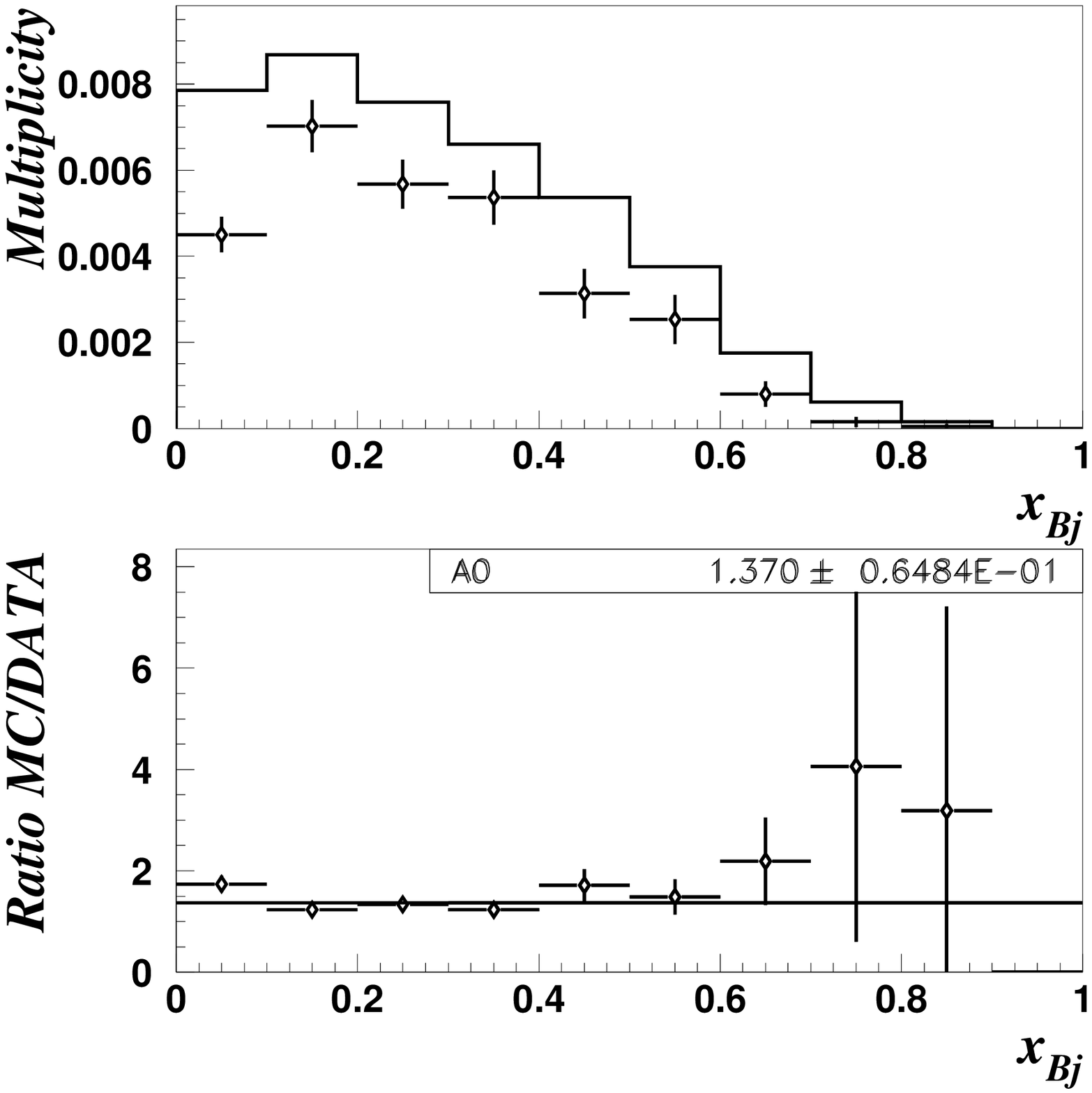,width=0.5\linewidth}}\\
\end{tabular}
}
\protect\caption{\it 
Yields
in the default MC (histogram)
and in the data (points with error bars) for $\alam$ as a function of
$E_\nu$, $W^2$, $Q^2$
and $x_{Bj}$. The MC/Data ratios and their fit to a constant are also shown.
}
\label{fig:antilambda_yields}
\end{figure}

The measured 
yields
of 
$\ko$, $\lam$ and $\alam$ particles
are compared 
in Figs.~\ref{fig:k0_yields}, \ref{fig:lambda_yields},
\ref{fig:antilambda_yields} to the
predictions of the default NOMAD MC simulation (see section~\ref{sec:nomad}).
A reasonable agreement in shapes is observed, 
while the discrepancy in the overall 
normalization is about a factor of 1.3 to 1.5.

The kinematic variables $E_\nu$, $W^2$, $Q^2$ and $x_{Bj}$ are not
independent. So, for example, discrepancies between the data and MC at
high $W^2$ are reflected in discrepancies at low $x_{Bj}$.

\section{PRODUCTION PROPERTIES\label{sec:prod_properties}}

We have also performed a detailed analysis of kinematic quantities 
describing the behaviour of neutral strange particles ($\ko$, $\lam$ and $\alam$)
inside the hadronic jet. This study allows an investigation of the
dynamics of fragmentation.
The differences in the production properties of $\ko$, $\lam$ and
$\alam$ are seen most clearly here.
The following distributions are of interest: $x_F = 2 p_L^*/W$ 
(Feynman-$x$ is the longitudinal momentum fraction in the hadronic center
of mass system), the transverse momentum squared,
$p_T^2$, of a particle with respect to the current (hadronic jet)
direction and 
the fraction $z = E_{lab}(V^0)/E_{lab}(\mbox{all hadrons})$ of
the total hadronic energy carried away by the neutral strange particle
in the laboratory system.

\subsection{$x_F$ distributions}

\begin{figure}[htb]
\begin{center}
\begin{minipage}{0.48\linewidth}
\mbox{\epsfig{file=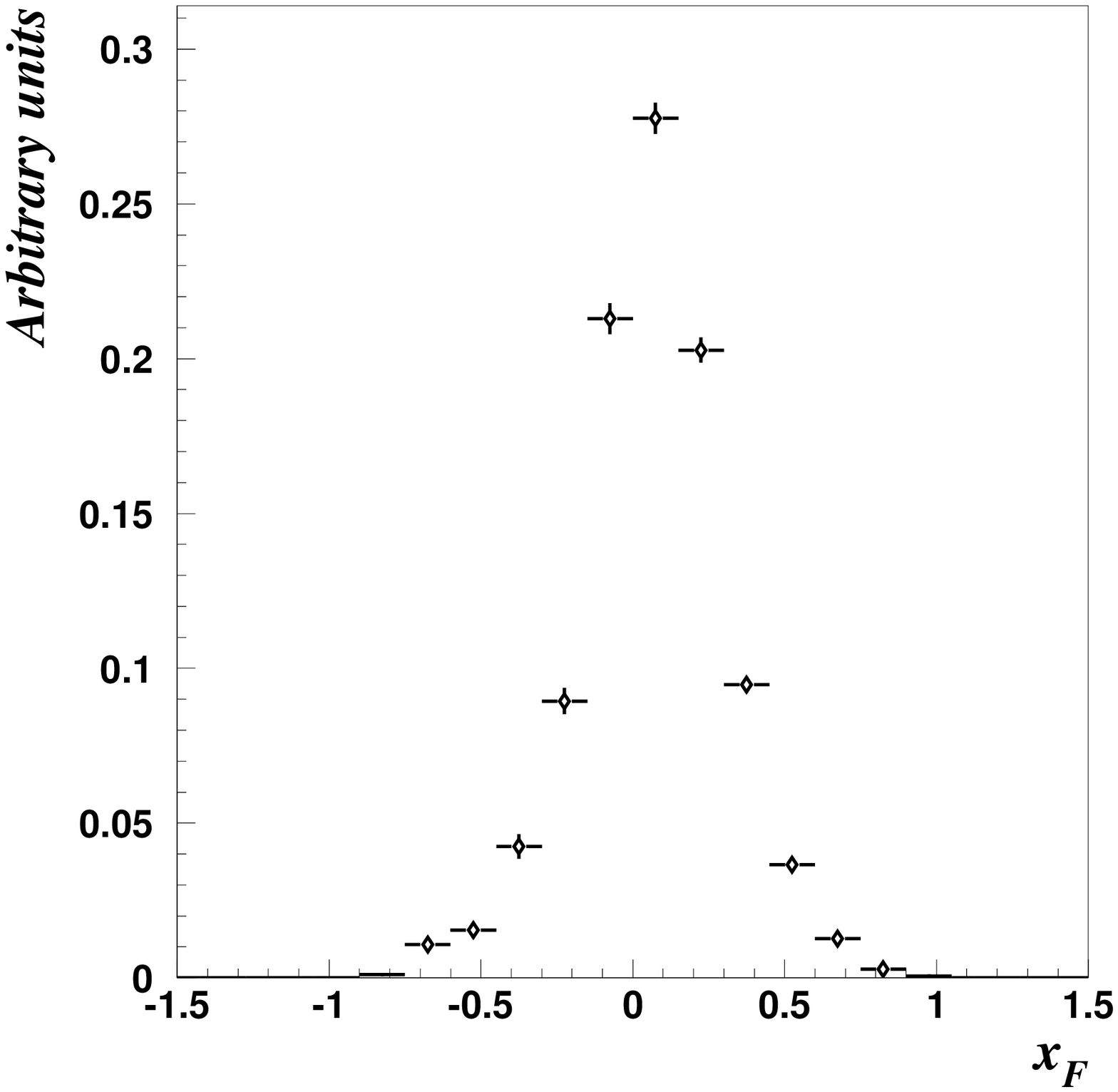,width=0.95\linewidth}}
\protect\caption{\label{fig:xf_k0}\it
Efficiency corrected $x_F$ distribution for $\ko$ mesons.}
\end{minipage}
\hfill
\begin{minipage}{0.48\linewidth}
\mbox{\epsfig{file=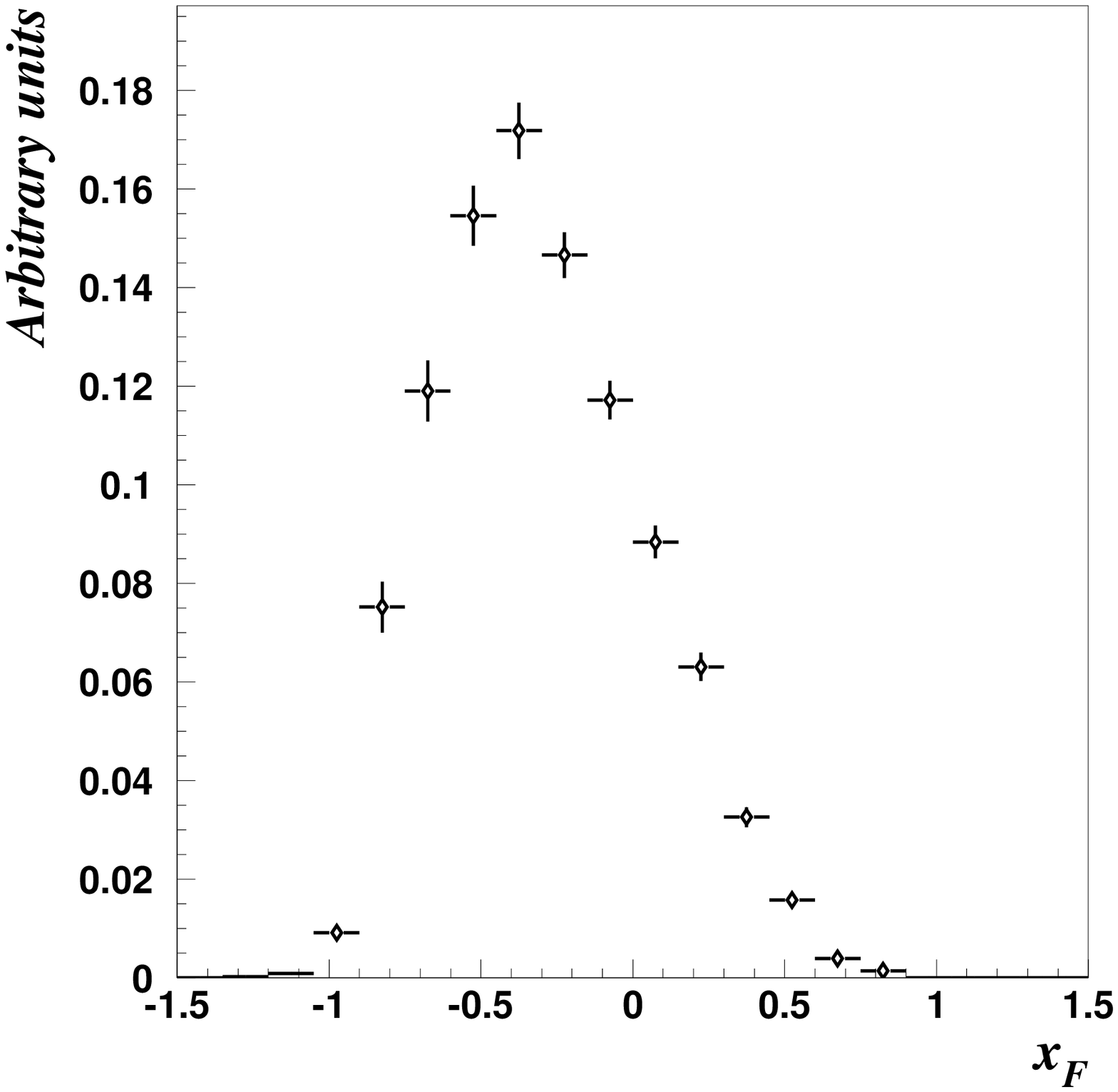,width=0.95\linewidth}}
\protect\caption{\label{fig:xf_lam}\it  
Efficiency corrected $x_F$ distribution for $\lam$ hyperons.}
\end{minipage}
\end{center}
\end{figure}

\begin{figure}[htb]
\begin{center}
\mbox{\epsfig{file=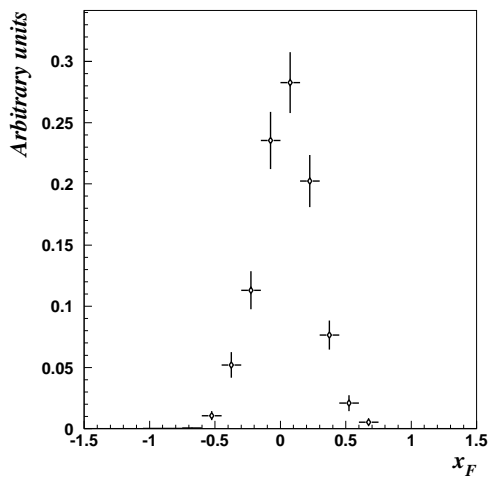,width=0.5\linewidth}}\\
\protect\caption{\label{fig:xf_alam}\it  
Efficiency corrected $x_F$ distribution for $\alam$ hyperons.}
\end{center}
\end{figure}
The efficiency corrected $x_F$ distributions observed in the data for 
neutral strange particles
are shown in Figs.~\ref{fig:xf_k0}, \ref{fig:xf_lam}, \ref{fig:xf_alam}.
These distributions indicate that $\lam$ are produced
mainly in the target fragmentation region ($x_F < 0$), while
$\ko$ are peaked in the central region with an asymmetry in the 
forward direction.
$\alam$ are produced in the central $x_F$
region ($|x_F| < 0.5$). One way to quantify differences in the $x_F$
distributions is to define an asymmetry parameter 
$A = (N_F - N_B)/(N_F + N_B)$, where $N_F$ and $N_B$ are the numbers
of particles produced forward and backward, respectively, in the hadronic
centre of mass. The asymmetry parameters $A$ and mean values of $x_F$ in
both data and MC are listed in Table~\ref{tab:xf_corrected}.
They are 
consistent with previous 
measurements~\cite{grassler,Bosetti,allasia,FNAL,Willocq,BEBC,Prospo}.
The observed disagreement between data and MC 
is probably due to the fact that the MC simulation 
does not properly describe the relative contributions of different $\vo$ 
production mechanisms.

\begin{table}[htb]
\begin{center}
\caption{\label{tab:xf_corrected} 
\it Mean values $\langle x_F \rangle$ 
and asymmetry parameters $A$ of the $x_F$ distributions
for $\ko$, $\lam$, $\alam$ in both MC and data.}
\vspace*{0.5cm}
\begin{tabular}{||c|c|c|c|c||}
\hline\hline
$\vo$ &\multicolumn{2}{|c|}{MC} & \multicolumn{2}{|c||}{DATA}\\
\cline{2-5}
&$\langle x_F \rangle$ & A         & $\langle x_F \rangle$ & A\\
\hline\hline
$\ko$     &$0.055 \pm 0.001$     & $ 0.152 \pm 0.002$
          &$0.064 \pm 0.001$     & $ 0.256 \pm 0.004$\\
$\lam$    &$-0.296 \pm 0.001$    & $-0.649 \pm 0.002$
          &$-0.295 \pm 0.002$    & $-0.589 \pm 0.004$\\
$\alam$   &$0.006 \pm 0.002$     & $-0.03 \pm 0.01$
          &$0.04 \pm 0.004$      & $0.18 \pm 0.02$\\
\hline\hline
\end{tabular}
\end{center}
\end{table}

\subsection{$z$ distributions}

\begin{figure}[htb]
\begin{center}
\begin{minipage}{0.48\linewidth}
\mbox{\epsfig{file=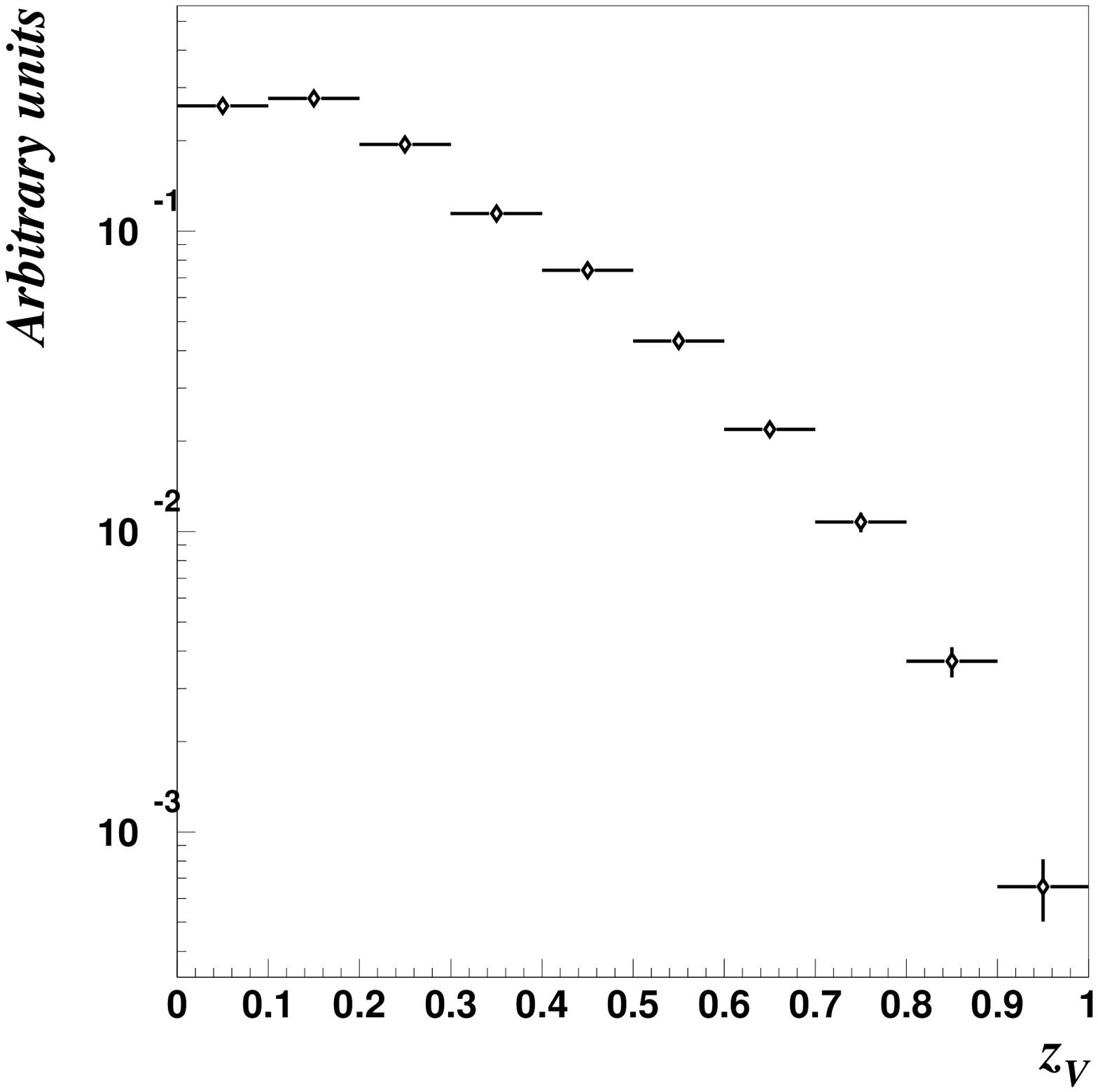,width=0.95\linewidth}}
\protect\caption{\label{fig:z_k0}\it 
Efficiency corrected $z$ distribution for $\ko$.}
\end{minipage}
\hfill
\begin{minipage}{0.48\linewidth}
\mbox{\epsfig{file=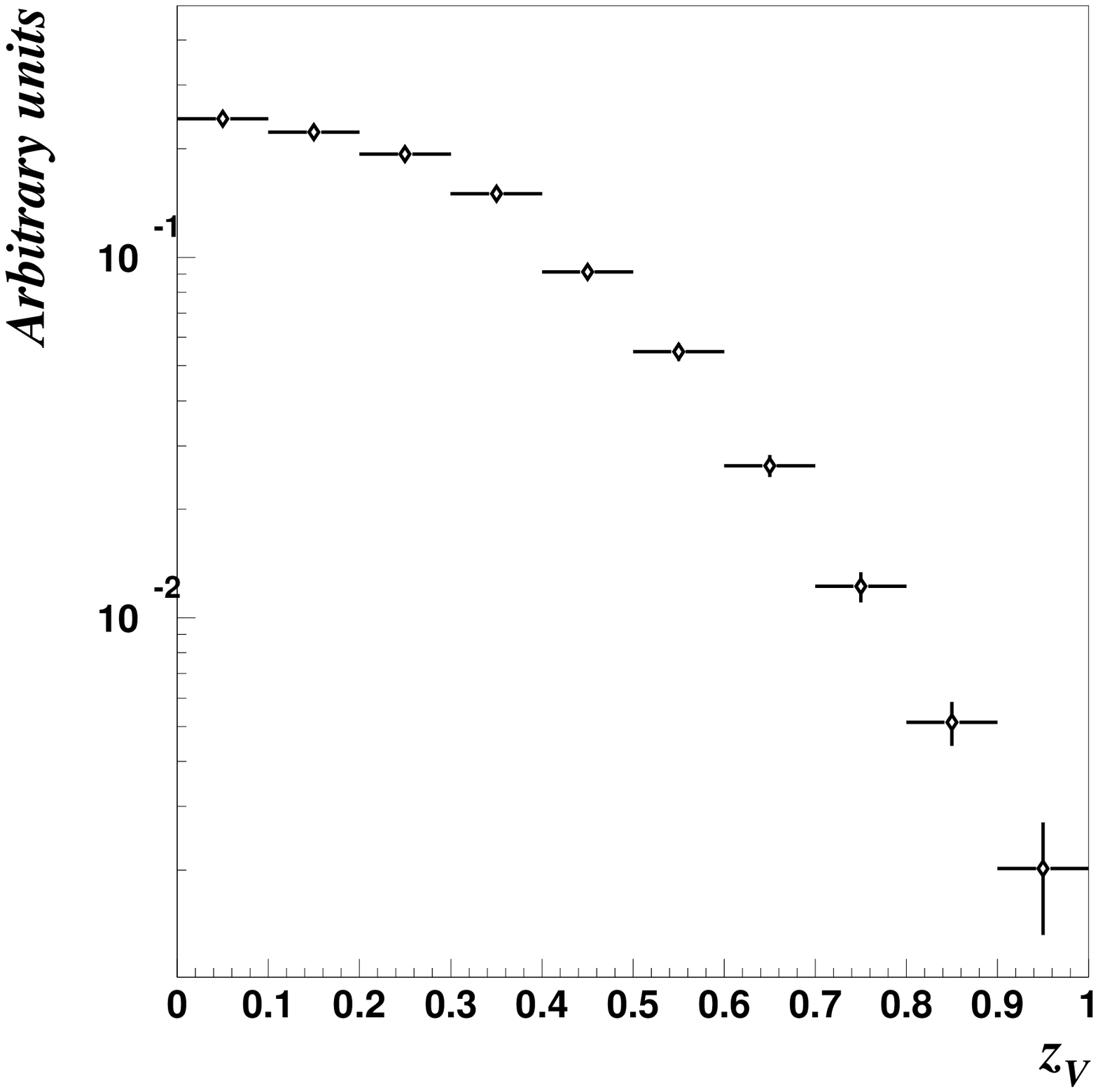,width=0.95\linewidth}}
\protect\caption{\label{fig:z_lam}\it 
Efficiency corrected $z$ distributions for $\lam$.}
\end{minipage}
\end{center}
\end{figure}

\begin{figure}[htb]
\begin{center}
\mbox{\epsfig{file=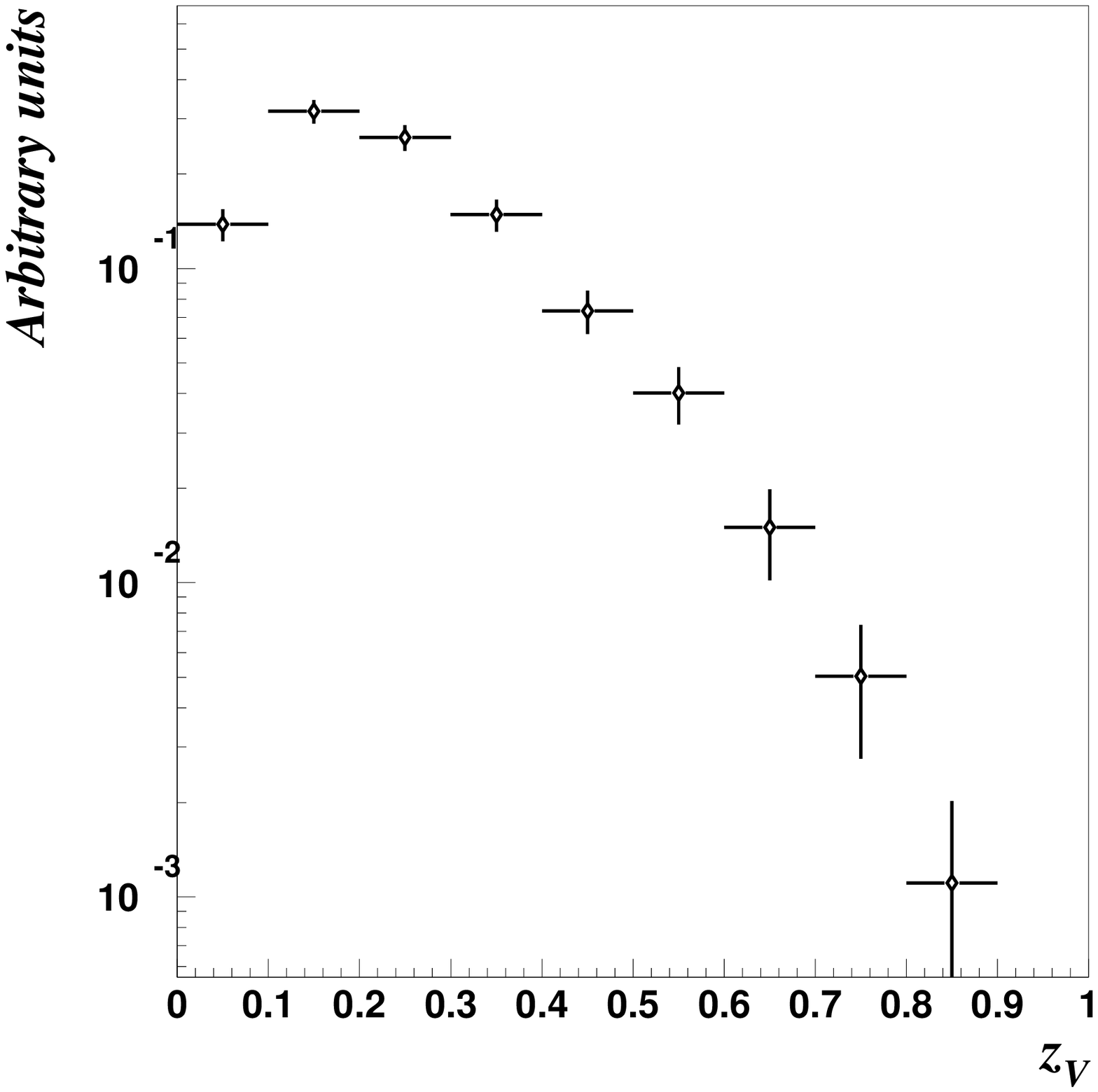,width=0.5\linewidth}}\\
\protect\caption{\label{fig:z_alam}\it 
Efficiency corrected $z$ distributions for $\alam$.}
\end{center}
\end{figure}

Efficiency corrected $z$ distributions for $\ko$, $\lam$ and $\alam$ 
are shown in Figs.~\ref{fig:z_k0}, \ref{fig:z_lam}, \ref{fig:z_alam}.
A turn-over at small values of $z$ can be seen for $\alam$ hyperons,
but not for $\ko$ and $\lam$.
A turn-over at small values of $z$ for $\ko$ and $\lam$
was observed in some of the previous neutrino experiments~\cite{Baker,grassler,allasia,ammosov3} and
was not observed in others~\cite{Brock,Bosetti}.
We note that uncorrected $z$ distributions show 
such a turn-over for all $\vo$ types in our experiment as well, 
due to a less efficient $\vo$ reconstruction at low momenta.

Below we study separately $z$ distributions of neutral strange
particles produced at $x_F<0$ (see \S~\ref{sec:z_distributions_target})
and at $x_F>0$ (see \S~\ref{sec:z_distributions_current}).
Detailed information on the mean values of $z$ distributions 
for both data and MC is given in Table~\ref{tab:mean_values_z}.

\begin{table}[htb]
{\footnotesize
\begin{center}
\caption{\label{tab:mean_values_z} 
\it Mean values of $z$ distributions for $\ko$,
  $\lam$, $\alam$ measured 
for the full sample 
and for $x_F<0$ and $x_F>0$ 
regions
in both MC and data.}
\vspace*{0.5cm}
\begin{tabular}{||c|c|c|c|c||}
\hline\hline
$\vo$ & & full sample & $x_F<0$ & $x_F>0$ \\
\hline
\raisebox{-1.5ex}{$\ko$}  & MC & $0.218 \pm 0.001$ & $0.092 \pm 0.001$ & $0.312 \pm 0.001$ \\         
       & DATA & $0.226 \pm 0.001$&$0.105 \pm 0.001$&$0.299 \pm
       0.001$\\ \hline

\raisebox{-1.5ex}{$\lam$} & MC &$0.227 \pm 0.001$&$0.179 \pm 0.001$&$0.462 \pm 0.003$  \\        
       & DATA &$0.250 \pm 0.002$&$0.206 \pm 0.001$&$0.434 \pm 0.005$\\ \hline

\raisebox{-1.5ex}{$\alam$} & MC &$0.215 \pm 0.003$&$0.139 \pm 0.002$&$0.296 \pm 0.004$ \\         
        & DATA &$0.242 \pm 0.005$&$0.147 \pm 0.005$&$0.308 \pm 0.008$\\
\hline\hline
\end{tabular}
\end{center}
}
\end{table}

\begin{figure}[htb]
\center{%
\begin{tabular}{cc}
\mbox{\epsfig{file=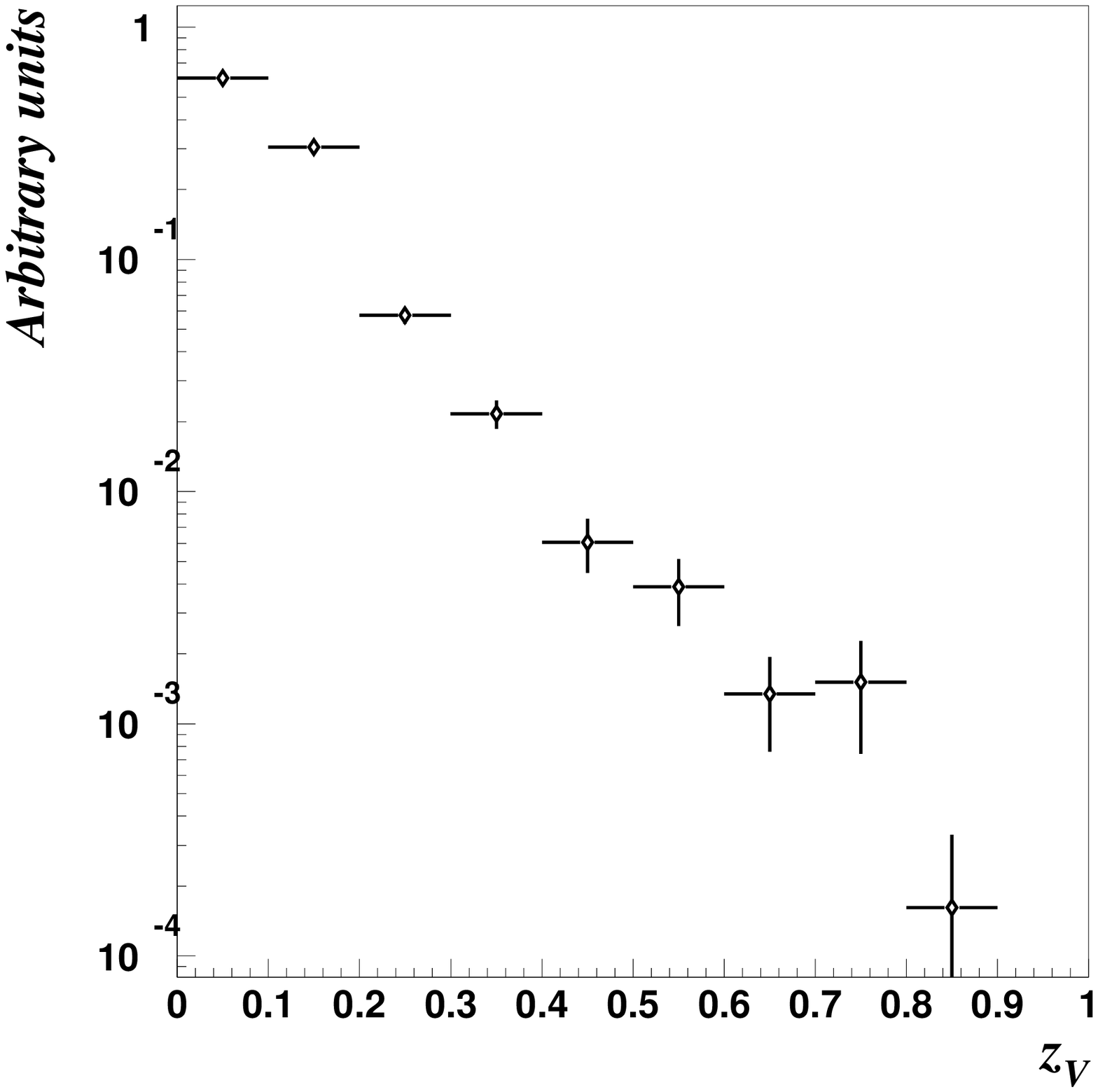,width=0.5\linewidth}}&
\mbox{\epsfig{file=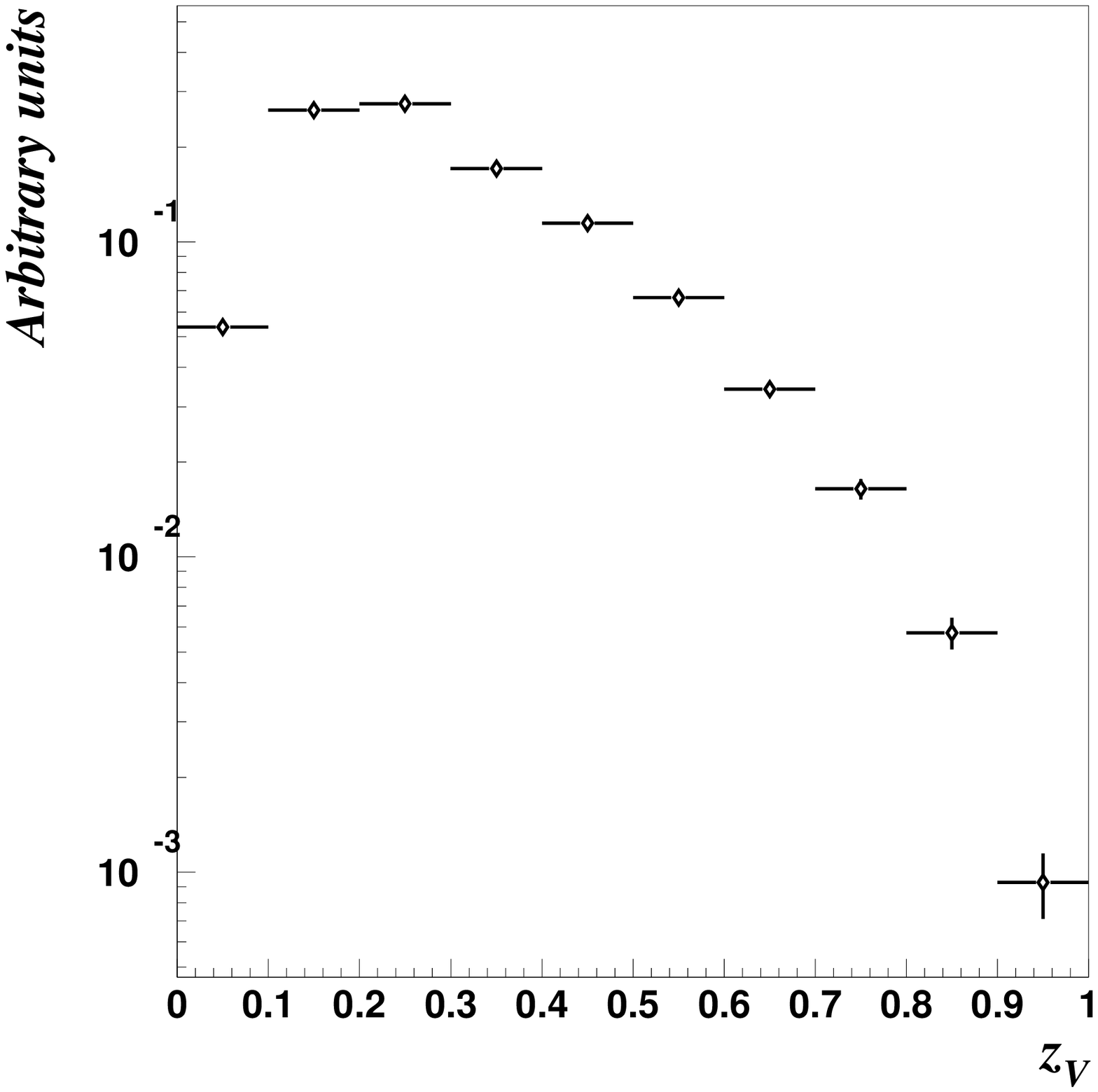,width=0.5\linewidth}}\\
\mbox{\epsfig{file=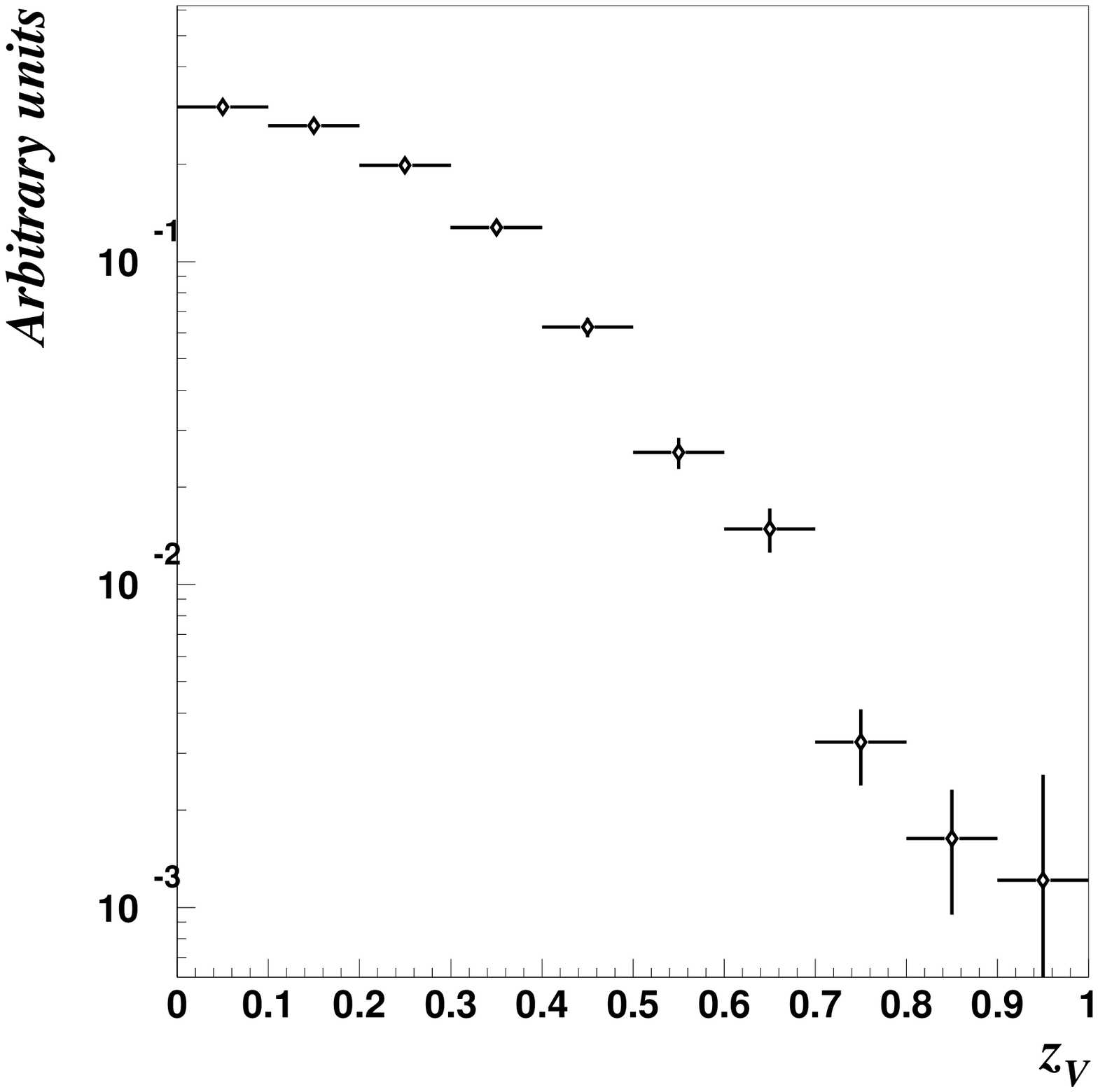,width=0.5\linewidth}}&
\mbox{\epsfig{file=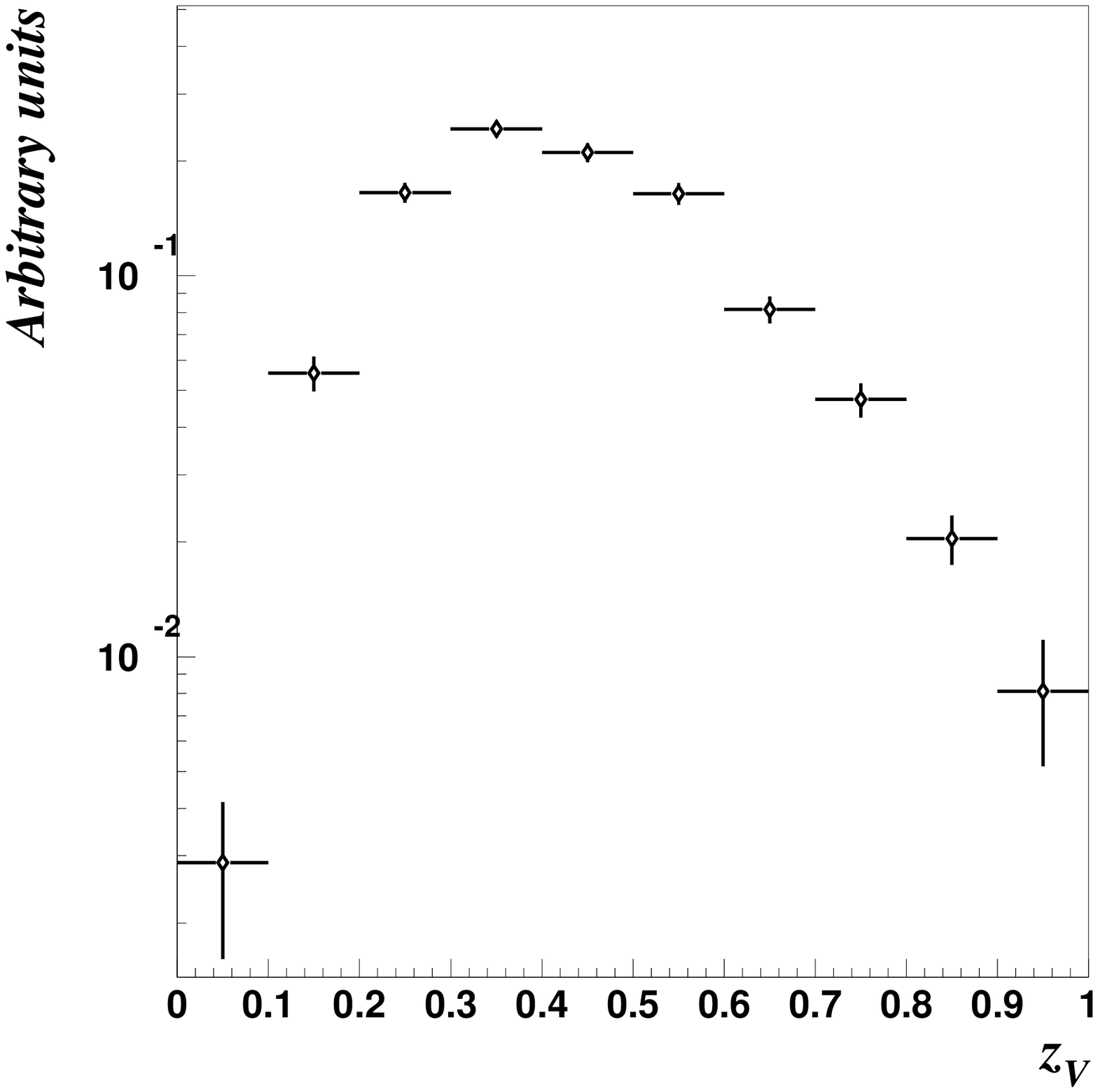,width=0.5\linewidth}}\\
\mbox{\epsfig{file=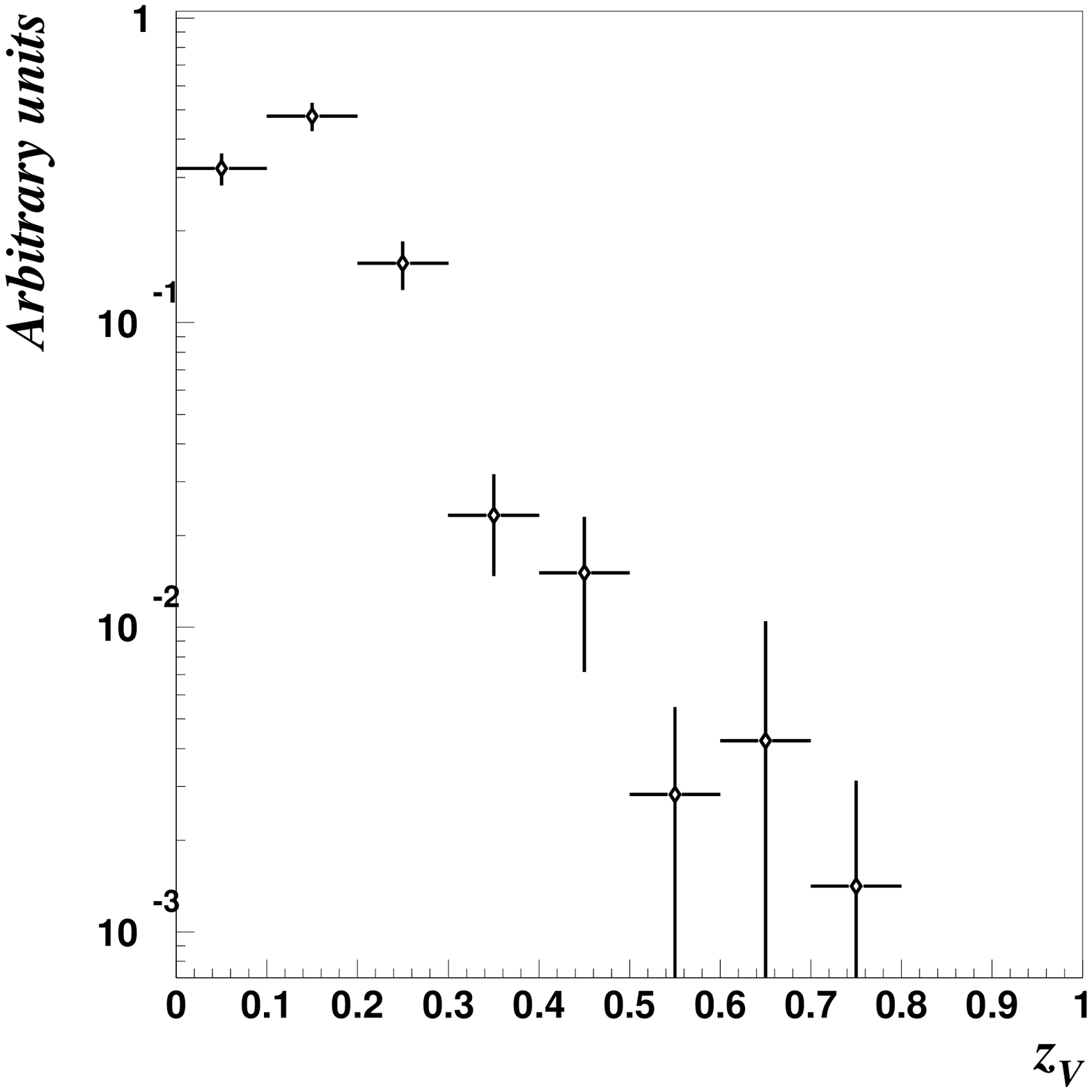,width=0.5\linewidth}}&
\mbox{\epsfig{file=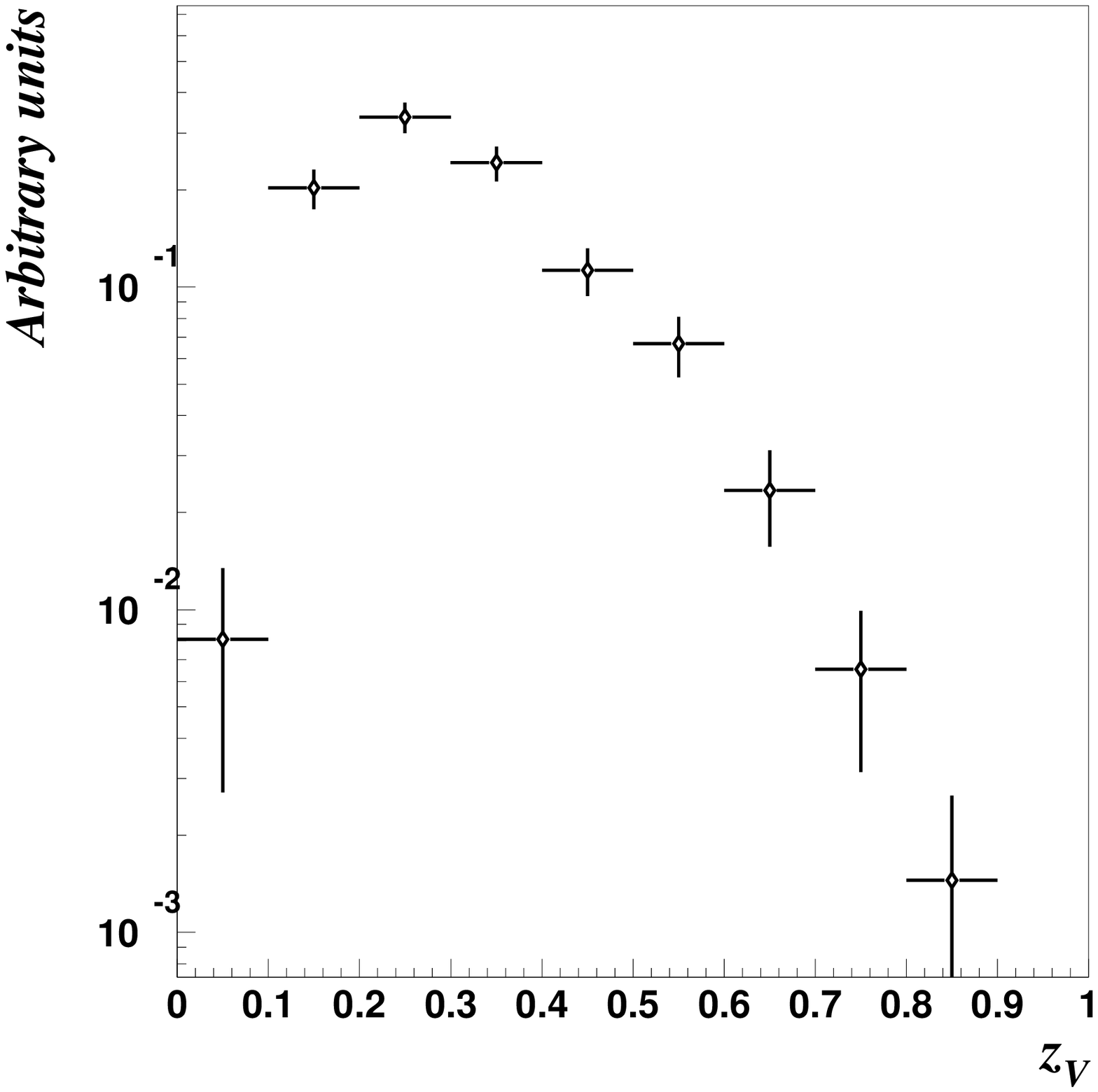,width=0.5\linewidth}}\\
\end{tabular}
}
\protect\caption{\it 
Efficiency corrected $z$ distributions for $x_F<0$ (left) and for $x_F>0$ (right) for  
$\ko$ (top), $\lam$ (centre) and $\alam$ (bottom).
}
\label{fig:z_xf_target_current}
\end{figure}

\subsubsection{$z$ distributions in the target fragmentation region}
\label{sec:z_distributions_target}

The efficiency corrected $z$ distributions
of $\ko$, $\lam$ and $\alam$ measured  in the target fragmentation
region are shown in Fig.~\ref{fig:z_xf_target_current} (left). 
One can see that the $z$ distribution of $\ko$ mesons has a maximum at
$z\to 0$ and decreases faster at larger $z$ (compared with
Fig.~\ref{fig:z_k0}) and that 
the $\ko$ mesons produced in the target fragmentation region carry 
in general
a small fraction of the hadronic jet energy. $\lam$ hyperons are believed
to be produced mostly from the remnant di-quark fragmentation and the shape of
the $z$ distribution is similar to that shown in Fig.~\ref{fig:z_lam}.
The turn-over in the $z$ distribution is
observed for $\alam$ hyperons produced in the target fragmentation region.

\subsubsection{$z$ distributions in the current fragmentation region}
\label{sec:z_distributions_current}

The efficiency corrected $z$ distributions
of $\ko$, $\lam$ and $\alam$ measured  in the current fragmentation
region are shown in Fig.~\ref{fig:z_xf_target_current} (right). 
This kinematical region is interesting because of the $u$ or $\bar
d$ (anti)quark fragmentation into $\lam$ or $\alam$ hyperons. 
All three $z$ distributions show similar 
behaviour
but with different
mean values of $z$. 
The $z$ distribution of $\lam$ hyperons at $x_F>0$ is drastically
different from that in the target fragmentation region. This is
evidence for the fragmentation of the outgoing $u$ quark into a $\lam$
hyperon. In fact, the $z$ distribution of $\lam$ in the current
fragmentation region 
is a measure of the
$D_u^\Lambda(z)$ fragmentation
function (normalized to unity in Fig.~\ref{fig:z_xf_target_current}).
The $z$ distribution of $\alam$ hyperons is sensitive to 
the $D_{\bar d}^{\bar\Lambda}(z)$ fragmentation function with a possible
contribution from the $D_u^{\bar\Lambda}(z)$ process. One can see that
it is harder than the one measured in the target fragmentation region. 

\subsection{Discussion}

There are different mechanisms responsible for $\ko$, $\lam$ and
$\alam$ production in the neutrino CC DIS process which are expected to
give different $x_F$ and $z$ distributions for these particles.

\begin{itemize}
\item[--] The $x_F$ distribution of $\ko$ mesons produced promptly in
  the $W^+ d \to u$ process that requires at least two quark-antiquark pairs to
  be created  $(d\bar d$ and $s\bar s)$
  is expected to be central. A contribution from heavier strange
  particle decays (mainly from ${K^\star}^+$) produced from the fragmentation
  of the outgoing $u$ quark can result in a forward $x_F$ distribution
  for $\ko$ mesons. Also $\ko$ mesons 
from a fragmentation
  of the outgoing (anti)quark in $W^+ d \to c \to s$ and $W^+ \bar u
  \to \bar d$ processes are expected to be produced in the forward
  $x_F$ region  and to carry a larger fraction of the jet energy.
  
\item[--] $\lam$ hyperons can be produced from the fragmentation of
  the nucleon di-quark remnant promptly and via the decay of heavier strange
  baryons at $x_F < 0$. $\lam$ hyperons can be produced also at $x_F
  > 0$ from the outgoing $u$ quark fragmentation.

\item[--] The production of $\alam$ hyperons in neutrino scattering from
  a valence quark requires three quark-antiquark pairs to
  be created  ($u\bar u$, $d\bar d$ and $s\bar s$) and is expected to
  populate the central region of the 
  $x_F$ distribution. There could also be a
  contribution from the 
outgoing
  antiquark fragmentation into
  a $\alam$ hyperon (in the $W^+ \bar u \to \bar d \to \alam$ process)
  which can produce these baryons in the forward $x_F$ region.
\end{itemize}

\subsection{$p_T^2$ distributions}

\begin{figure}[htb]
\begin{center}
\begin{minipage}{0.48\linewidth}
\mbox{\epsfig{file=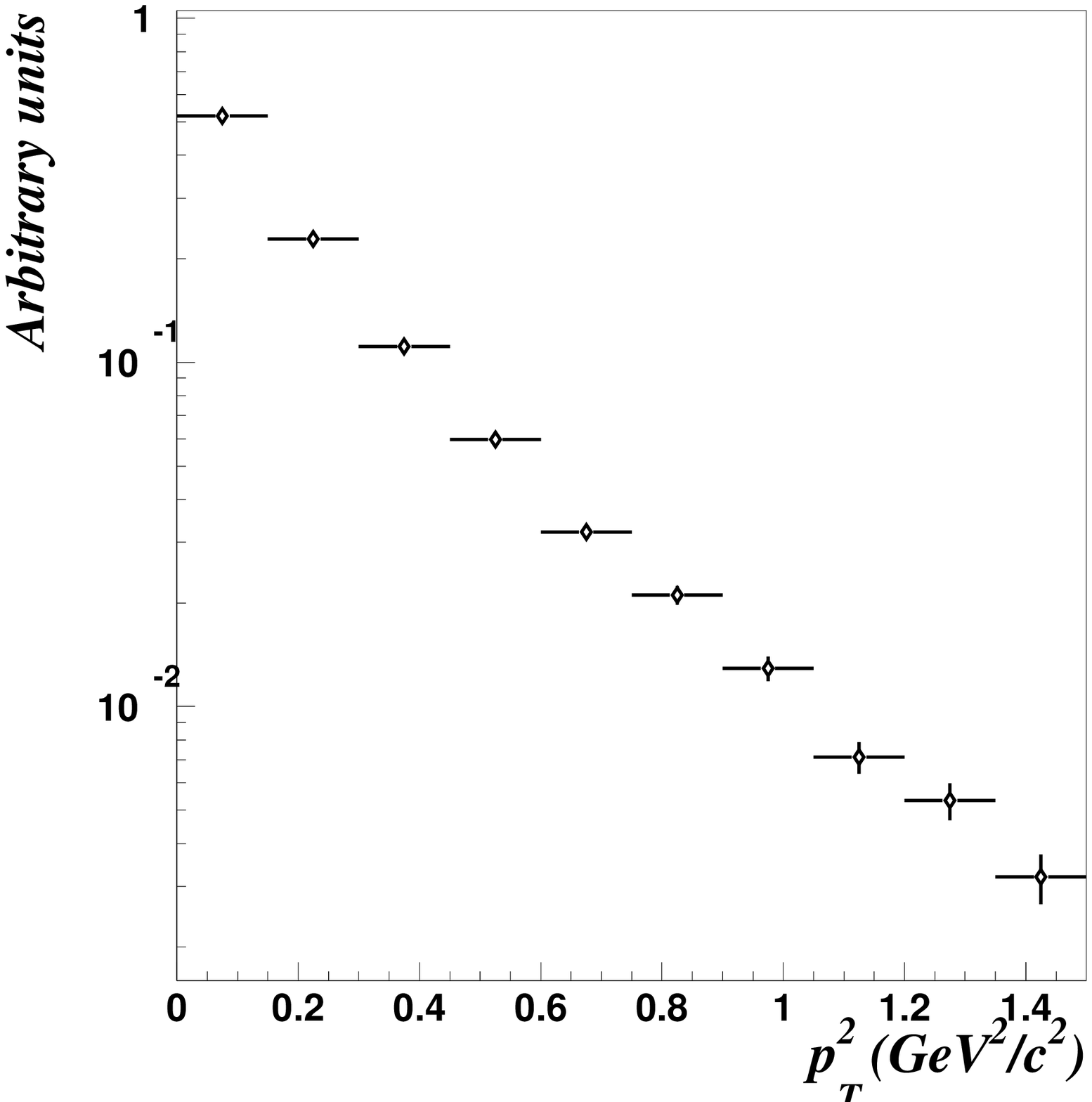,width=0.95\linewidth}}
\protect\caption{\label{fig:pt2_k0}\it 
Efficiency corrected $p_T^2$ distribution for $\ko$.}
\end{minipage}
\hfill
\begin{minipage}{0.48\linewidth}
\mbox{\epsfig{file=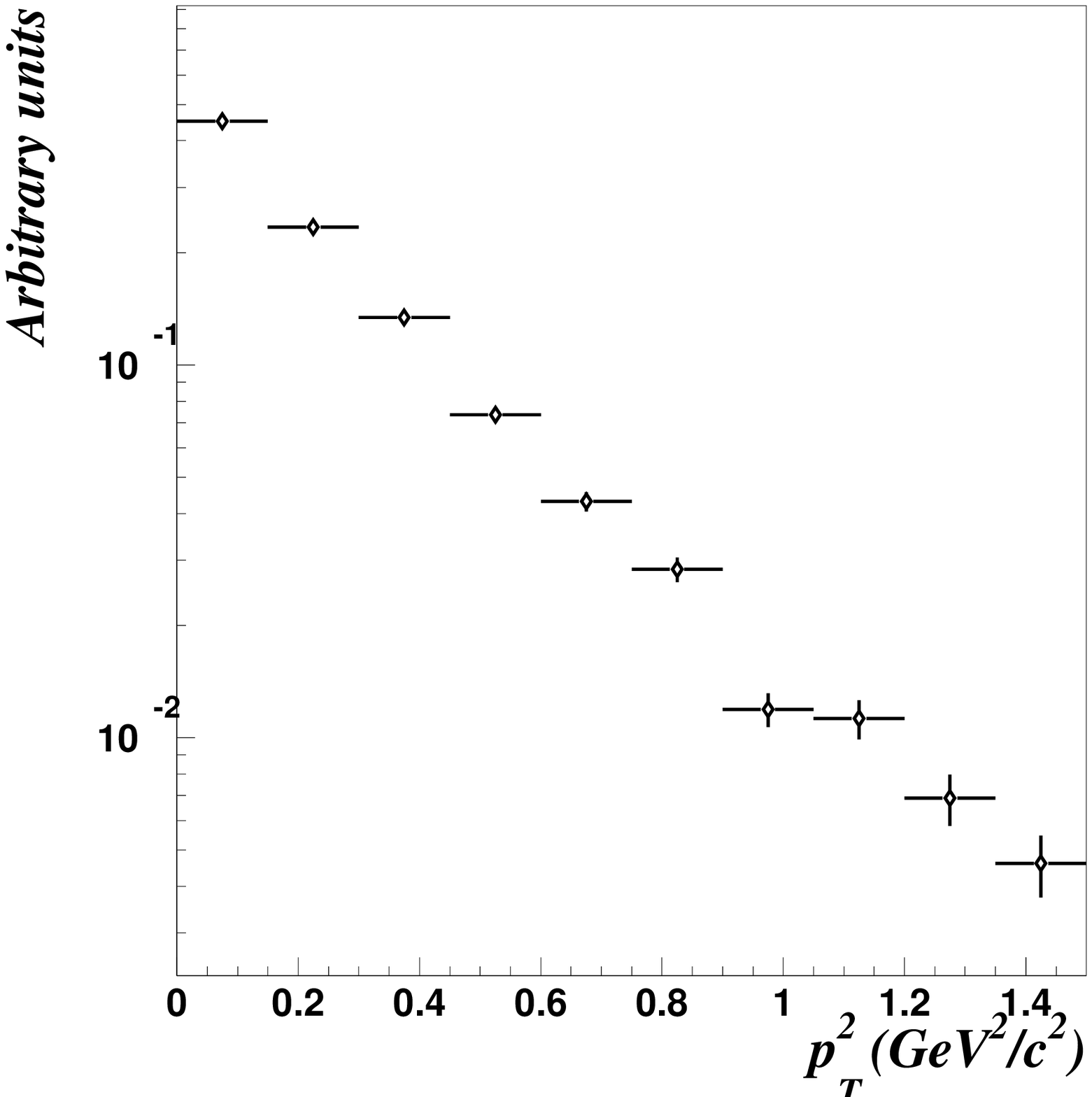,width=0.95\linewidth}}
\protect\caption{\label{fig:pt2_lam}\it 
Efficiency corrected $p_T^2$ distribution for $\lam$.}
\end{minipage}
\end{center}
\end{figure}

\begin{figure}[htb]
\begin{center}
\mbox{\epsfig{file=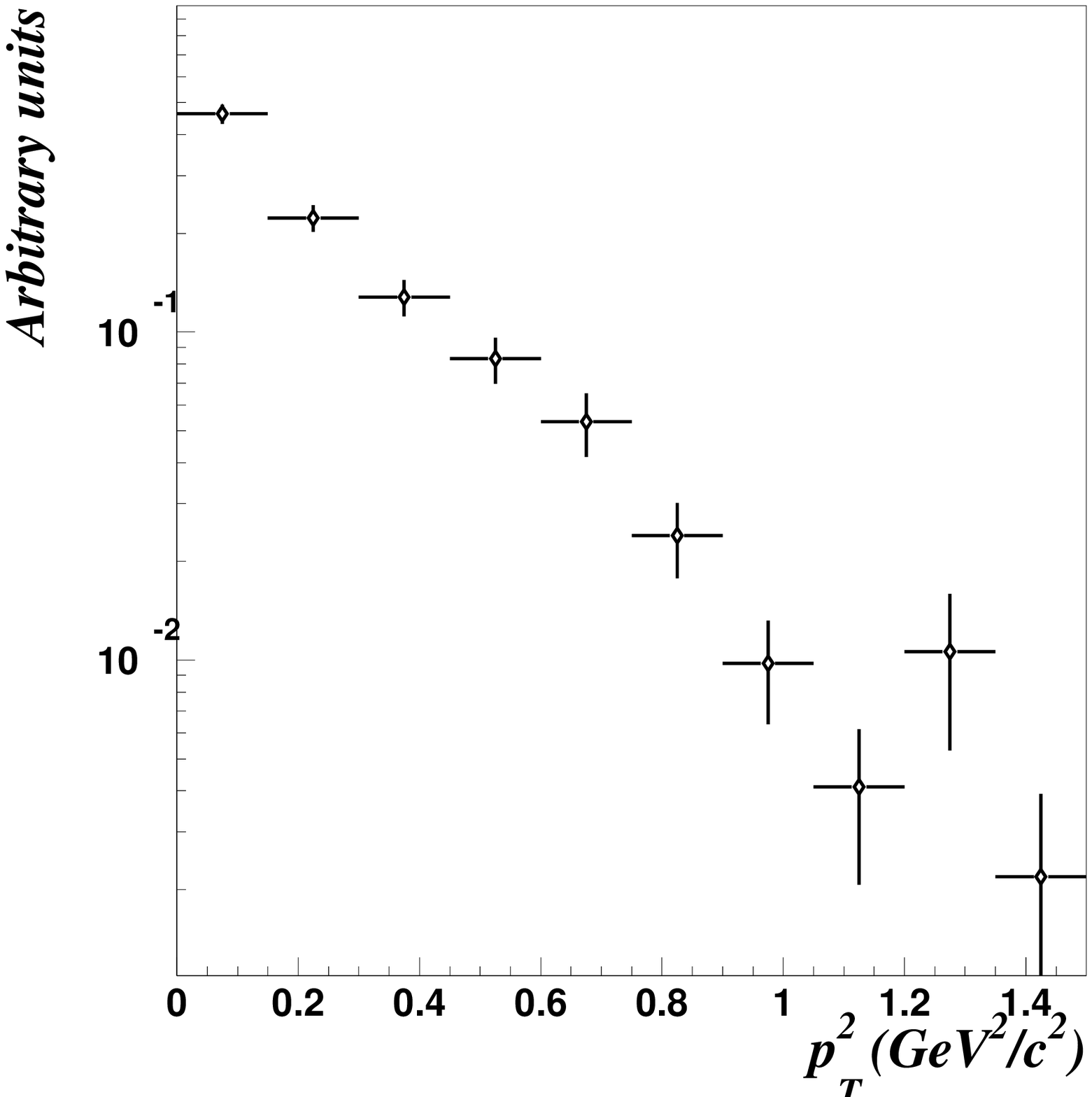,width=0.5\linewidth}}
\protect\caption{\label{fig:pt2_alam}\it 
Efficiency corrected $p_T^2$ distribution for $\alam$.}
\end{center}
\end{figure}

The efficiency corrected $p_T^2$ distributions of $\ko$, $\lam$ and
$\alam$ in the data are 
presented
in Figs.~\ref{fig:pt2_k0}, \ref{fig:pt2_lam}, \ref{fig:pt2_alam}.
They
show
an exponential behaviour of the form $C exp (-B \cdot p_T^2)$
in the region $\rm p_T^2 \lesssim 0.5$ $\rm (GeV^2/c^2)$ and 
a deviation from this dependence at higher $\rm p_T^2$.
We measured the slope parameter $B$ 
in the region $\rm 0 < p_T^2 < 0.5 \ GeV^2/c^2$
for each $\vo$ category in three kinematic regions: 
all
$x_F$, $x_F<0$ and $x_F>0$.
The results
are listed in Table~\ref{tab:pt2_corrected}.
The values of the slope parameter found in different $x_F$ regions are
similar,
except for $\alam$ hyperons.
\begin{table}[htb]
{\small
\begin{center}
\caption{\label{tab:pt2_corrected} 
\it The slope parameter 
  $B (GeV/c)^{-2}$ of the $p_T^2$ distribution for $\ko$, $\lam$, $\alam$
   measured separately for the full sample and for
  $x_F<0$ and $x_F>0$ 
regions
in both MC and in data.}
\vspace*{0.5cm}
\begin{tabular}{||c|c|c|c|c|c|c||}
\hline\hline
$\vo$ &\multicolumn{3}{|c|}{MC} & \multicolumn{3}{|c||}{DATA}\\
\hline\hline
        & full sample & $x_F<0$ & $x_F>0$ 
        & full sample & $x_F<0$ & $x_F>0$ \\
\hline
$\ko$   &$5.72 \pm 0.03$&$5.61 \pm 0.04$&$5.79 \pm 0.03$          
        &$5.21 \pm 0.10$&$5.40 \pm 0.21$&$5.15 \pm 0.11$\\

$\lam$  &$4.35 \pm 0.03$&$4.30 \pm 0.03$&$4.58 \pm 0.06$          
        &$4.12 \pm 0.13$&$4.18 \pm 0.15$&$4.07 \pm 0.27$\\

$\alam$ &$3.89 \pm 0.10$&$4.10 \pm 0.13$&$3.70 \pm 0.14$          
        &$4.42 \pm 0.47$&$6.59 \pm 0.74$&$3.30 \pm 0.64$\\
\hline\hline
\end{tabular}
\end{center}
}
\end{table}

\subsection{$\langle p_T^2 \rangle$ versus $x_F$ distributions}

\begin{figure}[htb]
\begin{center}
\begin{minipage}{0.48\linewidth}
\mbox{\epsfig{file=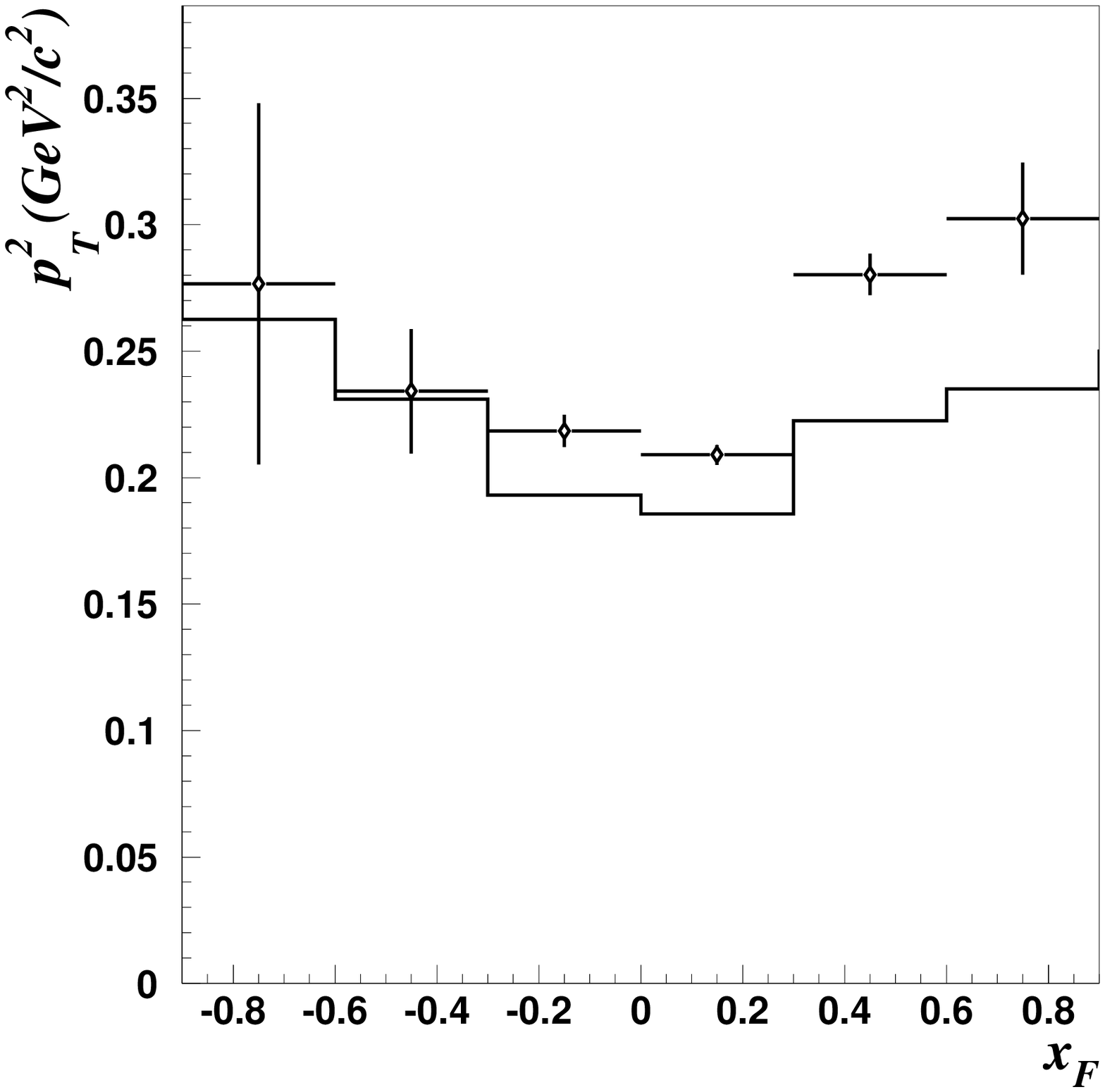,width=0.95\linewidth}}
\protect\caption{\label{fig:pt2_vs_xf_k0}\it 
Efficiency corrected $\langle p_T^2 \rangle$ versus $x_F$ 
distribution for $\ko$ in data (points with error bars) and MC (histogram).}
\end{minipage}
\hfill
\begin{minipage}{0.48\linewidth}
\mbox{\epsfig{file=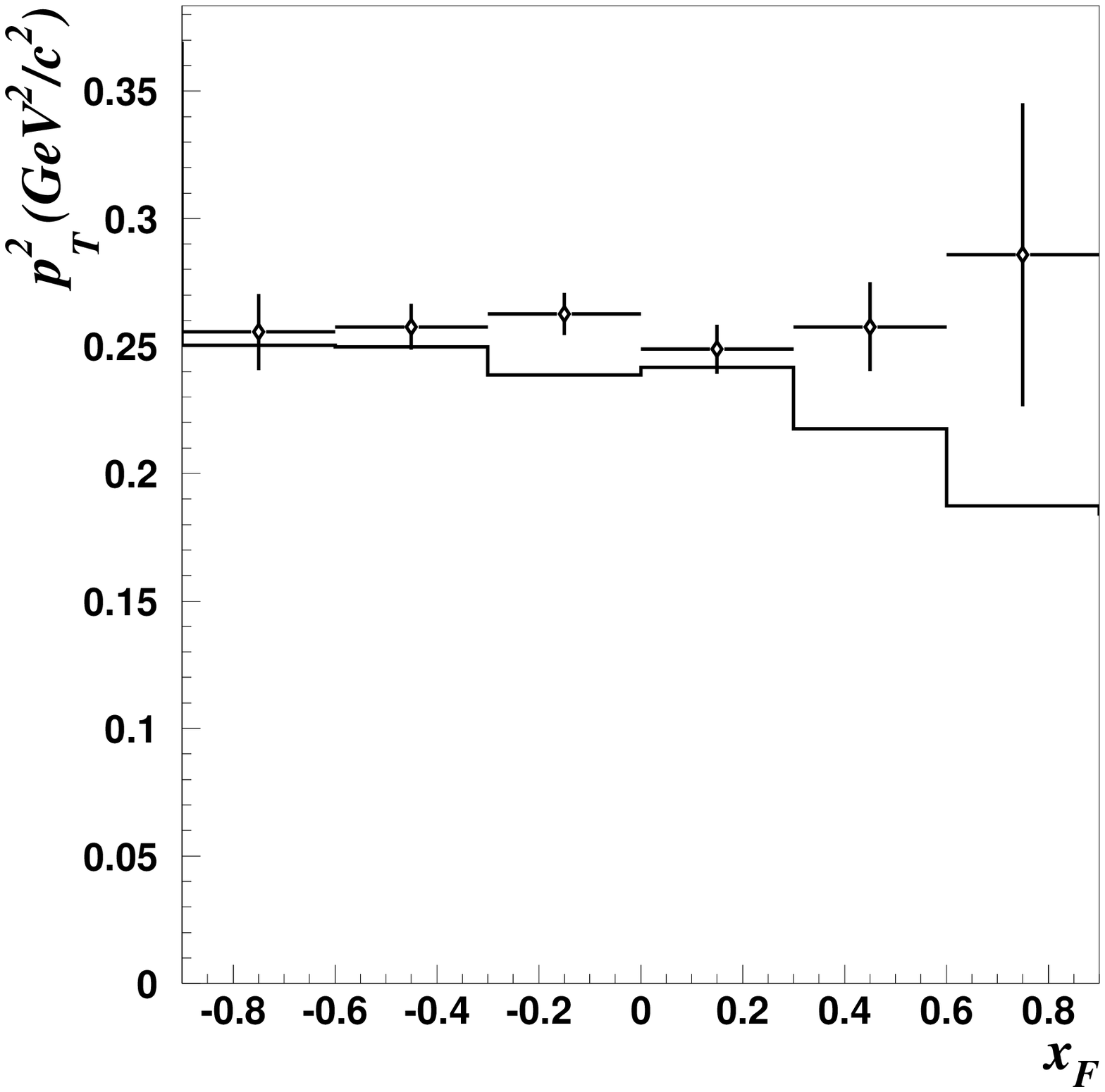,width=0.95\linewidth}}
\protect\caption{\label{fig:pt2_vs_xf_lam}\it 
Efficiency corrected $\langle p_T^2 \rangle$ versus $x_F$ 
distribution for $\lam$ in data (points with error bars) and MC (histogram).}
\end{minipage}
\end{center}
\end{figure}

\begin{figure}[htb]
\begin{center}
\mbox{\epsfig{file=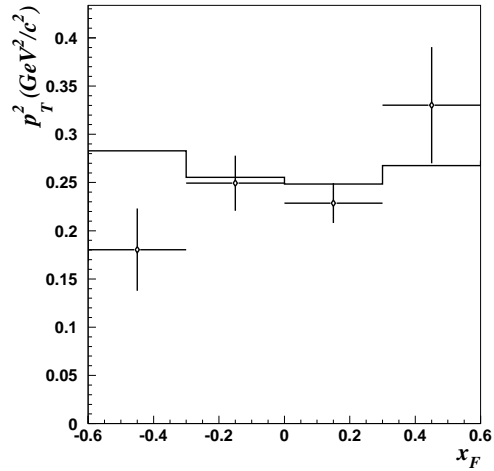,width=0.5\linewidth}}
\protect\caption{\label{fig:pt2_vs_xf_alam}\it 
Efficiency corrected $\langle p_T^2 \rangle$ versus $x_F$ 
distribution for $\alam$ in data (points with error bars) and MC (histogram).}
\end{center}
\end{figure}

We have also studied the dependence of the average $\langle p_T^2 \rangle$ 
on $x_F$ for $\ko$, $\lam$ and $\alam$ (see Figs.~\ref{fig:pt2_vs_xf_k0}, 
\ref{fig:pt2_vs_xf_lam}, \ref{fig:pt2_vs_xf_alam}). 
For the first time in a neutrino experiment 
the good quality of event reconstruction combined
with the large statistics of the data collected
allows the study of these distributions for neutral
strange particles\footnote{Similar distributions obtained for charged 
particles in bubble chamber neutrino experiments~\cite{pt2_xf_bebc} 
have been used to tune Monte Carlo simulation programs.}. 
The observed discrepancy between the data and simulated 
events in the region $x_F \gtrsim 0.3$ 
could be attributed to the absence of QCD effects in our Monte Carlo
simulation program:
the so-called soft-gluon effect could change the leading particle 
($x_F \rightarrow 1$) behaviour
inside the hadronic jet since the forward scattered quark is strongly 
accelerated and is therefore expected to radiate gluons, thus broadening 
the forward $p_T^2$ distribution. It has also been verified that there is
no accumulation of misidentified
$\vo$ in the region where the disagreement between the MC simulation
and the data is observed.

\section{STRANGE RESONANCES AND HEAVIER HYPERONS\label{sec:resonances}}

Apart from many good physics reasons
a study of the production of resonances and heavy hyperons
is also of great 
importance for tuning the LUND model parameters 
and for the theoretical interpretation of the $\lam$ and $\alam$
polarization 
measurements reported in our previous articles~\cite{NOMAD_lam,NOMAD_alam}.
This is essential because $\lam$ hyperons originating from the decays 
$\Sigma^\star \rightarrow \Lambda \pi$,
$\Sigma^0 \rightarrow \Lambda \gamma$ and $\Xi \rightarrow \Lambda \pi$
inherit a polarization from the parent 
particles and this polarization is different from that of a 
directly produced $\lam$.
Information about 
$\Sigma^0$, $\Xi$,
$\Sigma^\star$, $\Xi^\star$
and $K^\star$ 
yields 
can be obtained from an analysis of their decays  into channels 
containing
identified neutral strange particles~\cite{Dima_thesis,Cyril_thesis}.

Previous bubble chamber experiments with (anti)neutrino beams suffered from 
a lack of statistics. For example, the BEBC WA21 Collaboration~\cite{BEBC} reported 
the observation of $149 \pm 29$ ${K^\star}^+$, $42 \pm 19$ ${K^\star}^-$, 
$134 \pm 19$ ${\Sigma^\star}^+$ and less than 10 ${\Sigma^\star}^-$ in
$\nu_\mu p$ CC interactions, 
while the Fermilab 15-ft bubble chamber E380 Collaboration~\cite{FNAL} found 
$94 \pm 25$ $\Sigma^0$ and 4 $\Xi^-$ in $\nu_\mu$Ne CC events.

\subsection{A procedure for signal extraction} \label{sec:signal}

Our aim was to extract the fraction of neutral strange particles which are
decay products of resonances and heavier hyperons from the corresponding 
invariant mass distributions.
To construct such distributions we 
combine the neutral strange particle
with all possible charged tracks (of appropriate sign) emerging from the 
primary vertex except those identified as muons or electrons.
We have also studied the ($\lam$ $\gam$) combinations, where photons
are identified as conversions in the detector fiducial volume via our $\vo$
identification procedure.
The resulting distributions are fitted by a function describing
both the combinatorial background and the resonance signal.

The combinatorial background (BG) can be approximated by any function of
the form:
\begin{equation}
BG = P_n(m-M_{th}) \cdot Tail(m),
\end{equation}
where $P_n(m-M_{th})$ is a polynomial of order $n$ 
vanishing at $m=M_{th}$, 
$M_{th} = M_{V^0} + m_{\pi}$, and $Tail(m)$ is
any function vanishing at $m \to \infty$ faster than $P_n$ increases.

We have chosen the following BG parametrization:
\begin{equation}
BG = a_1 \Delta^{a_2} e^{-(a_3\Delta + a_4\Delta^2)},
\end{equation}
where $\Delta = m-M_{th}$.

For a resonance signal the standard relativistic Breit-Wigner (BW) function~\cite{BW} is used:
\begin{equation}
BW(m) = 
\frac{\Gamma}{(m^2 - M_0^2)^2 + M_0^2\Gamma^2} \left( \frac{m}{q}
\right), \ \ \ 
\mbox{with} \ \ \ \Gamma = \Gamma_0 \left( \frac{q}{q_0} \right) \frac{M_0}{m}
\label{eq:BW}
\end{equation}
where $M_0$, $\Gamma_0$ are the resonance mass and width,
respectively, and $q$ 
is the momentum of the decay product in the resonance rest frame
($q_0$ corresponds to $M_0$).

Finally, 
we have fitted the invariant mass distributions by: 
\begin{equation}
\frac{dN}{dm} = BG(\Delta) + a_5 BW^\prime(m),
\label{fit_new}
\end{equation}
for all combinations except ($\Lambda$ $\pi^-$), where two peaks due to
$\Sigma^{*-} \to \Lambda \pi^-$ and $\Xi^- \to \Lambda \pi^-$ decays 
are expected.
Here $BW^\prime(m)$ is the Breit-Wigner function of equation (\ref{eq:BW})
normalized to unity.
Such a fit is valid in all cases when the experimental mass resolution
is small compared with the natural width of the resonance.

Similarly, for the ($\Lambda$ $\pi^-$) case we have used
\begin{equation}
\frac{dN}{dm} = BG(\Delta) + a_5 BW^\prime_{\Sigma^{*-}}(m) +
a_6 BW^\prime_{\Xi^{-}}(m) 
\label{fit_xi_new}
\end{equation}
where the invariant mass resolution is used for the width $\Gamma_0$
in the Breit-Wigner function corresponding to the $\Xi^-$ decay. 
In the above formulae $a_1$ to $a_6$ are parameters of the fit.

In such 
an approach
using the HESSE and MINOS procedures of MINUIT~\cite{MINUIT}, 
the parameter $a_5$($a_6$) gives the number of signal events 
with the corresponding error which takes into account possible correlations
between different parameters.

As a consistency check we have also tried an alternative approach
which was to fit the invariant mass distributions with:
\begin{equation}
\frac{dN}{dm} = (1 + a_5 BW(m)) BG(\Delta),
\label{fit}
\end{equation}
and, similarly, with
\begin{equation}
\frac{dN}{dm} = (1 + a_5 BW_{\Sigma^{*-}}(m) +
a_6 BW_{\Xi^{-}}(m) ) BG(\Delta),
\label{fit_xi}
\end{equation}
and extracted the corresponding number of signal events. 

The results obtained using these two approaches were found to be similar.
In what follows we present our results using the first method.

\subsection{Results}

The yields of resonances and heavy hyperons have been studied in different 
kinematic regions and for neutrino interactions on different target
nucleons.

In NOMAD it is to some extent possible to separate neutrino
interactions on the neutrons 
and protons by imposing a cut on the sum of charges ($Q_{tot}$) 
of all the outgoing tracks at the primary neutrino interaction vertex.

We select $\nu p$ events requiring $Q_{tot} \ge 1$. According to the MC
simulation, in this proton-like sample 76\% of the events
are true $\nu p$ interactions.
The $\nu n$ events are selected by the requirement $Q_{tot} \le 0$. The 
purity of the corresponding neutron-like sample is about 85\%.

Since the hadron production mechanisms 
in the target and in the current fragmentation regions 
are expected to be different, it is
important to study separately the yields of resonances 
and heavy hyperons at $x_F<0$ and $x_F>0$. Such a study is 
also
necessary
for a correct theoretical interpretation of the $\lam$ ($\alam$)
polarization measurements reported in our previous
papers~\cite{NOMAD_lam,NOMAD_alam}.

In the following we denote as MC(pred.) the true number of heavy strange
particles 
reconstructed in
the MC, and MC(meas.) the number of heavy strange particles
extracted from the MC sample using our fitting procedure. Both quantities are 
normalized to the number of $\nu_\mu$ CC events in the data.

Note that MC(pred.) and MC(meas.) can be slightly different 
due to limitations of the signal extraction procedure
described in section~\ref{sec:signal}. The threshold and smearing
effects in the invariant mass distributions are at the origin of this discrepancy.
The ratio MC(pred.)/MC(meas.) will therefore be used to
correct the yields of heavy strange particles extracted from the data
(see section~\ref{sec:resonance_yields}).

The following resonances and heavier hyperons have been studied in the
present analysis.

\subsubsection{$\rm {K^\star}^\pm$}

\begin{figure}[htb]
\center{%
\begin{tabular}{cc}
\mbox{\epsfig{file=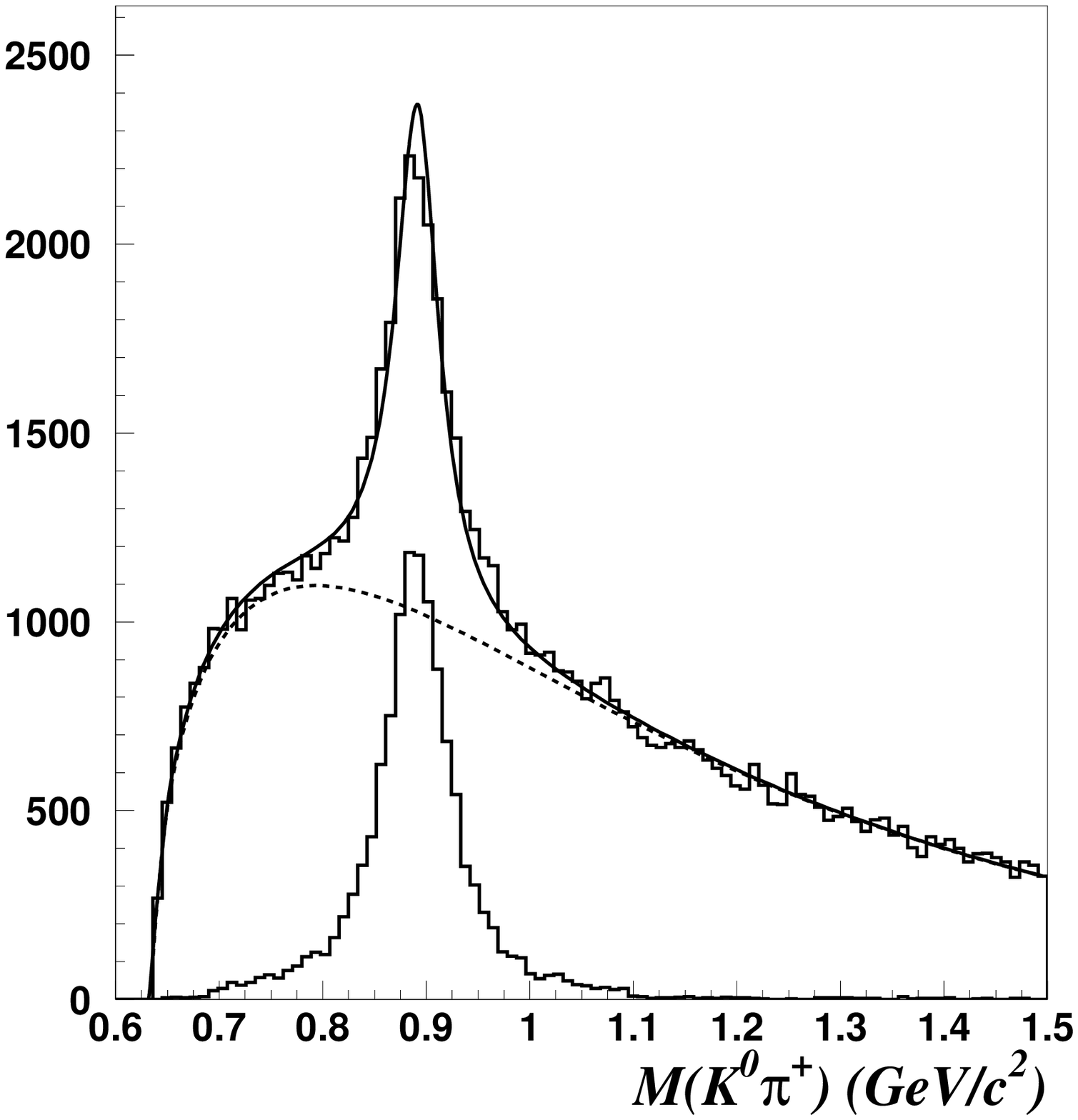,width=0.49\linewidth}}&
\mbox{\epsfig{file=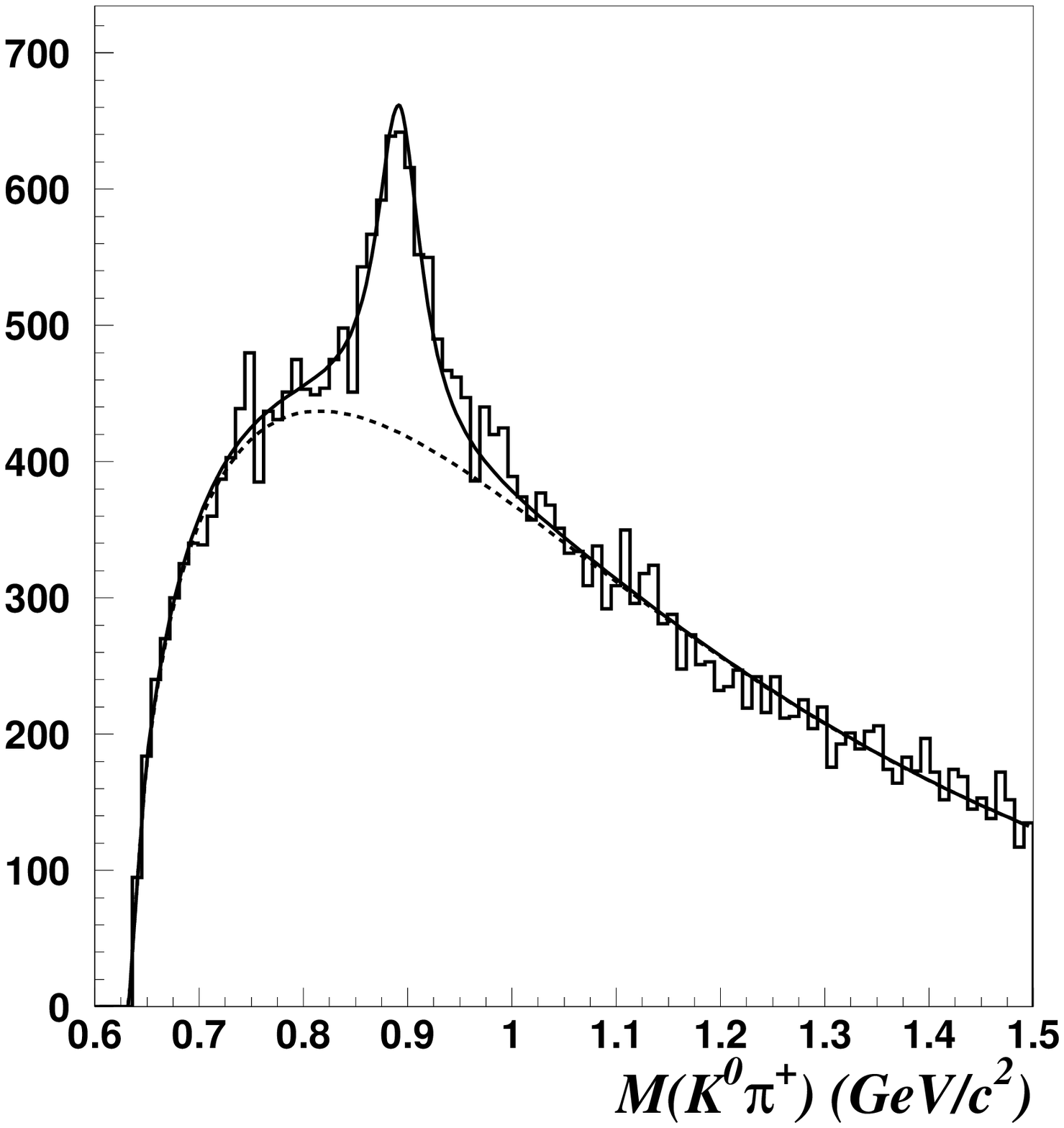,width=0.49\linewidth}}\\
\mbox{\epsfig{file=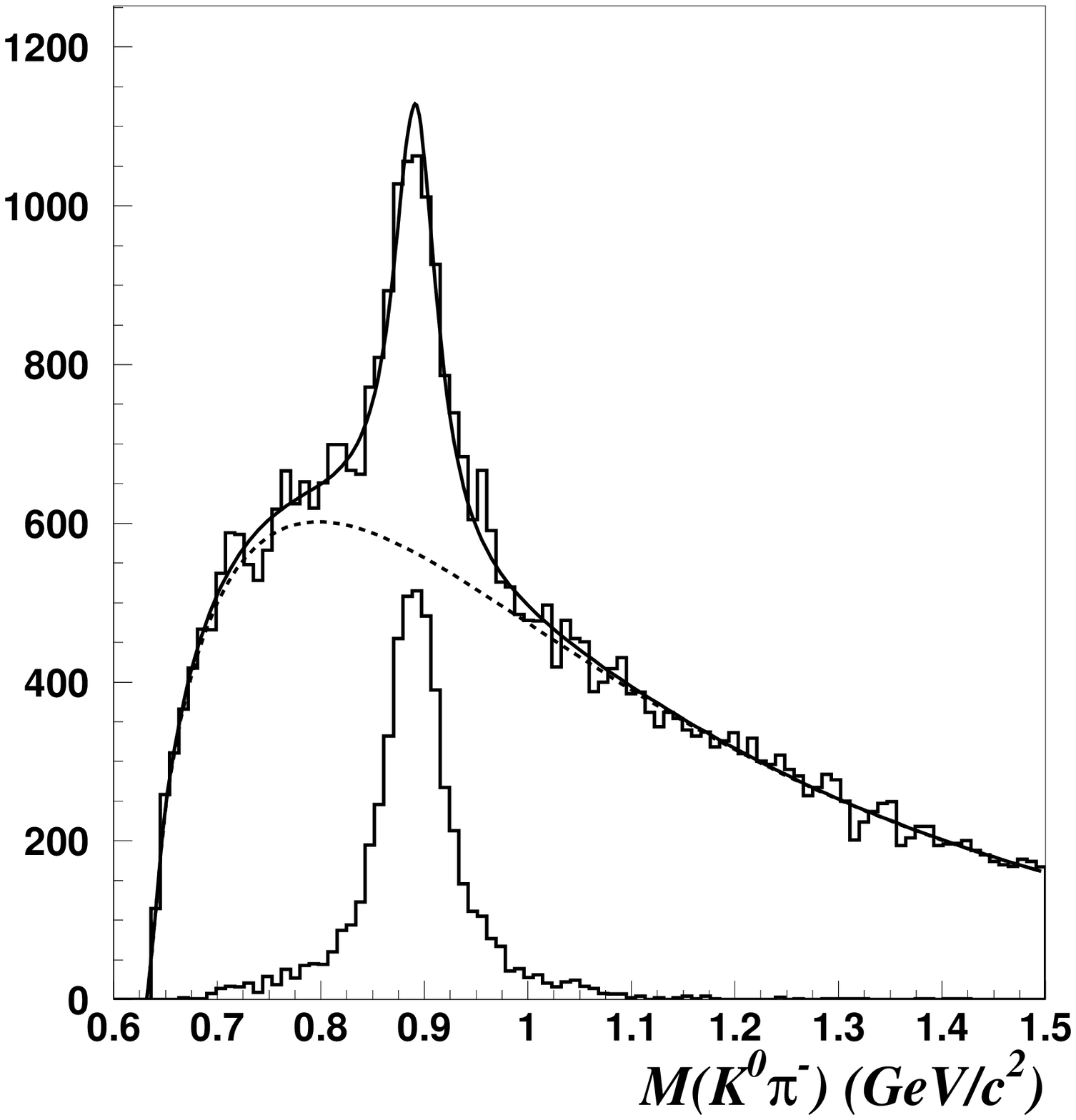,width=0.49\linewidth}}&
\mbox{\epsfig{file=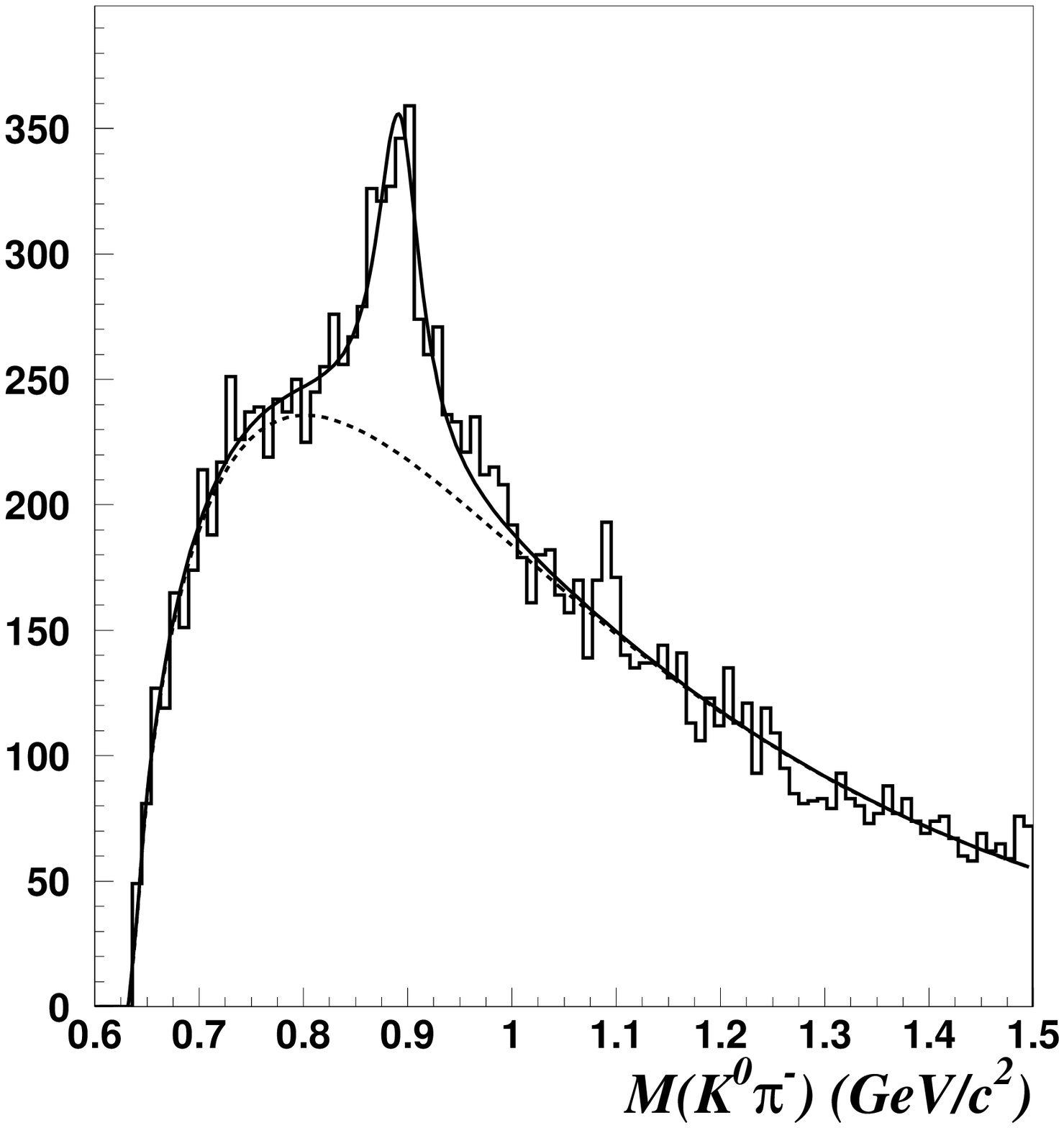,width=0.49\linewidth}}\\
\end{tabular}
}
\protect\caption{\it $\rm \ko \pi^+$ (top) and $\rm \ko \pi^-$ (bottom) 
invariant mass distributions for both MC (left) and data (right).
The solid lines are the results of the fit, while the dotted lines
describe the background term. In the MC plots the additional
histograms refer to the reconstructed true heavy strange particles.
}
\label{fig:kstar}
\end{figure}

The fitted ($\ko$ $\rm \pi^\pm$) invariant mass distributions are shown in 
Fig.~\ref{fig:kstar} for both MC and data samples. 
Detailed information
on the 
number of extracted
$\rm {K^\star}^\pm$ 
events
and the $\rm {K^\star}^\pm/\ko$ ratio 
is given in Tables~\ref{tab:k*+} and~\ref{tab:k*-}.
For the $\rm {K^\star}^\pm$ mass and width we have used 891.66 MeV 
and 50.8 MeV respectively. The $q_0$ value is 291 MeV/c~\cite{PDG}. 

%
%
\begin{table}[htb]
\begin{center}
\caption{\it $\rm {K^\star}^{+} \to \ko \pi^+$ summary}
\begin{tabular}{||c|c|c|c|c|c||}
\hline
\hline
$\rm N({K^\star}^+)$&full sample &\multicolumn{2}{c|}{$\ko$ fragmentation region} &
\multicolumn{2}{c||}{type of target nucleon}\\
\hline
&                          &$x_F<0$       &$x_F>0$      &$\nu p$       &$\nu n$ \\
\hline
DATA      & $2036 \pm 121$ &$315 \pm 56$  &$1731 \pm 108$&$1006 \pm 87$ &$1032 \pm 84$\\
MC(meas.) & $5373 \pm 104$ &$726 \pm 47$  &$4744 \pm 93$ &$1963 \pm 67$ &$3516 \pm 80$\\
MC(pred.) & $5953$         &$886$         &$5067$        &$2206$        &$3748$ \\

\hline
\hline
{\small $\rm N({K^\star}^+)/N(\ko)$} &\multicolumn{5}{c||}{uncorrected}  \\
\hline
DATA (\%)     & $13.5 \pm 0.8$&$9.7 \pm 1.7$&$14.6 \pm 0.9$&$15.7 \pm 1.4$&$11.9 \pm 1.0$\\
MC (\%)       & $27.3 \pm 0.5$&$14.5\pm 0.9$&$31.6 \pm 0.6$&$29.0 \pm 1.0$&$26.5 \pm 0.6$\\
\hline
\hline
\end{tabular}

\label{tab:k*+}
\end{center}
\end{table}

%
%
\begin{table}[htb]
\begin{center}
\caption{\it $\rm {K^\star}^{-} \to \ko \pi^-$ summary}
\begin{tabular}{||c|c|c|c|c|c||}
\hline
\hline
$\rm N({K^\star}^-)$&full sample &\multicolumn{2}{c|}{$\ko$ fragmentation region} &
\multicolumn{2}{c||}{type of target nucleon}\\
\hline
&                          &$x_F<0$       &$x_F>0$       &$\nu p$       &$\nu n$ \\
\hline
DATA      & $1146 \pm 89$ &$288 \pm 44  $&$865  \pm 78 $&$377  \pm 52 $&$775  \pm 73 $\\
MC(meas.) & $2304 \pm 74$  &$639 \pm 38$  &$1664 \pm 63$ &$729 \pm 39$ & $1576 \pm 63$\\
MC(pred.) & $2467$         &$723$         &$1743$        &$734$        & $1733$ \\
\hline
\hline
{\small $\rm N({K^\star}^-)/N(\ko)$} &\multicolumn{5}{c||}{uncorrected}  \\
\hline
DATA (\%)     & $7.6  \pm 0.6$&$8.9  \pm 1.3$&$7.3  \pm 0.7$&$5.9  \pm 0.8$&$8.9  \pm 0.8$\\
MC (\%)       & $11.5 \pm 0.4$&$12.7 \pm 0.8$&$11.1 \pm 0.4$&$10.8 \pm 0.6$&$11.9 \pm 0.5$\\
\hline
\hline
\end{tabular}
\label{tab:k*-}
\end{center}
\end{table}

It is interesting to note a more abundant ${K^\star}^+$ than ${K^\star}^-$ 
production in $\nu_\mu$ CC DIS.
This can be explained by the 
fact that the outgoing $u$ quark can fragment directly 
into a ${K^\star}^+$, while both
$\bar u$ and $s$ quarks needed to produce a ${K^\star}^-$ meson have to
be created in the fragmentation process.
 
One can see that there is a significant difference between $\rm {K^\star}^{\pm}$
yields in the default MC simulation and the NOMAD data (about a factor of 2).

\subsubsection{$\rm {\Sigma^\star}^\pm$}

For the $\rm {\Sigma^\star}^\pm$ mass and width we have taken the values 
from~\cite{PDG}: $\rm m({\Sigma^\star}^+)$=1382.8 MeV, 
$\rm \Gamma({\Sigma^\star}^+)$=35.8 MeV, $\rm m({\Sigma^\star}^-)$=1387.2 MeV, 
$\rm \Gamma({\Sigma^\star}^-)$=39.4 MeV. The $q_0$ value is 208 MeV/c.

\begin{figure}[htb]
\center{%
\begin{tabular}{cc}
\mbox{\epsfig{file=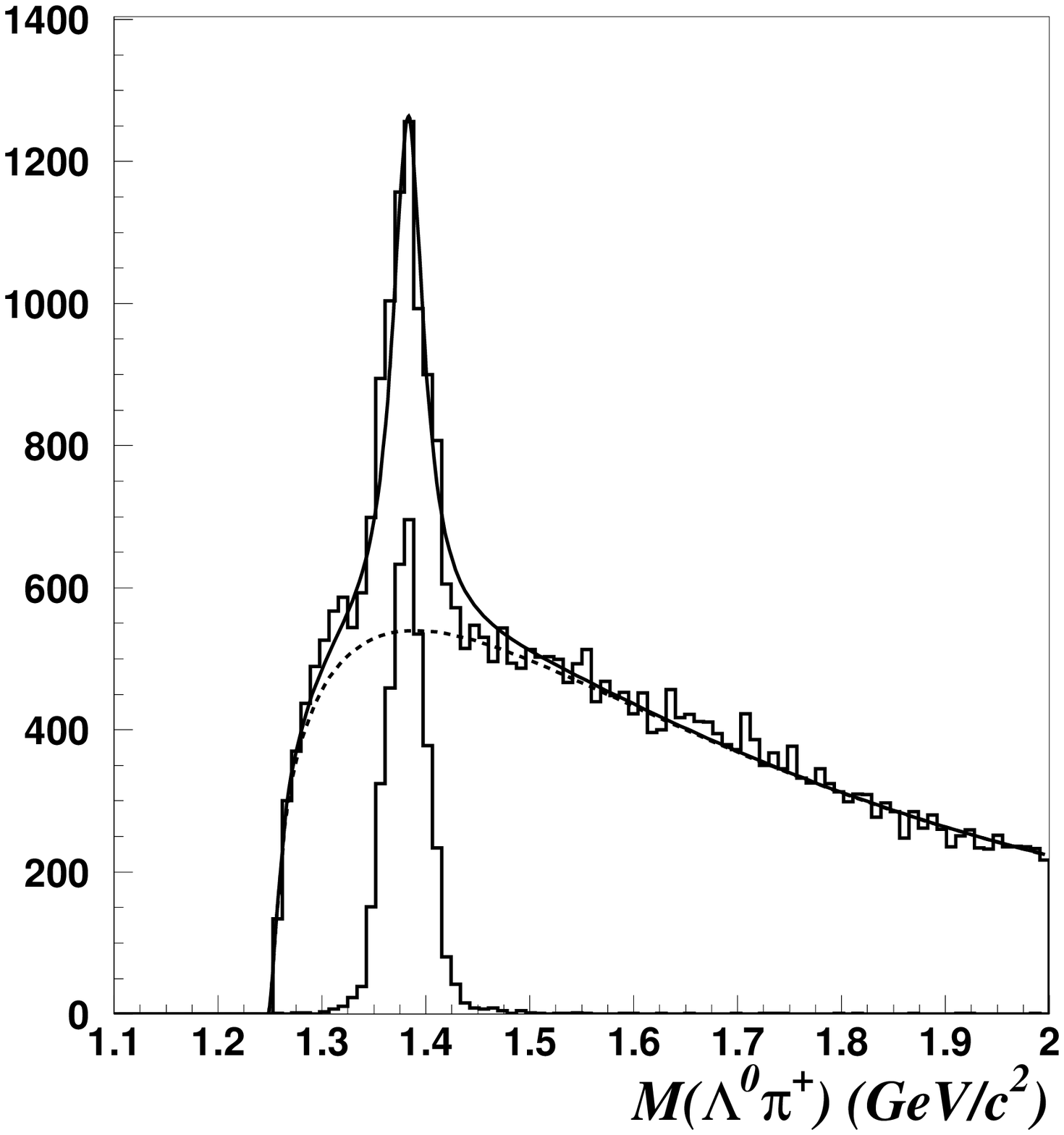,width=0.49\linewidth}}&
\mbox{\epsfig{file=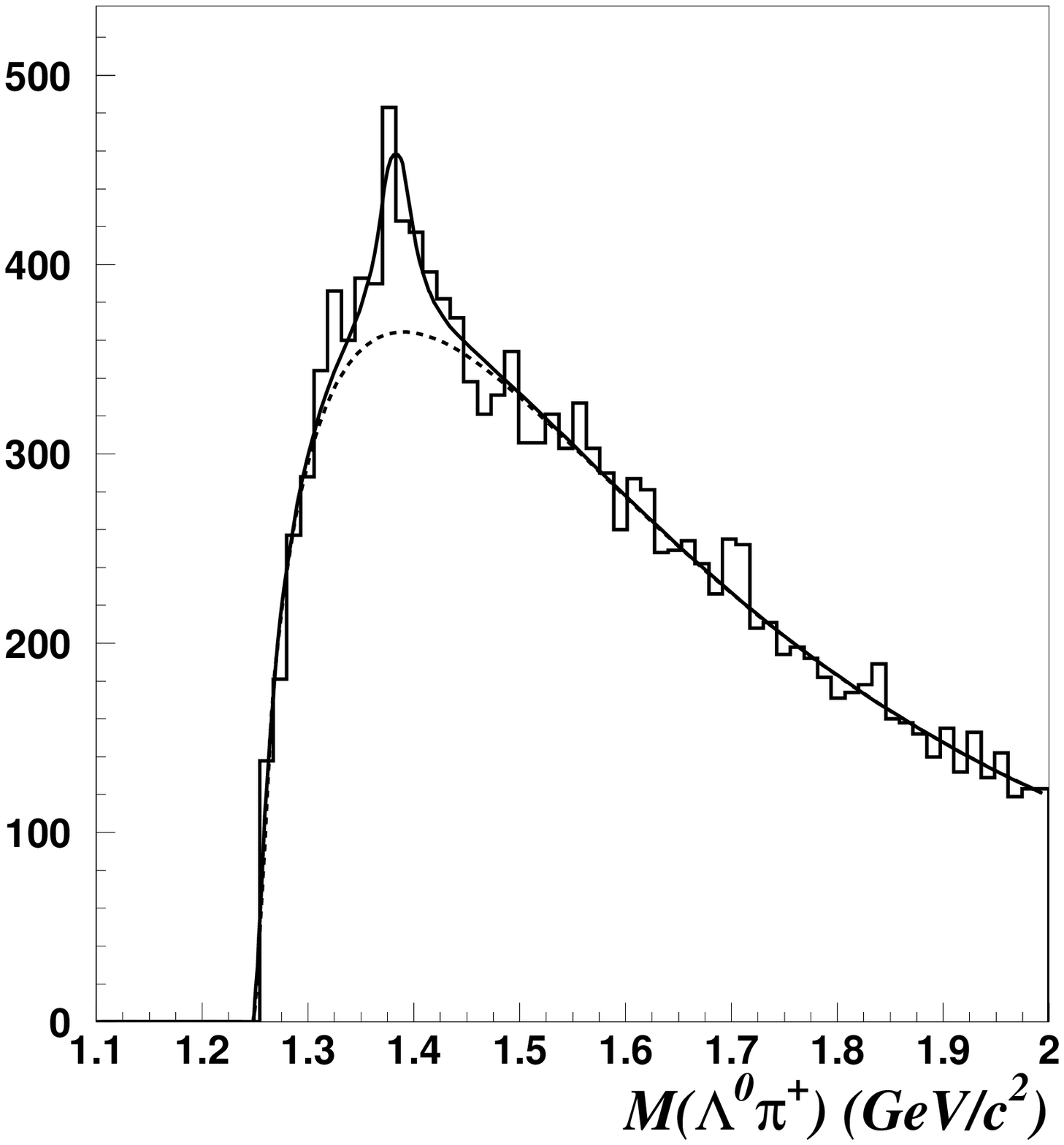,width=0.49\linewidth}}\\
\mbox{\epsfig{file=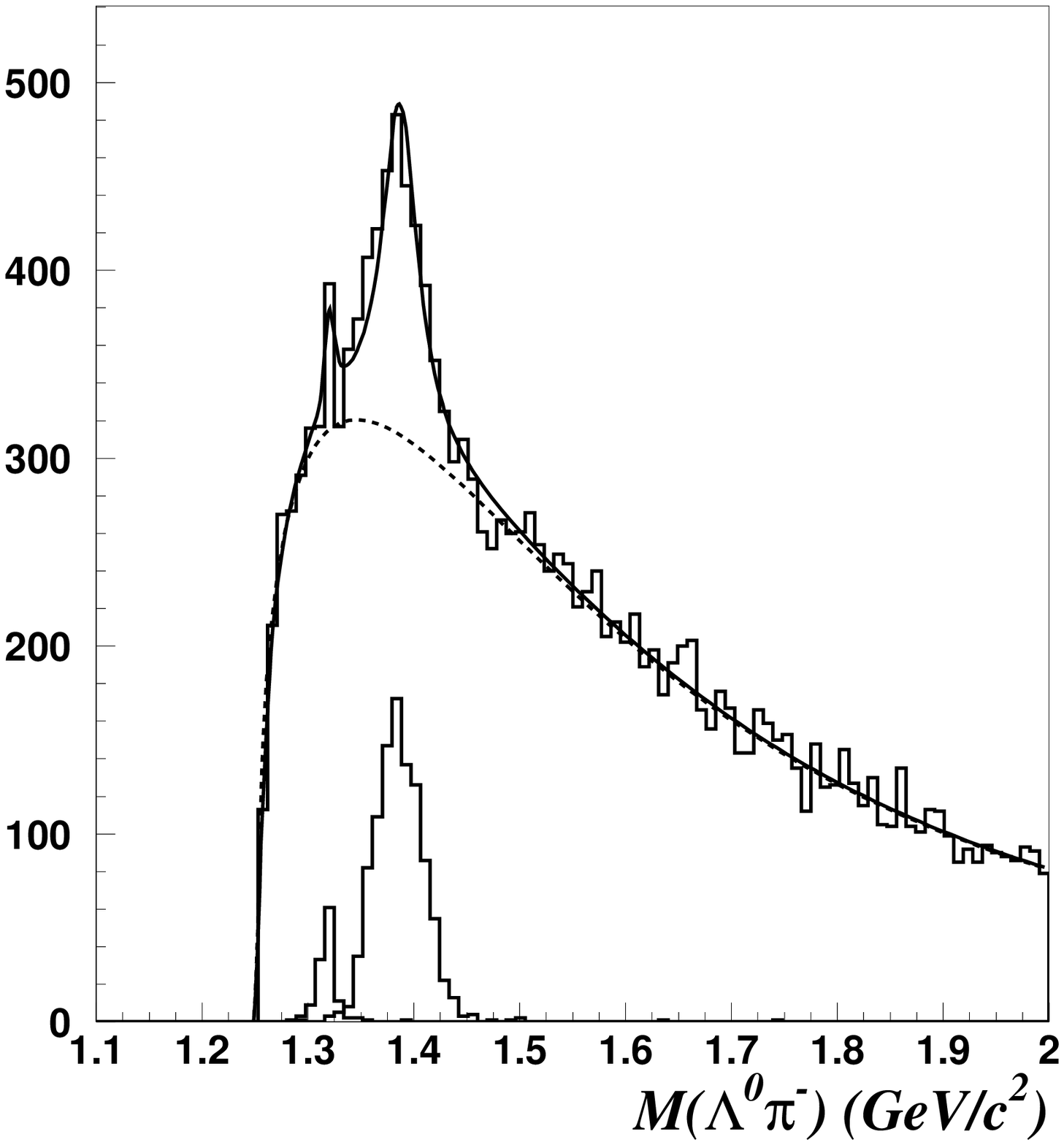,width=0.49\linewidth}}&
\mbox{\epsfig{file=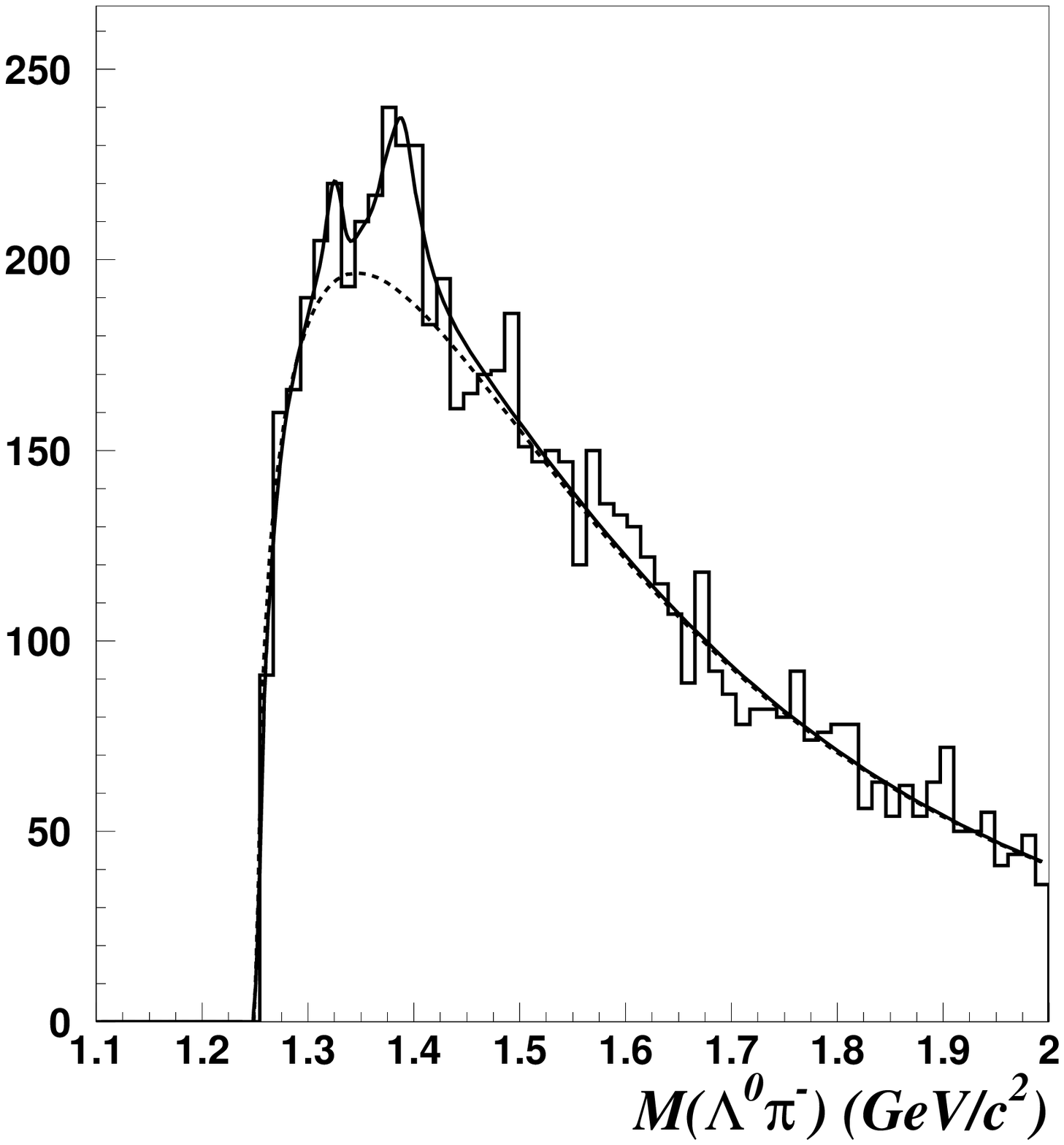,width=0.49\linewidth}}\\
\end{tabular}
}
\protect\caption{\it $\rm \lam \pi^+$ (top) and $\rm \lam \pi^-$ (bottom) 
invariant mass distributions for both MC (left) and data (right).
The solid lines are the results of the fit, while the dotted lines
describe the background term. In the MC plots the additional
histograms refer to the reconstructed true heavy strange particles.
}
\label{fig:sigmastar}
\end{figure}

The fitted invariant mass distributions for ($\lam$ $\rm \pi^\pm$) combinations 
in both MC and data samples are shown in Fig.~\ref{fig:sigmastar}. 
Detailed information on 
the number of extracted $\rm {\Sigma^\star}^\pm$ 
events and the $\rm {\Sigma^\star}^\pm/\lam$ ratio is given in 
Tables~\ref{tab:sigma*+} and~\ref{tab:sigma*-}.

%
%
\begin{table}[htb]
\begin{center}
\caption{\it $\rm {\Sigma^\star}^{+} \to \lam \pi^+$ summary}
\begin{tabular}{||c|c|c|c|c|c||}
\hline
\hline
$\rm N({\Sigma^\star}^+)$&full sample &\multicolumn{2}{c|}{$\lam$ fragmentation region} &
\multicolumn{2}{c||}{type of target nucleon}\\
\hline
&                          &$x_F<0$       &$x_F>0$      &$\nu p$       &$\nu n$ \\
\hline
DATA      & $416 \pm 80$   &$358 \pm 65$  &$63 \pm 47$  &$297 \pm 61$  &$120 \pm 51$  \\
MC(meas.) & $2070 \pm 68$ &$1427 \pm 57$  &$649 \pm 37$ &$1321 \pm 49$ &$754 \pm 46$\\
MC(pred.) & $1783$        &$1254$         &$529$        &$1150$        &$634$ \\
\hline
\hline
$\rm N({\Sigma^\star}^+)/N(\lam)$&\multicolumn{5}{c||}{uncorrected}  \\
\hline
\hline
DATA (\%)     & $5.2 \pm 1.0 $ &$6.4 \pm 1.2 $&$2.5  \pm 1.9$&$8.6 \pm 1.8 $&$2.6 \pm 1.1$\\
MC  (\%)      & $17.0 \pm 0.6$ &$15.9 \pm 0.6$&$20.6 \pm 1.2$&$32.6 \pm 1.2$&$9.3 \pm 0.6$\\
\hline
\hline
\end{tabular}

\label{tab:sigma*+}
\end{center}
\end{table}

%
%
\begin{table}[htb]
\begin{center}
\caption{\it $\rm {\Sigma^\star}^{-} \to \lam \pi^-$ summary}
\begin{tabular}{||c|c|c|c|c|c||}
\hline
\hline
$\rm N({\Sigma^\star}^-)$&full sample &\multicolumn{2}{c|}{$\lam$ fragmentation region} &
\multicolumn{2}{c||}{type of target nucleon}\\
\hline
&                          &$x_F<0$       &$x_F>0$      &$\nu p$       &$\nu n$ \\
\hline
DATA      & $206 \pm 63$   &$121 \pm 51$  &$93 \pm 37$  &$100 \pm 35$&$111 \pm 52$  \\
MC(meas.) & $551 \pm 48$   &$410 \pm 42$  &$145 \pm 25$ &$18 \pm 22$ &$528 \pm 43$\\
MC(pred.) & $489$          &$362$         &$126$        &$33$        &$456$ \\
\hline
\hline
$\rm N({\Sigma^\star}^-)/N(\lam)$&\multicolumn{5}{c||}{uncorrected}  \\
\hline
\hline
DATA (\%)     & $2.6 \pm 0.8$&$2.2 \pm 0.9$&$3.7 \pm 1.5$&$2.9 \pm 1.0$&$2.4 \pm 1.1$\\
MC (\%)       & $4.5 \pm 0.4$&$4.6 \pm 0.5$&$4.6 \pm 0.8$&$0.4 \pm 0.5$&$6.5 \pm 0.5$\\
\hline
\hline
\end{tabular}
\label{tab:sigma*-}
\end{center}
\end{table}

A striking difference between the $\rm {\Sigma^\star}^{\pm}$ yields in 
the default MC simulation and in the NOMAD data (a factor of about 3 
or even larger) is observed.

\subsubsection{$\rm \Xi^-$}

For the $\rm {\Xi}^-$ mass we have used 1321.32 MeV~\cite{PDG}, 
for the width the experimental resolution of 10 MeV has been taken.  
The $q_0$ value is 139 MeV/c. 

The bottom plots in Fig.~\ref{fig:sigmastar} show evidence for
$\rm \Xi^- \to \lam \pi^-$ decays.  
Detailed information on the number of extracted $\rm \Xi^-$ 
events and the $\rm \Xi^-/\lam$ ratio
is given in Table~\ref{tab:xi-}.

%
%
\begin{table}[htb]
\begin{center}
\caption{\it $\rm {\Xi}^{-} \to \lam \pi^-$ summary}
\begin{tabular}{||c|c|c|c|c|c||}
\hline
\hline
$\rm N(\Xi^-)$&full sample &\multicolumn{2}{c|}{$\lam$ fragmentation region} &
\multicolumn{2}{c||}{type of target nucleon}\\
\hline
&                         &$x_F<0$      &$x_F>0$      &$\nu p$       &$\nu n$ \\
\hline
DATA      & $42 \pm 30$   &$21 \pm 24$  &$18 \pm 17$ &$54 \pm 18$&$-11 \pm 24$  \\
MC(meas.) & $43 \pm 18$   &$33 \pm 15$  &$13 \pm 9$  &$9 \pm 8$  &$36 \pm 16$\\
MC(pred.) & $60$          &$47$         &$15$        &$14$       &$47$ \\
\hline
\hline
$\rm N(\Xi^-)/N(\lam)$&\multicolumn{5}{c||}{uncorrected}  \\
\hline
DATA (\%)     & $0.5 \pm 0.4$&$0.4 \pm 0.4$&$0.7 \pm 0.7$&$1.6 \pm 0.5$&$-0.2   \pm 0.5$  \\
MC  (\%)      & $0.4 \pm 0.2$&$0.4 \pm 0.2$&$0.4 \pm 0.3$&$0.2 \pm 0.2$&$0.4 \pm 0.2$\\
\hline
\hline
\end{tabular}

\label{tab:xi-}
\end{center}
\end{table}

\subsubsection{$\rm {\Sigma^0}$}

For the $\rm {\Sigma^0}$ mass we have taken the value
from~\cite{PDG}: $\rm m({\Sigma^0})$=1192.6 MeV, while
for the width the experimental resolution of 9 MeV has been used.  
The $q_0$ value is 74 MeV/c. 
The $\rm {\Sigma^0}$ peak has been fitted by a Gaussian function.

\begin{figure}[htb]
\center{%
\begin{tabular}{cc}
\mbox{\epsfig{file=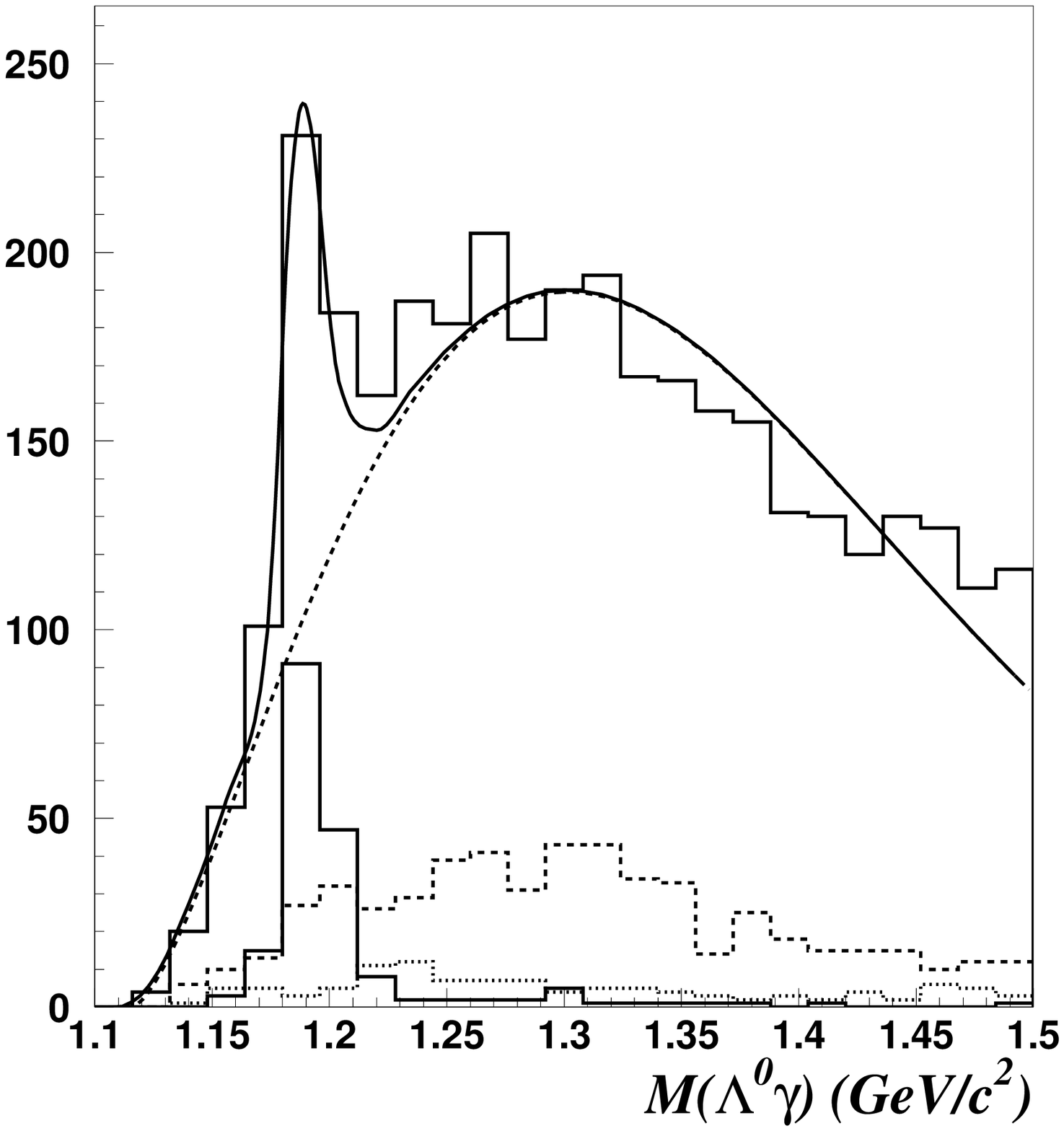,width=0.49\linewidth}}&
\mbox{\epsfig{file=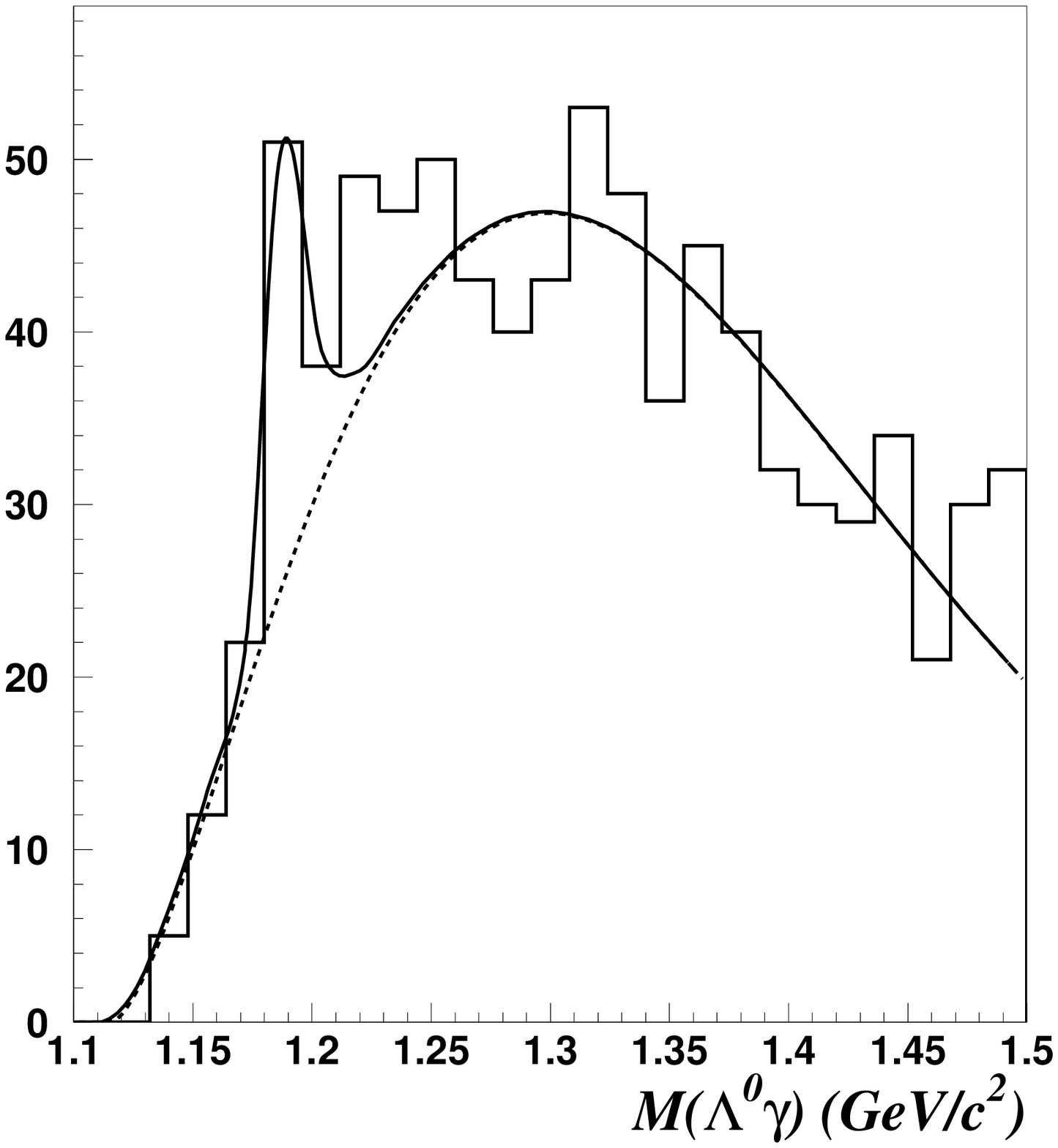,width=0.49\linewidth}}\\
\end{tabular}
}
\protect\caption{\it $\lam \gam$ invariant mass distributions 
for both MC (left) and data (right). The MC plot shows the expected signal
peak and 
background contributions
from $\rm \Xi^0 \rightarrow \lam \pi^0$ 
and $\rm \Sigma^{\star 0} \rightarrow \lam \pi^0$ decays
with only one reconstructed photon.}
\label{fig:sigma0}
\end{figure}

\begin{figure}[htb]
\center{%
\begin{tabular}{cc}
\mbox{\epsfig{file=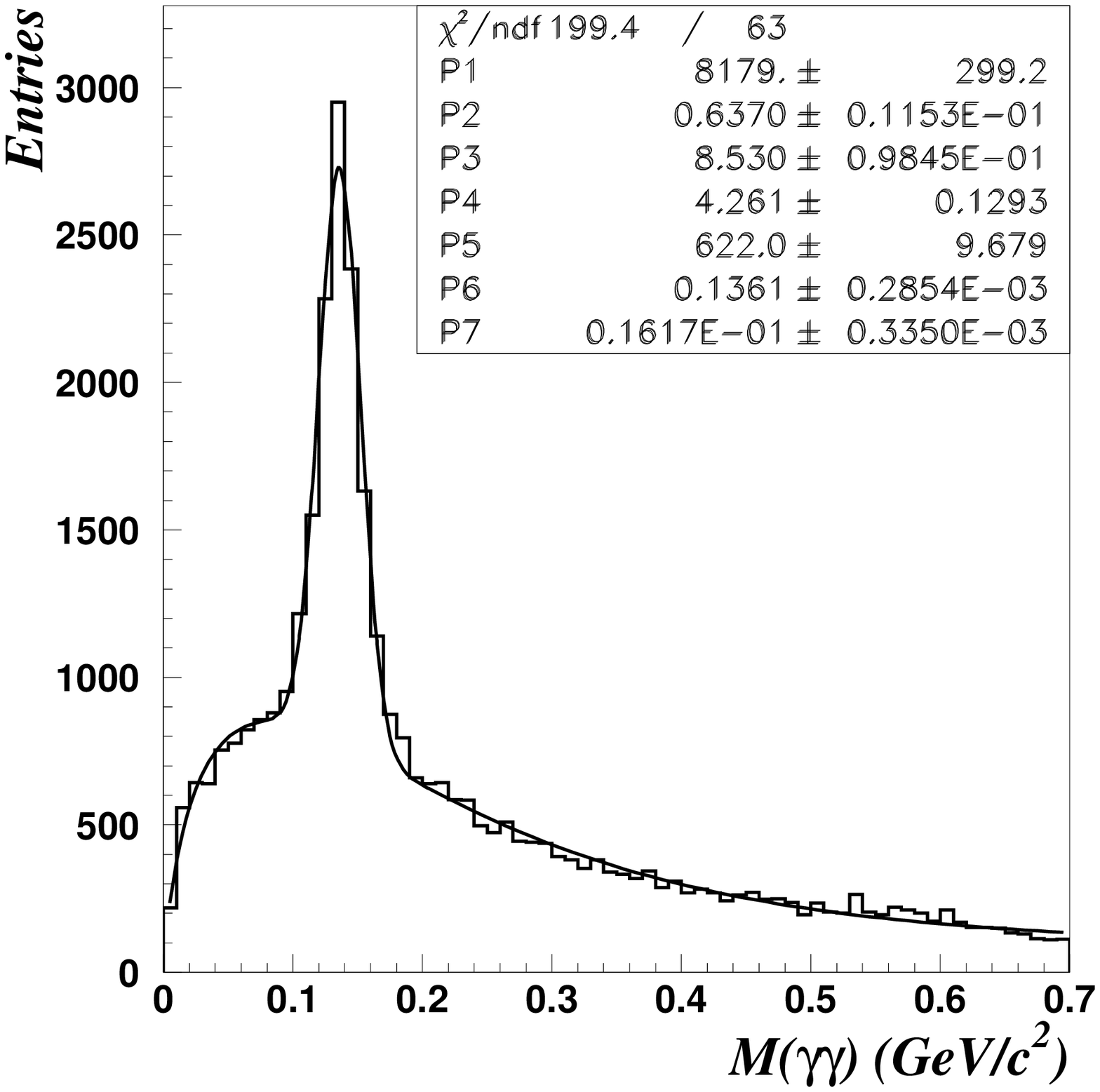,width=0.49\linewidth}}&
\mbox{\epsfig{file=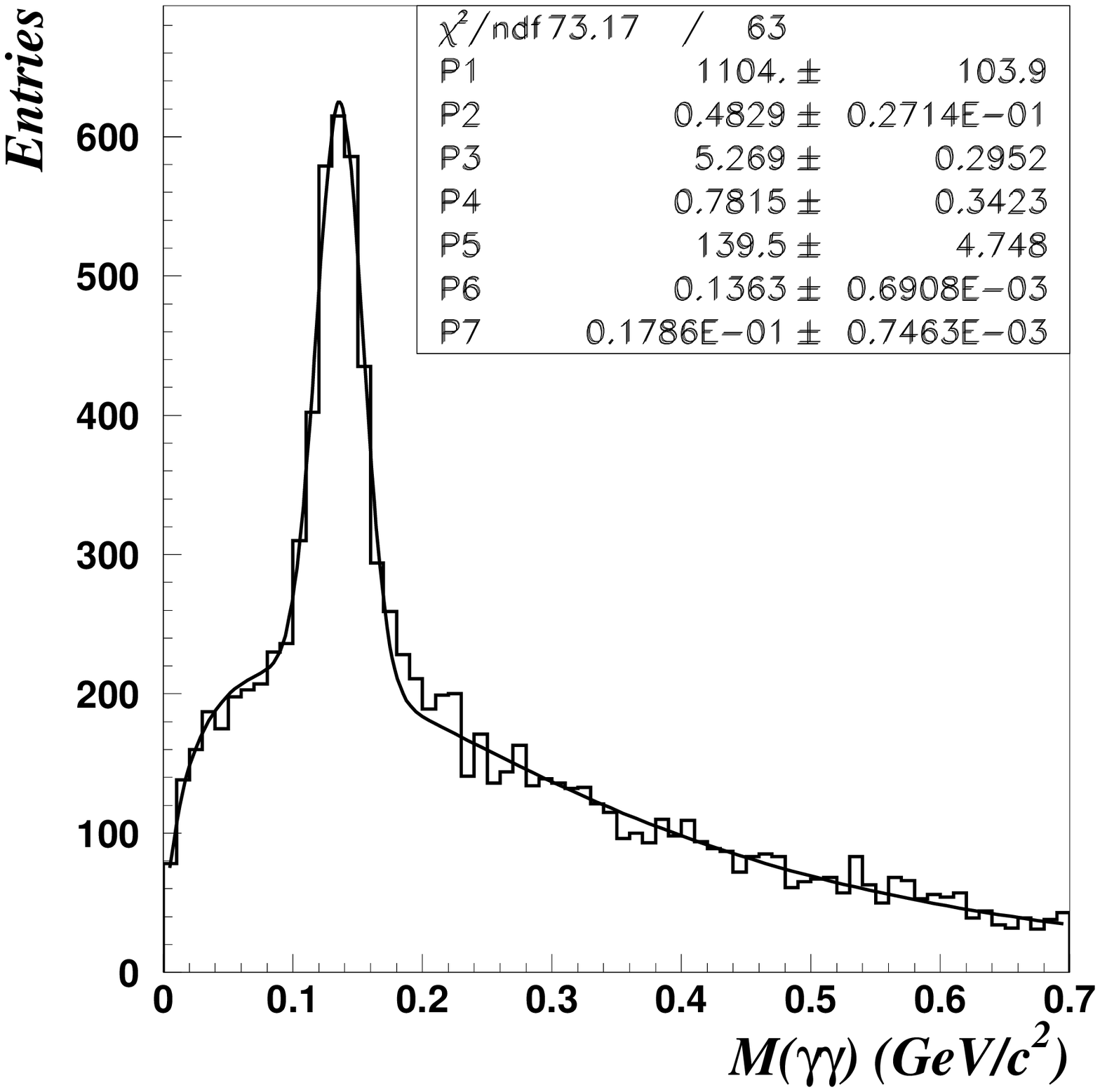,width=0.49\linewidth}}\\
\end{tabular}
}
\protect\caption{\it $\gam \gam$ invariant mass distributions 
for MC (left) and data (right). Both photons have been reconstructed 
as conversions in the DC fiducial volume and identified by our $\vo$ 
identification procedure. A clear peak corresponding to the $\rm \pi^0$ signal 
is visible in both distributions. The parameters $P_6$ and $P_7$ show 
the mass and the width of the Gaussian function after the fit.
}
\label{fig:pi0}
\end{figure}

Fig.~\ref{fig:sigma0} shows the 
fitted invariant mass distributions for ($\lam$ $\gam$) combinations 
in both Monte Carlo and data samples. The corresponding 
photons have been reconstructed 
as conversions in the DC fiducial volume and identified by our $\vo$ 
identification procedure. The quality of the photon
reconstruction is illustrated by the ($\gam \gam$) invariant mass distributions
shown in Fig.~\ref{fig:pi0} for both MC and data:
a peak corresponding to the $\rm \pi^0$ signal is evident.

A summary of the number of extracted $\rm {\Sigma^0}$ 
events and the $\rm {\Sigma^0}/\lam$ ratio is given in 
Table~\ref{tab:sigma0}.

\begin{table}[htb]
\begin{center}
\caption{\it $\rm \Sigma^0 \to \lam \gam$ summary}
\begin{tabular}{||c|c|c|c|c|c||}
\hline
\hline
$\rm N({\Sigma}^0)$&full sample &\multicolumn{2}{c|}{$\lam$ fragmentation region} &
\multicolumn{2}{c||}{type of target nucleon}\\
\hline
&                          &$x_F<0$       &$x_F>0$      &$\nu p$       &$\nu n$ \\
\hline
DATA      & $29 \pm 10$    &$17 \pm 9$    &$16 \pm 7$   &$16 \pm 7$   &$13 \pm 7$  \\
MC(meas.) & $82 \pm 12$    &$50 \pm 9$    &$37 \pm 8$    &$19 \pm 7$    & $61 \pm 10$\\
MC(pred.) & $80$           &$57$          &$22$          &$22$         &$57$ \\
\hline
\hline
$\rm N(\Sigma^0)/N(\lam)$&\multicolumn{5}{c||}{uncorrected}  \\
\hline
\hline
DATA (\%)     & $0.4 \pm 0.1$ &$0.3 \pm 0.2$&$0.7 \pm 0.3$&$0.5 \pm 0.2$&$0.3 \pm 0.2$\\
MC  (\%)      & $0.7 \pm 0.1$ &$0.6 \pm 0.1$&$1.2 \pm 0.3$&$0.5 \pm 0.2$&$0.8 \pm 0.1$\\
\hline
\hline
\end{tabular}

\label{tab:sigma0}
\end{center}
\end{table}

\subsection{Yields of strange resonances and heavy hyperons} \label{sec:resonance_yields}

\begin{table}[htb]
\caption{\label{tab:reson_yields}\it 
Corrected fractions (in \%) of observed $\ko$ and $\lam$ decays
that originate from the decays of 
strange resonances and heavy hyperons in the NOMAD data  
compared to the default MC predictions.
}
\begin{tabular}{||c|c|c|c|c|c|c||}
\hline\hline
& $\rm {K^\star}^+ \rightarrow \ko \pi^+$ & $\rm {K^\star}^- \rightarrow \ko \pi^-$ & 
$\rm {\Sigma^\star}^+ \rightarrow \lam \pi^+$ & $\rm {\Sigma^\star}^- \rightarrow \lam \pi^-$ & 
$\rm {\Sigma}^0 \rightarrow \lam \gamma$ & $\rm {\Xi}^- \rightarrow \lam \pi^-$    \\
\hline
DATA & $15.5 \pm 0.9$& $8.7 \pm 0.7$ & $5.8 \pm 1.1$      & $2.6 \pm 0.8$   & $7.3 \pm 2.4$ & $1.9 \pm 1.7$\\
MC & $31.4$ & $13.1$ & $16.6$ & $3.9$ & $12.7$ & $1.5$\\
\hline\hline
\end{tabular}
\end{table}

The integral yields of strange resonances and heavy hyperons produced in
$\nu_\mu$ CC are computed multiplying the results of the fits by the
ratio MC(pred.)/MC(meas.).

The results are presented in Table~\ref{tab:reson_yields} as fractions
of $\vo$ produced by heavy strange particles and resonances.

\subsection{Discussion}
 
The results of our study confirm discrepancies reported earlier~\cite{allasia}
in the description of the 
strange resonances and heavy hyperon
production in neutrino interactions
by the LUND model with default parameters~\cite{LEPTO,JETSET}.
These results could be potentially used to tune the MC parameters
responsible for the fragmentation into strange particles.
Moreover, an additional analysis of events with multiple production
of neutral strange particles could be very useful in this respect.
Such an analysis 
is currently in progress.

\section{CONCLUSION\label{sec:conclusion}}

We have reported the results of a study of strange particle production in
$\nu_\mu$ CC interactions using the data from the NOMAD experiment. 
Our analysis is based on a sample of 
$\nu_\mu$ CC events containing 15074 identified $\ko$, 8087 identified $\lam$
and 649 identified $\alam$ decays. This $\vo$ sample represents at
least a factor of
5 increase in statistics compared to previous (anti)neutrino
experiments performed with bubble chambers.
Yields
of neutral strange particles 
($\ko$, $\lam$, $\alam$) have been measured in this analysis
as a function of kinematic variables. 
For $\alam$ production such measurements 
are performed for the first
time in a neutrino experiment. 
The decays of resonances and heavy hyperons with identified 
$\ko$ and $\lam$
in the final state have been analyzed. 
Clear signals corresponding to $\rm {K^\star}^\pm$, $\rm {\Sigma^\star}^\pm$,
$\rm \Xi^-$ and $\rm \Sigma^0$ have been observed.
This study is potentially interesting for the tuning of Monte 
Carlo simulation programs and is also of special importance for a
quantitative theoretical interpretation of the $\lam$ and $\alam$ 
polarization measurements reported earlier~\cite{NOMAD_lam,NOMAD_alam}.

\vspace{.5 cm}
{\large \bf Acknowledgements}

\vspace{.5cm}
We gratefully acknowledge  the CERN SPS accelerator and beam-line staff
for the magnificent performance of the neutrino beam. 
We also thank the technical and secretarial staff of 
the collaborating institutes.
The experiment was  supported by  the following
funding agencies:
Australian Research Council (ARC) and Department of Industry, Science, and
Resources (DISR), Australia;
Institut National de Physique Nucl\'eaire et Physique des Particules (IN2P3), 
Commissariat \`a l'Energie Atomique (CEA),  France;
Bundesministerium f\"ur Bildung und Forschung (BMBF, contract 05 6DO52), 
Germany; 
Istituto Nazionale di Fisica Nucleare (INFN), Italy;
Joint Institute for Nuclear Research and 
Institute for Nuclear Research of the Russian Academy of Sciences, Russia; 
Fonds National Suisse de la Recherche Scientifique, Switzerland;
Department of Energy, National Science Foundation (grant PHY-9526278), 
the Sloan and the Cottrell Foundations, USA.

\end{document}